%% file: paperCS.tex
\documentclass[12pt,a4paper]{article}
\usepackage{axodraw}
\usepackage{url}
\usepackage{xspace}
\usepackage{graphicx}
\makeatletter
\newif\if@preliminary
\@preliminaryfalse
\def\preliminary{\@preliminarytrue}

\def\preprintno#1{\def\@preprintno{#1}}
\def\address#1{\def\@address{#1}}
\def\email#1#2{\thanks{\tt #1@{}#2}}
\def\abstract#1{\def\@abstract{#1}}
\renewcommand\abstractname{ABSTRACT}
\newlength\preprintnoskip
\newlength\abstractwidth
\@titlepagetrue
\renewcommand\maketitle{\begin{titlepage}%
  \let\footnotesize\small
  \hfill\parbox{\preprintnoskip}{%
  \begin{flushright}\@preprintno\end{flushright}}\hspace*{1cm}
  \vskip 60\p@
  \begin{center}%
    {\Large\bf\boldmath \@title \par}\vskip 1cm%
    {\sc\@author \par}\vskip 3mm%
    {\@address \par}%
    \if@preliminary
      \vskip 2cm {\large\sf PRELIMINARY DRAFT \par \@date}%
    \fi
  \end{center}\par
  \@thanks
  \vfill
  \begin{center}%
    \parbox{\abstractwidth}{\centerline{\abstractname}%
    \vskip 3mm%
    \@abstract}
  \end{center}
  \end{titlepage}%
  \setcounter{footnote}{0}%
  \let\thanks\relax\let\maketitle\relax
  \gdef\@thanks{}\gdef\@author{}\gdef\@address{}%
  \gdef\@title{}\gdef\@abstract{}\gdef\@preprintno{}
}%
\def\@citex[#1]#2{\if@filesw\immediate\write\@auxout{\string\citation{#2}}\fi
  \def\@citea{}\@cite{\@for\@citeb:=#2\do
    {\@citea\def\@citea{,\penalty\@m}\@ifundefined
       {b@\@citeb}{{\bf ?}\@warning
       {Citation `\@citeb' on page \thepage \space undefined}}%
\hbox{\csname b@\@citeb\endcsname}}}{#1}}
\def\citerange{\@ifnextchar [{\@tempswatrue\@citexr}{\@tempswafalse\@citexr[]}}
\def\@citexr[#1]#2{\if@filesw\immediate\write\@auxout{\string\citation{#2}}\fi
  \def\@citea{}\@cite{\@for\@citeb:=#2\do
    {\@citea\def\@citea{--\penalty\@m}\@ifundefined
       {b@\@citeb}{{\bf ?}\@warning
       {Citation `\@citeb' on page \thepage \space undefined}}%
\hbox{\csname b@\@citeb\endcsname}}}{#1}}
\long\def\@makecaption#1#2{%
  \vskip\abovecaptionskip
  \sbox\@tempboxa{#1: \emph{#2}}%
  \ifdim \wd\@tempboxa >\hsize
    #1: \emph{#2}\par
  \else
    \hbox to\hsize{\hfil\box\@tempboxa\hfil}%
  \fi
  \vskip\belowcaptionskip}
\def\fmslash{\@ifnextchar[{\fmsl@sh}{\fmsl@sh[0mu]}}
\def\fmsl@sh[#1]#2{%
  \mathchoice
    {\@fmsl@sh\displaystyle{#1}{#2}}%
    {\@fmsl@sh\textstyle{#1}{#2}}%
    {\@fmsl@sh\scriptstyle{#1}{#2}}%
    {\@fmsl@sh\scriptscriptstyle{#1}{#2}}}
\def\@fmsl@sh#1#2#3{\m@th\ooalign{$\hfil#1\mkern#2/\hfil$\crcr$#1#3$}}
\makeatother

\newcommand{\ie}{i.e.\ }
\newcommand{\eg}{e.g.\ }

\newcommand\as         {\ensuremath{\alpha_{\mathrm{s}}}}

\newcommand{\CF}{C_{\mathrm{F}}}
\newcommand{\Nc}{N_{\mathrm{c}}}
\newcommand{\CA}{C_{\mathrm{A}}}
\newcommand{\TR}{T_{\mathrm{R}}}                                               
\newcommand{\bV}{{\bf V}}
\newcommand{\bT}{{\bf T}}

\renewcommand\d {{\mathrm d}}
\newcommand{\shat}{\hat s}
\newcommand{\that}{\hat t}
\newcommand{\uhat}{\hat u}
\newcommand{\kperp}{k_\perp}
\newcommand{\kperpbf}{{\bf{k}}_\perp}
\newcommand{\kperpbfsq}{{\bf{k}}_{\perp}^{\,\!2}}
\newcommand{\kperpzero}{{\bf{k}}_{\perp,0}}
\newcommand{\kperpmax}{{\bf{k}}_{\perp,{\rm max}}}
\newcommand{\nn}{\nonumber}

\newcommand{\GeV}{{\ensuremath\rm GeV}}

\newcommand{\W}{\mathrm{W}}
\newcommand{\Z}{\mathrm{Z}}
\newcommand{\q}{\mathrm{q}}
\newcommand{\Q}{\mathrm{Q}}
\newcommand{\qbar}{\mathrm{\overline{q}}}

\newcommand{\g}{\mathrm{g}}
\newcommand{\mW}{m_{\mathrm{W}}}

\newcommand{\Delphi}{D\scalebox{0.8}{ELPHI}\xspace} 
\newcommand{\Dzero}{D\O\xspace} 
\newcommand{\Sherpa}{S\scalebox{0.8}{HERPA}\xspace} 
\newcommand{\Alpgen}{A\scalebox{0.8}{LPGEN}\xspace} 
\newcommand{\Madgraph}{M\scalebox{0.8}{AD}G\scalebox{0.8}{RAPH}\xspace} 
\newcommand{\Madevent}{M\scalebox{0.8}{AD}E\scalebox{0.8}{VENT}\xspace} 
\newcommand{\Helac}{H\scalebox{0.8}{ELAC}\xspace} 
\newcommand{\Ariadne}{A\scalebox{0.8}{RIADNE}\xspace} 
\newcommand{\Pythia}{P\scalebox{0.8}{YTHIA}\xspace} 
\newcommand{\Herwig}{H\scalebox{0.8}{ERWIG}\xspace} 
\newcommand{\Herwigpp}{H\scalebox{0.8}{ERWIG++}\xspace} 
\newcommand{\Compare}{C\scalebox{0.8}{OMPARE}\xspace} 
\newcommand{\Root}{R\scalebox{0.8}{OOT}\xspace} 
\newcommand{\bt}{\begin{tabular}}
\newcommand{\et}{\end{tabular}}
\def\be{\begin{equation}}
\def\ee{\end{equation}}
\def\bc{\begin{center}}
\def\ec{\end{center}}
\newcommand{\nnb}{\nonumber}
\newcommand{\bea}{\begin{eqnarray}}
\newcommand{\eea}{\end{eqnarray}}
\newcommand{\bit}{\begin{itemize}}
\newcommand{\eit}{\end{itemize}}

\begin{document}
\title{A parton shower algorithm based on Catani-Seymour dipole factorisation}
\preprintno{DCPT/07/86\\ IPPP/07/43}

\author{%
 S.~Schumann\email{steffen}{theory.phy.tu-dresden.de}$^a$,
 F.~Krauss\email{frank.krauss}{durham.ac.uk}$^b$
}
\address{\it%
$^a$Institut f\"ur Theoretische Physik, TU Dresden, D--01062 Dresden, Germany\\
$^b$Institute for Particle Physics Phenomenology, Durham University, Durham DH1 3LE, UK
\\[.5\baselineskip]
}
\abstract{
        In this publication the implementation of a new parton shower model 
        based on the Catani-Seymour dipole factorisation, as first suggested by
        \cite{Nagy:2005aa,Nagy:2006kb}, is discussed.  First results obtained 
        with the new algorithm are compared with experimental data.  
}
\maketitle

\tableofcontents



\input{csintro.tex}
\input{csconstruction.tex}
\input{cskinematics.tex}
\input{cschecks.tex}

\input{csapplications.tex}

\input{cssummary.tex}

\input{csacknowledgments.tex}

\input{bibliography.tex}
\end{document}

%% file: csintro.tex
\section{Introduction}


In the past decades, parton shower Monte Carlo programs, such as \Pythia 
\cite{Sjostrand:1993yb,Sjostrand:2006za} or \Herwig 
\cite{Corcella:2000bw,Corcella:2002jc} have been indispensable tools for planning 
and analysing particle physics experiments at different colliders.  It can be 
anticipated that they will play a similarly prominent r\^ole in the LHC era.  

There are a number of reasons for the success of these workhorses.  One of the 
most important ones rests in their ability to bridge the gap between few-parton 
final states, as described by fixed-order perturbative calculations, and the real 
world, where a multitude of hadrons etc.\ fills the detectors of the experiments.  
The transformation of the partons of perturbation theory into the visible hadrons,
hadronisation, is a direct consequence of the confinement property of QCD.  
Currently, this phenomenon can be described in terms of phenomenological models 
only, which depend on various phenomenological parameters tuned to data.  These 
parameters and hence the validity of the models in turn depends on the properties 
(such as the flow of energy and other QCD quantum numbers) of the parton ensemble; 
therefore it is important that these properties are kept under control.  It is the 
merit of parton showers that they provide a well-understood, theoretically sound
and universal framework of translating the few-parton states of fixed-order 
perturbation theory, calculated at some high scale, with multi-parton states at 
much lower scales, of the order of a few $\Lambda_{\rm QCD}$, where hadronisation 
sets in.  In so doing the parton showers help guarantee the validity of the tuned 
parameters of the hadronisation models.  

To achieve this translation of few-parton to multi-parton states, the parton shower 
programs rely on correctly describing QCD particle production in the dominant soft 
and collinear regions of phase space, giving rise to the bulk of radiation.  It is in
this region, where the complicated radiation pattern of multiple particle emission 
factorises into nearly independent - up to ordering in terms of a suitably chosen
parameter - individual emissions of single partons.  This approximation, namely 
expanding around the soft or collinear limit, ultimately leads to the resummation of
the corresponding leading logarithms, which are then typically encapsulated in 
exponential form in the Sudakov form factors.  Their probabilistic interpretation
in fact is the central feature allowing for a straightforward implementation in an
event generator, producing unweighted events.  Due to the resummation of leading 
logarithms it should thus not be too surprising that the parton shower programs more 
than often produce answers for QCD-related questions, which approximate exact results
very well.

However, the quality of the answers provided by the parton shower approximation alone 
relies on whether the question is related to the soft and/or collinear region in the
phase space of particle production.  If this is not the case, for example because of
the relevance of hard emissions or of non-trivial correlations of particles, the quality
of parton shower results tends to deteriorate.  In such cases, evidently a full quantum
mechanical treatment as provided by fixed-order calculations becomes mandatory.  
Therefore, the problem of systematically including higher-order effects into parton
shower programs has been in the center of research in the past few years.  In principle,
there have been two major avenues of investigation.  One dealt with the question of how
to include the correct QCD next-to leading order correction to total cross sections
\cite{Frixione:2002ik}-\cite{Nason:2006hf}, and has led to an implementation 
ready for use by the experiments in form of the MC@NLO code \cite{Frixione:2006gn}.  
The other considered the inclusion of tree-level multi-leg matrix elements into the 
simulation \cite{Catani:2001cc}-\cite{Mangano:2006rw}, and has lead to two types of 
algorithms being implemented.  One, based on \cite{Catani:2001cc,Krauss:2002up}, is 
the cornerstone of the event generator \Sherpa 
\cite{Gleisberg:2003xi}-\cite{Schalicke:2005nv} and an alternative formulation 
of the same algorithm, proposed in \cite{Lonnblad:2001iq}, has been implemented in 
\Ariadne \cite{Lonnblad:1992tz,Lavesson:2005xu}.  The other merging algorithm, based 
on \cite{Mangano:2001xp,Mangano:2006rw}, has been incorporated in 
\Alpgen \cite{Mangano:2002ea}, 
\Madgraph/\Madevent \cite{Stelzer:1994ta}-\cite{Alwall:2007st}, and 
\Helac \cite{Kanaki:2000ey,Papadopoulos:2005ky}.  Although it is not entirely clear how 
these two approaches relate in detail, some first comparisons \cite{Alwall:2007fs} 
show an interesting and assuring degree of agreement.

As one of the most recent outcomes of this line of research, it became apparent that
in order to systematically improve the event generators by including higher-order
corrections, also the parton shower algorithms themselves must be ameliorated.  Some
developments in this direction include an improved treatment of angular ordering and
massive partons in \Herwigpp \cite{Gieseke:2003rz} or the introduction of a new 
$k_\perp$-ordered shower in \Pythia \cite{Sjostrand:2004ef}.  
More recently, and motivated by the wish to include loop-level calculations in a 
more straightforward and systematic manner, the application of subtraction terms, 
prevalent in QCD next-to leading order calculations, has been proposed.  This paper 
reports on the construction of a parton shower based on such subtraction terms.  It 
uses the Catani-Seymour dipole formalism \cite{Catani:1996vz,Catani:2002hc}
and the corresponding subtractions as a starting point \footnote{
        This approach has also been employed in a parallel project, 
        \cite{Weinzierl}.}.  
This formulation of a parton shower has been proposed for the first time in 
\cite{Nagy:2005aa,Nagy:2006kb}.  A similar ansatz relies on antenna subtraction 
terms \cite{Kosower:1997zr,Campbell:1998nn} and has been presented 
recently in \cite{Giele:2007di}.
  
The paper is organised as follows: After briefly introducing the idea of parton
shower algorithms based on subtraction terms in Sec.~\ref{sec:psfromsubtraction} 
and a short review of the subtraction formalism of Catani-Seymour in 
Sec.~\ref{sec:CSreview}, Sec.~\ref{sec:CSconstruction} states the basic construction
principles of the proposed shower description.  In Sec.~\ref{sec:CSkinematics} 
the actual parton shower built on Catani-Seymour subtraction terms is constructed. 
The most general massive and the massless case for all the possible QCD splitting 
types are discussed in detail, and the modifications needed to include splittings 
of supersymmetric particles are discussed. The analytic expressions for the first 
shower emission from various core processes are compared with the 
corresponding exact tree-level matrix element calculations in 
Sec.~\ref{sec:MEPScomparison}.  In Sec.~\ref{sec:applications} predictions obtained  
with the developed shower formalism are confronted with experimental data and other 
calculations. The focus hereby is on hadron production in $e^+e^-$ collisions, 
and Drell-Yan and QCD jet production at the Fermilab Tevatron. 
Sec.~\ref{sec:cssummary} is devoted to the summary, conclusions, and an 
outlook on further developments.

\subsection{Parton showers based on subtraction methods}
\label{sec:psfromsubtraction}

Since its formulation almost a decade ago, the Catani-Seymour dipole formalism 
\cite{Catani:1996vz,Catani:2002hc} has been widely used in the calculation of 
next-to leading order (NLO) corrections in QCD, see for instance 
\cite{Campbell:1999ah}-\cite{Dittmaier:2007wz}.  

Such calculations typically face the problem of infrared divergences in both the 
real and the virtual parts of the NLO correction.  In principle, such divergences 
are not really a problem, since for physically meaningful observables, the 
Kinoshita-Lee-Nauenberg theorem \cite{Kinoshita:1962ur,Lee:1964is} guarantees their 
mutual cancellation.  To technically perform this cancellation, however, the 
divergences need to be regularised, which is usually performed by dimensional 
regularisation, \ie continuing the calculation to $d$ dimensions.  
There, the infrared divergences manifest themselves as poles in $1/(4-d)$ or 
$1/(4-d)^2$.  To deal with the poles and achieve the cancellation, subtraction methods 
may be used.  In general, they rely on the fact that the infrared divergences in the 
real correction part follow an universal pattern.  This allows to construct simplified 
terms in a process-independent way that encapsulate all infrared divergences 
occurring in the full matrix element.  Then, subtracting these terms from the 
real-correction matrix elements will yield an infrared-finite result, such that this 
subtracted matrix element can be safely integrated numerically in four dimensions.  
In addition, the subtraction terms are chosen such that they can be analytically 
integrated in $d$ dimensions over the phase space of the additional soft or collinear 
particle causing the divergences.  This yields the poles in $1/(4-d)$ or $1/(4-d)^2$, 
which are then added to the virtual part of the correction, and thus cancel the poles 
there.

The catch with the subtraction methods is that the subtraction terms can be constructed 
locally from the (colour-ordered) Born matrix element.  In the Catani-Seymour method,
for instance, pairs of particles are interpreted as emitting particle and spectator and
are subjected to a splitting kernel creating a third particle.  In this splitting 
process, one of the particles actually splits, while the recoil is compensated for by 
the spectator, which may be interpreted as its colour partner.  At the same time, the 
phase space factorises exactly into a phase space over the original particles, already 
present at the Born level, and into a phase space of the additional particle emerging 
in the splitting.  This exact factorisation corresponds to an exact mapping of the two 
original momenta (emitter and spectator) onto three four-momenta.  At each point of the 
procedure all particles remain on their respective mass shell.  

This is why constructing parton showers based on such methods currently is being pursued 
by different groups.  It is clear that these showers, in full conformance with original 
formulations employing the splitting of individual, single partons, are based on the 
universal soft and collinear dominance of QCD radiation.  Similar to the original shower 
algorithms, the emerging large logarithms occurring with each individual parton emission
can be resummed in a straightforward way through a Markovian process.  This, in 
principle, renders both formulations formally equivalent.  On the other hand, however, 
showers based on subtraction terms have the practical advantage that the conservation 
of four-momentum is built in with particles that remain on their mass shell at any 
given point
\footnote{It is interesting to note that the latest refinements of the parton showers
        in \Herwig and \Pythia also put more emphasis on the notion of a colour-connected
        partner compensating recoils etc.\ \cite{Gieseke:2003rz,Sjostrand:2004ef}.}.  
It can be anticipated that these features ultimately will allow for a more transparent 
merging with multi-leg matrix elements and a drastically alleviated matching with full 
NLO calculations.

\subsection{Short review of the Catani-Seymour subtraction method}
\label{sec:CSreview}

The Catani-Seymour subtraction method has been introduced in \cite{Catani:1996vz} 
for massless partons and it has been extended to massive partons in \cite{Catani:2002hc}.  
To fix the notation for the rest of the paper, it will be briefly reviewed here.

The essence of this method is embedded in the dipole factorisation formula
\bea\label{eq:CSMaster1}
|{\cal M}_{m+1}|^2=
 \sum_{i,j}\sum_{k\neq i,j}{\cal D}_{ij,k}
+\sum_{i,j}\sum_{a}{\cal D}_{ij}^a
+\sum_{a,i}\sum_{k\neq i}{\cal D}_k^{ai}
+\sum_{a,i}\sum_{b\neq a}{\cal D}^{ai,b}
+\dots\;.
\eea
The individual dipole contributions ${\cal D}$ provide the correct approximation of 
the $(m+1)$-parton matrix element squared in the different singular regions of phase
space \footnote{Note that squared matrix elements shall always be understood as properly 
normalised with respect to the  colour degrees of freedom of incoming particles.}.  
In each term $i$, $j$ and $k$ denote final-state partons and $a$ and $b$ stand for 
initial-state partons. The first sum always runs over the two particles to be combined, 
whereas the second sum takes care of the spectators.  Accordingly, the four terms correspond 
to the splitting of a final-state parton accompanied by a final-state or initial-state 
spectator and emissions off incoming particles in the presence of a final-state or an 
initial-state spectator, respectively.  Finally, the dots in the equation above denote 
some potential finite terms which do not exhibit any divergence.        

For the case of final-state emitters with a final-state spectator, for instance, 
the individual dipole contributions read \cite{Catani:1996vz}
\bea\label{eq:FFdipolecontrib}
{\cal D}_{ij,k} = -\frac{1}{2p_ip_j}\hspace*{4mm}
\langle_{\!\!\!\!\!\!\!\!\!\!\;m}\,\, 1,\,\dots,\,\widetilde{ij}\,\dots,\,\tilde k,\,\dots,\,m+1|
             \frac{{\bf T}_k\cdot{\bf T}_{ij}}{{\bf T}^2_{ij}}{\bf V}_{ij,k}
             |1,\,\dots,\,\widetilde{ij}\,\dots,\,\tilde k,\,\dots,\,m+1
                   \rangle_m\,,
\eea
when all the involved partons are assumed to be massless.
The occurring $m$-parton states are constructed from the original $(m+1)$-particle matrix 
element by replacing the partons $i$ and $j$ with the new parton $\widetilde{ij}$, 
the emitter, and the original parton $k$ with $\tilde k$, the spectator.  In the massless 
case, their momenta are given by
\bea
\tilde p_{ij}^\mu = p_i^\mu+p_j^\mu-\frac{y_{ij,k}}{1-y_{ij,k}}p_k^\mu 
\quad\mbox{\rm and}\quad
\tilde p_{k}^\mu = \frac{1}{1-y_{ij,k}}p_k^\mu\,,
\eea
where the dimensionless, Lorentz-invariant quantity $y_{ij,k}$ is given by
\bea\label{eq:yijk}
y_{ij,k} = \frac{p_ip_j}{p_ip_j+p_ip_k+p_jp_k}\,.
\eea
It is simple to show exact four-momentum conservation, \ie 
$\tilde p_{ij}^\mu + \tilde p_k^\mu = p_i^\mu + p_j^\mu + p_k^\mu$, with all 
particles on their mass shell.  In the matrix element on the right hand side of 
Eq.\ (\ref{eq:FFdipolecontrib}), the ${\bf T}_{ij}$, ${\bf T}_{k}$ are the colour 
charges of the emitter and spectator, respectively, and the ${\bf V}_{ij,k}$ are 
matrices in the emitter's spin and colour space, responsible for its branching.  
The operators ${\bf V}_{ij,k}$ also depend on the dimensionless, Lorentz-invariant 
quantities 
\bea\label{eq:zizj}
\tilde z_i = \frac{p_ip_k}{p_ip_k+p_jp_k} = 
\frac{p_i\tilde p_k}{\tilde p_{ij}\tilde p_k}
\;\;\;\mbox{\rm and}\;\;\;
\tilde z_j = \frac{p_jp_k}{p_ip_k+p_jp_k} = 
\frac{p_j\tilde p_k}{\tilde p_{ij}\tilde p_k} = 1-\tilde z_i\,.
\eea
For instance, for the case of a quark splitting in the final state with a final-state 
spectator, \ie $q_{ij}\to q_i+g_j$, where $s$ and $s'$ denote the spins of 
$\widetilde{ij}$ and $i$, respectively, and where the subscripts label the momenta,
\bea
\langle s|{\bf V}_{q_ig_j,k}(\tilde z_i,y_{ij,k})|s'\rangle = 
8\pi\mu^{2\epsilon}\as \CF
\left[\frac{2}{1-\tilde z_i(1-y_{ij,k})}-(1+\tilde z_i)-
      \epsilon(1-\tilde z_i)\right]\delta_{ss'}\,.
\eea
Here, $\epsilon = 4-d$, with $d$ the number of dimensions.  Similar expressions emerge 
for the other QCD splittings or when masses are included.  However, as a general 
property, the matrices ${\bf V}_{ij,k}$ do not become singular, if any of the scalar 
products $p_ip_j$, $p_ip_k$ or $p_jp_k$ vanishes, and therefore the only soft or 
collinear divergences in the dipole terms ${\cal D}_{ij,k}$ are related to 
$p_ip_j\to 0$.

The collinear limit of the two final-state partons $i$ and $j$ originating from a 
splitting $\widetilde{ij} \to i + j$ is defined through their relative transverse 
momentum $k_\perp\to 0$.  This limit can be investigated by decomposing the momenta as
\bea
p^\mu_i &=& \hphantom{(1-))}zp^\mu + \frac{-\kperp^2}{\hphantom{1(}z\hphantom{lll}}
        \frac{n^\mu}{2pn}+k^\mu_\perp\,,\\
p^\mu_j &=& (1-z)p^\mu + \frac{-\kperp^2}{1-z}\frac{n^\mu}{2pn}-k^\mu_\perp\,,
\eea
where the lightlike $p^\mu$ defines the collinear direction and $n^\mu$ is an auxiliary 
lightlike vector that specifies the spacelike transverse momentum $\kperp^\mu$, 
with $\kperp^2 = -{\kperpbfsq}$, through $p\kperp=n\kperp=0$. Then, in the collinear 
limit, the scalar product $p_ip_j$ reads
\bea\label{eq:FSmasslesscollinearlimit}
p_ip_j = -\frac{k^2_\perp}{2z\,(1-z)}\,,\quad \kperp^2 \to 0\,,
\eea
and the dipole variables are given by
\bea
y_{ij,k} &\to& -\frac{k_\perp^2}{2z(1-z)pp_k}\;,\;\;\;
\tilde z_i = 1-\tilde z_j \to z\;,\nnb\\
\tilde p_k^\mu &\to& p_k^\mu\;\;\;\mbox{\rm and}\;\;\;
\tilde p_{ij}^\mu \to p^\mu\,.
\eea
It can then be shown that in this limit the matrices ${\bf V}_{ij,k}$ become 
proportional to the Altarelli-Parisi splitting kernels,
\bea
{\bf V}_{ij,k} \to 
8\pi\mu^{2\epsilon}\as\,  \hat P_{(ij),i}(z,\,k_\perp;\,\epsilon)\,.
\eea
In this limit the only remaining dependence of the dipole contributions
${\cal D}_{ij,k}$ on the spectator $k$ resides in its colour factor ${\bf T}_k$ 
and it can be shown that Eq.~(\ref{eq:FFdipolecontrib}) reproduces the well-known universal collinear 
behaviour of the $(m+1)$-parton matrix element,
\bea
\lefteqn{\langle_{\!\!\!\!\!\!\!\!\!\!\!\!\!\!\!\!m+1}\,\, 1,\,\dots,\,i,\,\dots,\,j,\,\dots,\,m+1|
             |1,\,\dots,\,i,\,\dots,\,j,\,\dots,\,m+1\rangle_{m+1}}\nnb\\
&&\stackrel{k_\perp\to 0}\longrightarrow
\frac{4\pi\mu^{2\epsilon}\as}{p_ip_j}\hspace*{4mm}
\langle_{\!\!\!\!\!\!\!\!\!\!\;m}\,\, 1,\,\dots,\,ij,\,\dots,\,m+1|\hat P_{(ij),i}(z,\,k_\perp;\,\epsilon)
             |1,\,\dots,\,ij,\,\dots,\,m+1\rangle_{m}\,,
\eea
where again, the kernel $\hat P$ is a $d$-dimensional Altarelli-Parisi splitting 
function.  

In contrast, the limit where $p_j$ becomes soft is given by $p_j^\mu = \lambda q^\mu$
with $\lambda\to 0$ and $q^\mu$ some, in principle arbitrary, four-vector.  In this
limit, the dipole variables become
\bea
y_{ij,k} &\to& 0\;,\;\;\;
\tilde z_i = 1-\tilde z_j \to 1\;,\nnb\\
\tilde p_k^\mu &\to& p_k^\mu\;\;\;\mbox{\rm and}\;\;\;
\tilde p_{ij}^\mu \to p_i^\mu\,,
\eea
and ${\bf V}_{ij,k}$ tends to
\bea
\frac{1}{1-\tilde z_i(1-y_{ij,k})} \stackrel{\lambda\to 0}{\longrightarrow}
\frac{1}{\lambda}\cdot\frac{p_ip_k}{(p_i+p_k)q}\,.
\eea
Therefore,
\bea
\lambda{\bf V}_{ij,k} \stackrel{\lambda\to 0}{\longrightarrow}
16\pi\mu^{2\epsilon}\as {\bf T}_{ij}^2\frac{p_ip_k}{(p_i+p_k)q}\,.
\eea
It can thus be shown that the well-known soft limit of the $(m+1)$-parton matrix element 
is recovered, namely
\bea
\lefteqn{\langle_{\!\!\!\!\!\!\!\!\!\!\!\!\!\!\!\!m+1}\,\, 1,\,\dots,\,i,\,\dots,\,j,\,\dots,\,m+1|
             |1,\,\dots,\,i,\,\dots,\,j,\,\dots,\,m+1\rangle_{m+1}}\nnb\\
&&\stackrel{\lambda\to 0}\longrightarrow
-\sum\limits_{i, k\neq i}\frac{8\pi\mu^{2\epsilon}\as}{\lambda^2(p_iq)}\hspace*{4mm}
\langle_{\!\!\!\!\!\!\!\!\!\!\;m}\,\, 1,\,\dots,\,ij,\,\dots,\,m+1|
             \frac{{\bf T}_k\cdot{\bf T}_i (p_ip_k)}{(p_i+p_k)q}
             |1,\,\dots,\,ij,\,\dots,\,m+1\rangle_{m}\,.
\eea

Taken together, these considerations and similar reasoning for the other dipole 
contributions translate into the dipole formula, Eq.\ (\ref{eq:CSMaster1}), to 
provide a point-wise approximation to the full $(m+1)$-parton matrix element, 
which exactly recovers all the soft and collinear divergences.  

Before starting the discussion on the construction of a parton shower algorithm from 
the Catani-Seymour dipole formula in Sec.~\ref{sec:CSconstruction} the generalisation 
of Eq.~(\ref{eq:FSmasslesscollinearlimit}) to the massive case and the analogous result 
for the splitting of an initial-state parton shall be briefly repeated.

First, re-consider the splitting $\widetilde{ij} \to i + j$ from above. This time, 
however, both the emitter and the splitting products are allowed to be massive, 
the corresponding mass shell conditions read $p^2=m^2_{ij}$, $p^2_i=m^2_i$ and $p^2_j=m^2_j$. 
The momenta $p_i$ and $p_j$ can again be written in a Sudakov parametrisation 
according to 
\bea
p^\mu_i &=& \hphantom{(1-))}zp^\mu+\frac{\hphantom{(1}-\kperp^2-z^2m^2_{ij}+m^2_i\hphantom{())}}{z}\frac{n^\mu}{2pn}+k^\mu_\perp\,,\\
p^\mu_j &=& (1-z)p^\mu+\frac{-\kperp^2-(1-z)^2m^2_{ij}+m^2_j}{1-z}\frac{n^\mu}{2pn}-k^\mu_\perp\,,
\eea
with $n^2=0$ and $\kperp$ perpendicular to both $p$ and $n$. Identifying 
$\kperp^2 = -{\kperpbfsq}$ the invariant mass of partons $i$ and $j$ is now given
by
\bea\label{eq:FSmassivecollinearlimit}
(p_i+p_j)^2 = \frac{\kperpbfsq}{z\,(1-z)}+ \frac{m^2_i}{z}+\frac{m^2_j}{1-z}\,,\quad \kperpbfsq \to 0\,.
\eea
Accordingly, the collinear singularity is shielded when at least one of 
the two partons has a finite mass.  

Finally, consider the case when final-state parton $i$ becomes collinear to an 
initial-state parton $a$. This corresponds to the splitting $a \to \widetilde{ai}+i$, 
with $\widetilde{ai}$ the initial-state parton that enters the $m$-parton process. 
Considering only massless initial states, all the partons involved in the splitting 
are consistently taken to be massless. Decomposing the final-state momentum 
$p_i$ according to
\bea
p^\mu_i &=& (1-x)p^\mu_a+\frac{-\kperp^2}{1-x}\frac{n^\mu}{2p_an}+\kperp\,,
\eea
the collinear limit is reached for 
\bea\label{eq:IScollinearlimit}
p_ap_i = \frac{\kperpbfsq}{2(1-x)}\,,\quad \kperpbfsq \to 0\,,
\eea
with $\kperpbfsq$ the magnitude of the spacelike transverse momentum vector $\kperp$, 
namely $\kperp^2 = -{\kperpbfsq}$. The definitions Eq.~(\ref{eq:FSmassivecollinearlimit}) and 
Eq.~(\ref{eq:IScollinearlimit}) constitute the basic relations for identifying the 
transverse momentum vector for the different splitting types in terms of the respective 
splitting variables used to describe the branchings, see Sec.~\ref{sec:CSkinematics}.

%% file: csconstruction.tex
\section{Construction of the algorithm}
\label{sec:CSconstruction}

To formulate a parton shower algorithm based on the Catani-Seymour dipole formulae, 
the corresponding splitting operators ${\cal D}$ that describe the emission of an 
additional parton from an arbitrary $m$-parton state have to be analyzed and rewritten
in a suitable form, before they can be used for a showering algorithm.  To this end, a 
number of issues has to be resolved:
\begin{itemize}  
\item First of all, only the four-dimensional expressions of the splitting kernels 
	${\cal D}$ will enter the parton shower.  In addition, the splitting kernels 
	are employed in their spin-averaged form.  This manipulation is straightforward 
	and a detailed discussion is therefore not necessary.  The resulting splitting 
	kernels depend on the actual configuration of emitters and spectators in the 
        initial- and final state and they will be listed in the corresponding 
        parts of Sec.\ \ref{sec:CSkinematics}.
\item In order to keep the probabilistic notion enabling simulation, to use a Markovian 
	formulation for the showering process and to facilitate the hadronisation at the 
	end of the shower, issues concerning colour correlations have to be solved.  
	While the original Catani-Seymour dipole formulae consider all colour 
        correlations, the shower will account only for the leading terms in 
        $1/\Nc$.  This will be further discussed in Sec.\ \ref{sec:CScolours}.
\item Also, the phase space factorisation and the corresponding combination procedure 
        is effectively inverted to construct the kinematics of the individual 
        splittings.  This yields splitting kernels for $1 \to 2$ QCD branchings 
        that allow for the inclusion of finite parton masses in quite a general way.  
        Each splitting parton thereby is accompanied by a single colour-connected 
        spectator parton compensating the recoil of the splitting. The only exception 
        here are initial-state splittings in the presence of an initial-state spectator, 
        where the recoil is taken by all final-state partons of the event.  The 
        introduction of the spectator allows to assemble the shower kinematics such 
        that four-momentum conservation can be ensured after each individual branching 
        with all external partons on their mass-shell.  Accordingly, this parton 
        shower algorithm can be stopped at any intermediate stage as well as started 
        again for a partially evolved parton ensemble.  However, the exact procedure
        for reconstructing the kinematics of each splitting again depends on whether
        the emitter and spectator are in the initial- or final state, respectively.  
        The corresponding formulae are listed in Sec.\ \ref{sec:CSkinematics}.
\item The actual shower evolution variable specifying and ordering subsequent 
        emissions is chosen to be the transverse momentum between the splitting 
        products for branching final-state partons and the transverse momentum with 
        respect to the beam for emissions from the initial state, collectively 
        denoted by $\kperpbf$. The physics underlying this choice will be further 
        detailed in Sec.\ \ref{sec:CSordering}.
\item Furthermore, choices have to be made concerning the scales entering the QCD 
	running coupling constant, $\as$, and the parton distribution functions when 
	initial-state partons are present.  This will be discussed in 
	Sec.\ \ref{sec:CSscales}.
\item Based on these considerations, appropriate Sudakov form factors are constructed 
        that determine the probability for a certain branching process not to occur 
        for a given range of the evolution variable, $\kperpbf$.  These Sudakov form 
        factors constitute the basis of the actual Monte Carlo showering algorithm.  
        Again, their specific form depends on the details of emitter and spectator 
        parton and they will thus be given in corresponding parts of 
        Sec.\ \ref{sec:CSkinematics}, too.
\item This section closes with some general considerations concerning the treatment
        of parton masses, cf.\ Sec.\ \ref{sec:CSmassives}.
\end{itemize}

\subsection{Colour factors and spectators}
\label{sec:CScolours}

The starting point for every parton shower evolution is a given set of partons 
and their momenta from a fixed-order matrix element calculation.  In the large-$\Nc$ 
approximation a colour flow can be assigned to each parton configuration.  Since
in most cases the initial matrix element calculation is already summed and averaged 
over the colours of final and initial partons, the assignment typically is performed
a posteriori in different ways in different codes.  However, as a result the partons
entering the parton shower after this assignment have a well-defined colour, and, due
to the large-$\Nc$ limit, one or two uniquely assigned colour partners \footnote{  
        Representing the colour flow pictorially by coloured strings of partons,
        two configurations emerge, namely open or closed strings.  An open string 
        consists of a colour-triplet state followed by colour octets and ends
        with a colour anti-triplet. Mapping the colour flows, initial-state quarks 
        (colour triplets) correspond to final-state anti-quarks (colour anti-triplets), 
        whereas initial-state anti-triplets can be treated as final-state triplets. 
        A closed colour string corresponds to a configuration of colour-octet partons 
        only. Accordingly, the end of a closed string is colour-connected to its 
        beginning and therefore the whole colour string is invariant under cyclic 
        permutations of its individual constituents.}.
Motivated by considerations on the colour dynamics for soft emissions in the 
Catani-Seymour formalism, in a corresponding shower formulation the spectator parton 
accompanying a given splitting is colour-connected to the emitter parton.  
For the case of a splitting gluon/gluino then there are always two possible colour 
partners, whereas splitting (anti-)quarks/squarks will have only one spectator parton 
candidate.  Following this reasoning, the initial partons will enter the parton shower 
stage in well-defined pairs of potential emitters and spectators.  The subsequent
parton shower will not change this feature.      

To formalise the treatment of colour inside the parton shower presented here, consider 
the colour-operators present in the Catani-Seymour dipole contributions.  In the 
large-$\Nc$ limit, they are easily calculated for any $m$-parton state at the price 
of loosing colour correlations beyond $1/\ \!\Nc$.  However, in this limit only two 
cases need to be considered.  Independent of the actual spectator flavour, the colour 
algebra for a splitting (anti-)quark/squark yields, 
\bea
-\frac{\bT_k\cdot\bT_{ij}}{\bT^2_{ij}} \to 1 + {\cal O}(\frac{1}{\Nc^2})\,,
\eea
whereas a splitting gluon/gluino results in
\bea
-\frac{\bT_k\cdot \bT_{ij}}{\bT^2_{ij}} \to \frac12 + {\cal O}(\frac{1}{\Nc^2})\,.
\eea
For convenience, these two results can be combined by introducing ${\cal N}^{spec}_{ij}$,
the number of possible spectators the emitting parton possesses, then
\bea
-\frac{\bT_k\cdot\bT_{ij}}{\bT^2_{ij}} \to 
\frac{1}{{\cal N}^{spec}_{ij}} + {\cal O}(\frac{1}{\Nc^2})\,.
\eea

\subsection{Ordering parameter}
\label{sec:CSordering}

Having the individual splitting process under control, \ie having at hand the
corresponding splitting kernel with all relevant colour factors and the way the 
kinematics of the emission is constructed, the full showering algorithm with its 
sequence of splittings can be addressed.  While the individual splitting kernel
properly takes into account the soft and collinear divergent regions, in the parton
shower itself these regions are cut away and, formally speaking, combined with the 
virtual bits to yield a probabilistic description of the splitting process.  The cut 
on the soft and collinear region implies the emergence of corresponding logarithms 
of the cut parameter, which the parton shower aims to resum.  Technically, this 
resummation is achieved by arranging the individual emissions in a Markov chain, 
treating each emission on the same footing, and by ordering the emissions with some 
ordering parameter.  This has been detailed in textbooks such as \cite{Ellis:1991qj}.  
In different parton shower implementations, there are different ordering parameters 
realised, such as the invariant mass of the splitting particle 
\cite{Sjostrand:1985xi}-\cite{Bengtsson:1987kr}, the opening angle of the pair 
\cite{Gieseke:2003rz,Marchesini:1987cf}, or their relative transverse momentum 
\cite{Lonnblad:1992tz,Sjostrand:2004ef}. At the level of doubly leading logarithms, 
these choices are all equivalent, but there are substantial differences on the level 
of next-to leading logarithms, \ie on the level of single soft logarithms.  This 
is closely tied with the treatment of quantum coherence effects 
\cite{Ermolaev:1981cm}-\cite{Dokshitzer:1987nm}, which are properly taken into 
account by ordering subsequent emissions through their respective opening angles 
\cite{Marchesini:1987cf}.  In \cite{Gustafson:1986db} it has been shown that 
another way of properly accounting for coherence effects is evolving in a 
dipole-like picture with subsequent emissions ordered by transverse momenta.  

In the implementation presented here, the parton shower will be ordered by transverse
momenta, \ie by the $\kperpbf$ in Eqs.\ (\ref{eq:FSmassivecollinearlimit}) and 
(\ref{eq:IScollinearlimit}).  Apart from the proper treatment of quantum coherence 
effects, this choice has additional benefits: First of all, as will be discussed 
in the next section, cf.\ Sec.\ \ref{sec:CSscales}, by ordering with $\kperpbf$ 
the ordering parameter also enters as the relevant scale in the coupling constant 
and the parton distribution functions.  Second, the definition used here allows for 
a shower formulation on the basis of Lorentz-invariant quantities, see for instance 
\eg Eqs.\ (\ref{eq:yijk}) and (\ref{eq:zizj}).  Also, ordering by $\kperpbf$ 
immediately implies that the parton shower cut-off is related to some minimal 
transverse momentum necessary to resolve partons, which seems quite appealing in 
terms of the physical interpretation of such a resolution criterion.  Last but not 
least an ordering by transverse momenta appears to allow for quite a straightforward 
merging of the parton shower with multi-leg tree-level matrix elements in the spirit 
of \cite{Catani:2001cc,Krauss:2002up}.  The merging method presented there bases on 
Sudakov suppression weights for matrix elements, which are constructed from the 
transverse momenta of their nodes, and on a vetoed parton shower respecting the minimal 
scale of a $k_\perp$-jet definition.  

In the parton shower evolution each colour-singlet is separately evolved.  To this end, 
all emitter-spectator dipoles are iterated over and for each of those configurations
a $\kperpbf$ is chosen according to the corresponding Sudakov form factor.  The dipole
with the largest $\kperpbf$ is selected to split according to the kinematics detailed 
below.  As long as this largest $\kperpbf$-value is larger than the infrared cut-off
$\kperpzero$, the shower evolution will continue, and this largest $\kperpbf$ of the
current evolution step serves as the maximal scale for all dipoles in the colour-singlet 
in the next splitting step.    

\subsection{Scales to be chosen}
\label{sec:CSscales}

When discussing the details of a parton shower implementation, some care has to be
taken in the choice of various, in principle undetermined, occurring scales.  There
are a number of choices to be made, namely:
\begin{itemize}
\item The evolution variable and the related evolution cut-off:\\
        As already discussed in the previous section, in this implementation the
        relative transverse momentum of the produced parton w.r.t.\ its emitter
        has been chosen as the relevant evolution variable.  It is given by Eqs.\ 
        (\ref{eq:FSmassivecollinearlimit}) and (\ref{eq:IScollinearlimit}).  
        Correspondingly, a cut-off has to be set as a tuning parameter, to stay away 
        from phase-space regions where the perturbative expansion for the running 
        coupling is divergent. The choice of this cut-off is dictated by two aspects.  
        First of all, it seems to be more attractive to try to assign as much phase 
        space for particle creation to the, in principle, well-understood perturbative 
        parton shower rather than to a phenomenological hadronisation approach such as 
        the Lund string fragmentation \cite{Andersson:1983ia,Andersson:1998tv} or a 
        cluster model \cite{Webber:1983if}-\cite{Stephens:2004mf}.  This implies that 
        the cut-off should be as small as possible.  On the other hand, it is clear 
        that perturbative QCD breaks down and looses its predictive power at small 
        scales.  This is best exemplified by the infrared behaviour of the running 
        coupling which exhibits a Landau pole at $\Lambda_{\rm QCD}$.  As will be 
        discussed in the next item, since the running coupling in the shower is 
        evaluated at a scale related to $\kperpbf$, this feature of QCD prohibits 
        cut-offs in the region of $\Lambda_{\rm QCD}$.  Therefore, a suitable choice 
        seems to be a cut-off $\kperpzero$ of the order of $1$ GeV, sufficiently 
        separated from the Landau pole.
\item The argument of the running coupling constant, $\mu_R$:\\
        In the previous item it has been already hinted at the choice typically made in
        parton showers, to take the running coupling at scales of the order of
        $\kperpbf$.  The reason for this choice is that it incorporates and resums some 
        of the higher-order corrections to the splitting.  Specifically, in this 
        implementation the choice is to take $\mu^{\rm F.S.}_R = \mu_R = \kperpbf$ if the
        emitter is a final-state particle and $\mu^{\rm I.S.}_R = \mu_R = \kperpbf/2$
        if the emitter is a parton in the initial state.  
\item The argument of the parton density functions, $\mu_F$:\\
        Similar to the case of the running coupling constant, a choice has also been
        made at which scale to take the parton distribution functions, if necessary.
        In parton showers, there are typically two answers, namely to either again 
        take the transverse momentum or to use the virtual mass of the initial emitter.
        Here the choice again is to use $\mu_F = \kperpbf$.  
\end{itemize}

\subsection{General considerations on massive particles}
\label{sec:CSmassives}

Taking into account finite quark mass effects in the Standard Model (SM) clearly is 
of importance when producing heavy quarks, bottom or top quarks, in a hard scattering 
process.  In addition, many extensions of the SM introduce new strongly-interacting 
heavy particles, whose QCD radiation needs to be modeled to understand the patterns
of particle and energy flows in their production and eventual decays.  Prime examples 
are scalar quarks and gluinos in supersymmetric theories \cite{Haber:1984rc} or heavy 
excitations of the SM quark and gluon fields in models with additional 
space-time dimensions \cite{Appelquist:2000nn}.  While at lepton colliders heavy 
objects only appear in the process' final state, at hadron colliders charm and 
bottom quarks can also constitute the partonic initial state. An example where 
these are of phenomenological relevance is the associated production of heavy quarks 
and scalar Higgs particles in supersymmetric models, which is a promising channel 
to gain deeper insight into the mechanism of electroweak symmetry breaking, see for 
instance \cite{Djouadi:2005gj} and references therein.
 
In the following section, QCD splitting operators will be derived, that fully take 
into account finite masses of partons in the final state.  This includes both emission 
from heavy particles but also the splitting of gluons into heavy quarks such as 
charm or bottom.  Splittings of gluons into heavier objects or branchings of heavy 
states into other heavy objects are beyond the scope of this work as they are not 
well modeled by the soft or quasi-collinear approximation and should rather be 
described with full matrix elements.  For all the formulae presented in Sec.\
\ref{sec:CSkinematics}, the massless limit is smoothly obtained when setting the 
parton masses to zero.  This will be explicitly examined for some of the important 
results there. 

Throughout this work, incoming QCD partons will always be treated as massless.  The 
leading logarithms that arise for emissions off incoming heavy quarks, logarithms of the 
type  $(\as\log(Q^2/m^2_Q))^n$, with $Q^2$ the scale of the hard-scattering process 
and $m_Q$ the quark mass, are summed to all orders in QCD when using heavy-quark parton 
distribution functions at the factorisation scale $\mu_F \sim Q$ and considering the 
incoming quarks as massless \cite{Dicus:1988cx,Boos:2003yi}.  A scheme to consistently 
incorporate explicit masses for incoming heavy quarks, relying on modified heavy-quark 
density functions \cite{Collins:2002ey}, has recently been presented in 
\cite{Kersevan:2006fq}.


%% file: cskinematics.tex
\section{Kinematics of the individual splittings}
\label{sec:CSkinematics}

In the following sections, Secs.\ \ref{sec:kinematicsFF}-\ref{sec:kinematicsII}, 
the actual parton shower built on Catani-Seymour subtraction terms is constructed.  
To this end, all combinations of initial- and final-state emitter and spectator 
partons are considered in detail, following closely the original publications 
on the subtraction method \cite{Catani:1996vz,Catani:2002hc}.  First, the kinematic 
variables characterising the individual splitting under consideration are discussed.  
Then the explicit form of the phase space element for the three-parton state under 
consideration is re-expressed through the kinematic variables above, and their 
respective bounds are given.  In a next step, the polarisation-averaged splitting 
kernels for the respective emitter-spectator configuration are listed.  This 
allows to give the factorised form of matrix elements with one additional parton 
in the soft and collinear limits of its production and the factorised form of 
the corresponding differential cross section, which includes both matrix element 
and phase space factorisation.  From there, it is quite straightforward to deduce 
the actual Sudakov form factor for the emitter-spectator configuration.  Finally, 
the actual kinematics of the splitting is constructed, which may slightly differ  
from the evolution parameters due to mass effects.  For each case then also the 
more familiar massless limit is briefly discussed. In Sec.~\ref{sec:kinematicsSUSY} 
the QCD splitting functions for supersymmetric particles are presented.

\subsection{Final-state emitter and final-state spectator}
\label{sec:kinematicsFF}

\begin{figure}
\centerline{
  \begin{picture}(160,120)(0,0)
  \Line(80,50)(110, 70)
  \Line(110, 70)(130,100)
  \Line(80,50)(120, 20)
  \LongArrow( 90,30)(105, 18)
  \LongArrow(110,85)(120, 100)
  \LongArrow(120,60)(135, 60)
  \Line(110,70)(143,70)
  \Vertex(110,70){2.5}
  \GCirc(80,50){10}{1}
  \put( 90, 70){$\widetilde{ij}$}
  \put(140, 100){$i$}
  \put(145, 60){$j$}
  \put(125, 15){$k$}
  \put( 40, 50){$\bV_{ij,k}$}
  \put( 90, 14){$p_k$}
  \put(100, 99){$p_i$}
  \put(120, 50){$p_j$}
  \end{picture} }
\caption{Effective diagram for the splitting of a final-state parton  
        connected to a final-state spectator.  The blob denotes the $m$-parton 
        matrix element, and the outgoing lines label the final-state partons 
        participating in the splitting.}
\label{fig:split_FF}
\end{figure}
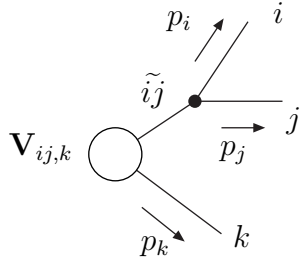

The first case to be investigated is when both the emitter and the spectator parton 
are in the final state, cf. Fig.\ \ref{fig:split_FF}.  Accordingly, the splitting 
$\{\widetilde{ij},\tilde{k}\} \to \{i,j,k\}$ has to be studied.  When considering 
processes without colour-charged initial-state particles, such as jet production in 
lepton-lepton collisions, this is the only QCD radiation process and thus constitutes 
the basis of a corresponding final-state parton shower.  However, the observed 
factorisation of the differential cross section for producing an additional parton 
also holds in the presence of initial-state partons, where only the additional 
branching channels discussed below then have to be taken into account as well.

\subsubsection{Massive case}
\label{sec:kinematicsFFmassive}

In the most general case all partons involved in the splitting can have arbitrary 
masses, \ie $\tilde p_{ij}^2=m^2_{ij}$, $\tilde p_{k}^2=p_k^2=m^2_{k}$, 
$p_{i}^2=m^2_{i}$ and $p_{j}^2=m^2_{j}$, respectively. In order to avoid on-shell 
decays, which should be described by their respective proper matrix element, only 
those situations are considered, where $m^2_{ij} \leq m_i^2+m_j^2$.

\begin{itemize}
\item \underline{Kinematics:}\\
        Exact four-momentum conservation is ensured by the requirement
        \bea\label{eq:FFMomCon}
        \tilde p_{ij}+\tilde p_k = p_i+p_j+p_k \equiv Q\,.
        \eea
        The splitting is characterised by the dimensionless variables $y_{ij,k}$, 
        $\tilde z_i$ and $\tilde z_j$.  They are given by
        \bea
        y_{ij,k}=\frac{p_ip_j}{p_ip_j+p_ip_k+p_jp_k}\,,\quad 
        \tilde z_i=1-\tilde z_j= \frac{p_ip_k}{p_ip_k+p_jp_k}\,.
        \eea
        With these definitions the invariant transverse momentum of partons $i$ and 
        $j$, defined in Eq.\ (\ref{eq:FSmassivecollinearlimit}), can be written as
        \bea\label{eq:FFmassivekperp}
        \kperpbfsq = (Q^2-m^2_i-m^2_j-m^2_k)y_{ij,k}\,\tilde z_i(1-\tilde z_i) - 
                        (1-\tilde z_i)^2m^2_i - \tilde z^2_im^2_j\,.
        \eea
        For convenience, the rescaled parton masses 
        \bea
        \mu_n = \frac{m_n}{\sqrt{Q^2}} \quad (n=i,j,k,ij)\,,
        \eea
        and the relative velocities between $p_i+p_j$ and $p_i$ ($p_k$), 
        $v_{ij,i}$ ($v_{ij,k}$), 
        \bea
        v_{ij,i} 
        &=& 
        \frac{\sqrt{(1-\mu^2_i-\mu^2_j-\mu^2_k)y^2_{ij,k}-4\mu^2_i\mu^2_j}}
                {(1-\mu^2_i-\mu^2_j-\mu^2_k)y_{ij,k}+2\mu^2_i}\,,\\
        v_{ij,k} 
        &=& 
        \frac{\sqrt{\left[2\mu^2_k+(1-\mu^2_i-\mu^2_j-\mu^2_k)(1-y_{ij,k})\right]^2
                                -4\mu^2_k}}
                {(1-\mu^2_i-\mu^2_j-\mu^2_k)(1-y_{ij,k})}\,,
        \eea
        as well as the velocity between $\tilde p_{ij}$ and $\tilde p_k$,
        \bea
        \tilde v_{ij,k} = 
        \frac{\sqrt{\lambda(1,\mu^2_{ij},\mu^2_k)}}{1-\mu^2_{ij}-\mu^2_k}\,,
        \eea
        are introduced.
\item \underline{Phase space:}\\
        In the case of a final-state emitter with a final-state spectator, the
        corresponding three-parton phase space $\d\Phi(p_i,p_j,p_k;Q)$ must be
        analyzed.  It exactly factorises into a two-parton contribution 
        $\d\Phi(\tilde p_{ij},\tilde p_k;Q)$ and a single-parton phase space factor 
        $[\d p_i(\tilde p_{ij},\tilde p_k)]$, 
        \bea
        \d\Phi(p_i,p_j,p_k;Q) = 
        \d\Phi(\tilde p_{ij},\tilde p_k;Q)\,
        [\d p_i(\tilde p_{ij},\tilde p_k)]\,\Theta(1-\mu_i-\mu_j-\mu_k)\,,
        \eea
        where the latter is given by
        \bea\label{eq:FF2BodyPS}
        [\d p_i(\tilde p_{ij},\tilde p_k)] = 
        \frac{(\tilde p_{ij}+\tilde p_k)^2}{16\pi^2}\,
        \frac{(1-\mu^2_i-\mu^2_j-\mu^2_k)^2}
                {\sqrt{\lambda(1,\mu^2_{ij},\mu^2_k)}}\,
        (1-y_{ij,k})\,\d y_{ij,k}\,\d\tilde z_i\frac{\d\phi}{2\pi}\,.
        \eea
        Here and in the following, $\lambda$ denotes the K{\"a}llen function, 
        \bea
        \lambda(x,y,z) = x^2+y^2+z^2-2(xy+xz+yz)\,.
        \eea
        The boundaries of the full, unconstrained, phase space read 
        $\phi \in [0,2\pi]$, whereas the lower and upper limits for $\tilde z_i$ 
        and $y_{ij,k}$ are 
        \bea
        z_\mp 
        &=& 
        \frac{2\mu^2_i+(1-\mu^2_i-\mu^2_j-\mu^2_k)y_{ij,k}}
                {2(\mu^2_i+\mu^2_j+(1-\mu^2_i-\mu^2_j-\mu^2_k)y_{ij,k})}
        (1\mp v_{ij,i}v_{ij,k})\,,
        \label{eq:FFmassiveboundszi}\\
        y_-   
        &=& 
        \frac{2\mu_i\mu_j}{1-\mu^2_i-\mu^2_j-\mu^2_k}\,,\;\;\mbox{\rm and}\;\;\; 
        y_+ 
        = 
        1-\frac{2\mu_k\,(1-\mu_k)}{1-\mu^2_i-\mu^2_j-\mu^2_k}\,,
        \label{eq:FFmassiveboundsyijk}
        \eea
        respectively. 
\item \underline{Splitting kernels:}\\
        The polarisation-averaged QCD splitting kernels $\langle \bV_{ij,k}\rangle$ 
        read
        \begin{eqnarray}
        \langle \bV_{\Q_i\g_j,k}(\tilde z_i,y_{ij,k})\rangle 
        &=& 
        \CF\left\{\frac{2}{1-\tilde z_i+\tilde z_iy_{ij,k}}-
                        \frac{\tilde v_{ij,k}}{v_{ij,k}}
                        (1+\tilde z_i + \frac{m^2_i}{p_ip_j})\right\}
        \label{eq:FFPqqmassive}\,,\\
        \langle \bV_{\g_i\g_j,k}(\tilde z_i,y_{ij,k})\rangle 
        &=& 
        2\CA\left\{\frac{1}{1-\tilde z_i+\tilde z_iy_{ij,k}}+
                        \frac{1}{\tilde z_i+y_{ij,k}-\tilde z_iy_{ij,k}}\right.\nn\\
        &&\hspace*{10mm}+\left.
                        \frac{\tilde z_i\,(1-\tilde z_i)-z_+z_--2}{v_{ij,k}}\right\}\,,
        \label{eq:FFPggmassive}\\
        \langle \bV_{\Q_i\Q_j,k}(\tilde z_i)\rangle 
        &=& 
        \TR\,\frac{1}{v_{ij,k}}\,
                \left\{1-2\left[\tilde z_i\,(1-\tilde z_i)-z_+z_-\right]\right\}
        \label{eq:FFPgqmassive}\,.
        \end{eqnarray}
        Here, Eq.\ (\ref{eq:FFPqqmassive}) describes the QCD splitting $Q \to Qg$, 
        of a massive quark $Q$, the case of a splitting anti-quark is formally 
        identical. The corresponding expressions for the splitting $g \to gg$, or 
        $g \to Q\bar Q $ are given in Eqs.\ (\ref{eq:FFPggmassive}) and 
        (\ref{eq:FFPgqmassive}), respectively.  Note that in the above splitting 
        kernels the free parameter $\kappa$ that occurs in the full NLO subtraction 
        scheme \cite{Catani:2002hc} has been set to zero to obtain the simplest 
        expressions for the different $\langle \bV_{ij,k}\rangle$.
  
        It should be stressed here that the scalar product $p_ip_j$ present in 
        Eq.\ (\ref{eq:FFPqqmassive}) can be written solely in terms of the splitting 
        variables and the scale $\kperpbfsq$:
        \bea\label{eq:FFSplitpipj}
        p_ip_j = 
        \frac{\kperpbfsq}{2\tilde z_i\,(1-\tilde z_i)} + 
        \frac{(1-\tilde z_i)m_i^2}{2\tilde z_i}+ 
        \frac{\tilde z_im_j^2}{2(1-\tilde z_i)}\,.
        \eea
        However, in Eq.\ (\ref{eq:FFPqqmassive}) the final-state gluon is massless 
        and correspondingly $m^2_j=0$ such that the last term of 
        Eq.\ (\ref{eq:FFSplitpipj}) vanishes in this specific case.
\item \underline{Matrix element:}\\
        Using the above splitting functions, the full $(m+1)$-parton matrix element 
        factorises in the soft and collinear limit according to
        \bea\label{eq:FFampl2massive}
        \left|{\cal M}_{m+1}\right|^2 
        = 
        \left|{\cal M}_{m}\right|^2\,
        \sum\limits_{ij}\sum\limits_{k\neq ij}\,
        \frac{1}{(p_i+p_j)^2-m^2_{ij}}\,\frac{1}{{\cal N}^{spec}_{ij}}\,
        8\pi\as\,\langle \bV_{ij,k}(\tilde z_i,y_{ij,k})\rangle\,,
        \eea
        cf.\ \cite{Catani:1996vz}, where the sum covers all the possible emitter-spectator 
        pairs. When combining this with the $(m+1)$-parton phase space a fully 
        factorised expression for the differential cross section is obtained,
        namely
        \bea\label{eq:FFSigmamassive}
        \d\hat\sigma_{m+1} 
        = 
        \d\hat\sigma_m\,\sum\limits_{ij}\sum\limits_{k\neq ij}\,
        \frac{\d y_{ij,k}}{y_{ij,k}}\,\d\tilde z_i\,\frac{d\phi}{2\pi}\,
        \frac{\as}{2\pi}\,\frac{1}{{\cal N}^{spec}_{ij}}\,
        J(y_{ij,k})\langle \bV_{ij,k}(\tilde z_i,y_{ij,k}) \rangle\,, 
        \eea
        where the Jacobian
        \bea\label{FFmassiveJ}
        J(y_{ij,k}) 
        = 
        \frac{1-\mu^2_i-\mu^2_j-\mu^2_k}{\sqrt{\lambda(1,\mu^2_{ij},\mu^2_k)}}\,
        \frac{1-y_{ij,k}}{1+\frac{\mu^2_i+\mu^2_j-\mu^2_{ij}}
                                {y_{ij,k}(1-\mu^2_i-\mu^2_j-\mu^2_k)}}
        \eea
        emerges from the phase-space factors of Eq.\ (\ref{eq:FF2BodyPS}) combined
        with the propagator term of Eq.\ (\ref{eq:FFampl2massive}).
\item \underline{Sudakov form factor:}\\
        A first step toward the construction of the corresponding Sudakov form factor
        is achieved by realising that the $y_{ij,k}$-integration in the equation above, 
        Eq.\ (\ref{eq:FFSigmamassive}), can be replaced by an integration over the 
        ordering parameter, the transverse momentum, according to
        \bea\label{eq:dkperpFF}
        \frac{\d y_{ij,k}}{y_{ij,k}}=\frac{\d \kperpbfsq}{\kperpbfsq}\,.
        \eea
        Cutting the available phase space through the requirement of a minimal 
        relative transverse momentum squared $\kperpbfsq>\kperpzero^2>0$ and some upper 
        limit $\kperpmax^2$ for the splitting products $i$ and $j$, the $\tilde z_i$ 
        integration boundaries become
        \bea\label{eq:FFPSlimits}
        z_-(\kperpmax^2,\kperpzero^2) 
        &=& 
        {\rm Max}
        \left(\frac12\left(1-\sqrt{1-\frac{\kperpzero^2}{\kperpmax^2}}\right)\,,\,
                z_-\right)\,,\\
        z_+(\kperpmax^2,\kperpzero^2) 
        &=& 
        {\rm Min}
        \left(\frac12\left(1+\sqrt{1-\frac{\kperpzero^2}{\kperpmax^2}}\right)\,,\,
                z_+\right)\,,
        \eea
        with $z_\mp$ taken from Eq.\ (\ref{eq:FFmassiveboundszi}).  Having chosen
        a valid pair for $\kperpbfsq$ and $\tilde z_i$ this can then easily be solved 
        for $y_{ij,k}$,
        \bea
        y_{ij,k} 
        = 
        \frac{1}{Q^2-m^2_i-m^2_j-m^2_k}
                \left(\frac{\kperpbfsq}{\tilde z_i(1-\tilde z_i)} +
        \frac{(1-\tilde z_i)m^2_i}{\tilde z_i}+
        \frac{\tilde z_im^2_j}{1-\tilde z_i}\right)\,.
        \eea
        If the calculated $y_{ij,k}$ fulfils the requirement $y_{ij,k} \in [y_-,y_+]$, 
        with $y_\mp$ defined in Eq.\ (\ref{eq:FFmassiveboundsyijk}), a valid splitting 
        has been constructed, \ie a physical branching allowed by phase space.  

        The Sudakov form factor corresponding to having no emission from one of the 
        process' final--final dipoles between the maximum transverse momentum squared 
        $\kperpmax^2$ and the infrared cut-off $\kperpzero^2$ reads
        \bea\label{eq:FFsudakov}
        \lefteqn{\Delta_{\rm FF}(\kperpmax^2,\kperpzero^2)}\nn\\
        &=& 
        \exp\left(-\sum\limits_{ij}\sum\limits_{k\neq ij}\,
                \frac{1}{{\cal N}^{spec}_{ij}}\,
                \int\limits_{\kperpzero^2}^{\kperpmax^2}
                \frac{\d \kperpbfsq}{\kperpbfsq}\,
                \int\limits_{z_-}^{z_+}\d\tilde z_i\,
                \frac{\as(\kperpbfsq)}{2\pi}\,J(y_{ij,k})
                \langle \bV_{ij,k}(\tilde z_i,y_{ij,k}) \rangle\right)\,.
        \eea
        As already advertised in Sec.\ \ref{sec:CSscales}, the scale of the running 
        coupling has thereby been chosen equal to the current transverse momentum 
        squared.
\item \underline{Physical kinematics:}\\
        Having a valid set of splitting variables, the actual physical branching
        kinematics must be constructed in order to fully specify the splitting 
        $\{\widetilde{ij},\tilde{k}\} \to \{i,j,k\}$.  In the most general case, both 
        the emitter and the spectator parton are massive, prohibiting a simple 
        Sudakov parametrisation of $p_i$ and $p_j$ in terms of light-like momenta 
        $\tilde p_{ij}$ and $\tilde p_k$.  Instead they must be expressed in 
        light-cone kinematics with massive base momenta.  In the emitter-spectator 
        centre-of-mass frame the new spectator momentum can be fixed to
        \bea
        p_k 
        &=& 
        \frac{\sqrt{\left[2\mu^2_k+(1-\mu^2_i-\mu^2_j-\mu^2_k)(1-y_{ij,k})\right]^2-
                        4\mu^2_k}}
             {\sqrt{\lambda(1,\mu^2_{ij},\mu^2_k)}}
        \left(\tilde p_k-\frac12\left[1+\mu^2_k-\mu^2_{ij}\right]Q\right)\nn\\
        &&+
        \left[\frac12(1-\mu^2_i-\mu^2_j-\mu^2_k)(1-y_{ij,k})+\mu^2_k\right]Q\,.
        \eea
        Then the situation is most easily discussed in a frame where $Q-p_k$ is at 
        rest and the momentum $p_k$ points along the $z$-direction.  In this frame, 
        the light-cone momenta of $Q-p_k$ and $p_k$ can be written as
        \bea
        Q-p_k = (M,M,\vec{0})\;\;\;\mbox{\rm and}\;\;\;
        p_k   = (m_k\,e^x,m_k\,e^{-x},\vec{0})\,.
        \eea
        The ansatz for the light-cone momenta of the new emerging final-state 
        partons reads
        \bea
        p_i = (m_{i,\perp}\,e^y,m_{i,\perp}\,e^{-y},\vec{l}_\perp)\,,\;\;\;
        p_j = (m_{j,\perp}\,e^z,m_{j,\perp}\,e^{-z},-\vec{l}_\perp)\,,
        \eea
        with $m_\perp$ being the transverse mass of the respective parton, defined 
        according to 
        \bea
        m_\perp = \sqrt{m^2+\vec{l}_\perp^{2}}\,.
        \eea
        The kinematics is fully determined through energy-momentum conservation and 
        the constraint
        \bea
        \tilde z_i = 1-\tilde z_j = \frac{p_ip_k}{p_ip_k+p_jp_k}\,.
        \eea
        Then,
        \bea
        \vec{l}_\perp^2 
        = 
        \left(\frac{M^2+m^2_i+m^2_j}{2M}\right)^2 - 
        m^2_i - 
        \left(\frac{M^2+m^2_i+m^2_j-2M^2\tilde z_i}{2M}\,
        \left(\frac{\cosh x}{\sinh x}\right)\right)^2\,,
        \eea
        and
        \bea
        \cosh y 
        &=& 
        \frac{M^2+m^2_i-m^2_j}{2Mm_{i,\perp}}\,,\quad 
        \sinh y 
        = 
        \frac{\cosh x}{\sinh x}
                \left(\cosh y-\frac{M\tilde z_i}{m_{i,\perp}}\right)\,,\\
        \cosh z 
        &=& 
        \frac{M^2-m^2_i+m^2_j}{2Mm_{j,\perp}}\,,\quad 
        \sinh z = \frac{\cosh x}{\sinh x}
                \left(\cosh z-\frac{M(1-\tilde z_i)}{m_{j,\perp}} \right)\,.
        \eea
        Expressed through ordinary four-vectors the parton momenta in this frame read
        \bea
        p_i 
        &=& 
        (m_{i,\perp}\cosh y,l_{\perp}\cos\phi,l_{\perp}\sin\phi,m_{i,\perp}\sinh y)\,,\\
        p_j 
        &=& 
        (m_{j,\perp}\cosh z,-l_{\perp}\cos\phi,-l_{\perp}\sin\phi,m_{j,\perp}\sinh z)\,,
        \eea
        with the angle $\phi$ not fixed by the splitting and therefore uniformly 
        distributed in the transverse plane.  The kinematics is completed by 
        rotating and boosting back the momenta $p_i$, $p_j$ and $p_k$ into the 
        laboratory frame.

        If the spectator is massless, the new final-state momenta can alternatively 
        be given in a simple Sudakov parametrisation in the centre-of-mass frame of 
        the emitter and the spectator:
        \bea
        p_i 
        &=& 
        \hphantom{(1-\tilde)}\tilde z_i \,\tilde p_{ij}+
        \frac{\hphantom{(1(}\kperpbfsq-\tilde z_i^2m_{ij}^2+m_i^2\hphantom{()))}}
                {\hphantom{(1-}\tilde z_i\,2\tilde p_{ij}\tilde p_k\hphantom{))}}\,
        \tilde p_k + 
        \kperp\,,
        \label{eq:FFpiKIN}\\
        p_j 
        &=& 
        (1-\tilde z_i) \,\tilde p_{ij}+
        \frac{\kperpbfsq-(1-\tilde z_i)^2m_{ij}^2+m_j^2}
                {(1-\tilde z_i)\,2\tilde p_{ij}\tilde p_k}\,\tilde p_k - 
        \kperp\,,
        \label{eq:FFpjKIN}\\
        p_k 
        &=& 
        \left(\frac{(1-\mu^2_i-\mu^2_j)(1-y_{ij,k})}{1-\mu^2_{ij}}\right)\tilde p_k
        \label{eq:FFpkKIN}\,, 
        \eea
        with the spacelike transverse-momentum vector $\kperp$ pointing in a 
        direction perpendicular to both the emitter and the spectator momentum. 
\end{itemize}

\subsubsection{Massless case}
\label{sec:kinematicsFFmassless}

The case of a final-final splitting is considerably simpler in the massless limit,
\ie where all occurring partons can be treated as massless, 
$\tilde p^2_{ij}=\tilde p^2_k=p^2_k=p_i^2=p_j^2=0$.  In this case, of course, the 
variables chosen to specify the splitting remain unchanged with respect to the 
fully massive case.  However, neglecting masses the ordering parameter reduces to
\bea\label{eq:FFKT}
\kperpbfsq 
= 
Q^2y_{ij,k}\,\tilde z_i(1-\tilde z_i)
= 
2\tilde p_{ij}\tilde p_{k}\,\,y_{ij,k}\,\tilde z_i\,(1-\tilde z_i)\,,
\eea
with the identification of $Q^2=2\tilde p_{ij}\tilde p_k$ this is identical with the 
transverse momentum defined in Eq.\ (\ref{eq:FSmasslesscollinearlimit}).  The full 
phase space for the emission of an extra parton extends to $\tilde z_i \in [0,1]$, 
$y_{ij,k} \in [0,1]$, whereas $\phi$ again uniformly covers the interval $[0,2\pi]$.  

In the massless limit also the spin averaged splitting kernels 
$\langle \bV_{ij,k}\rangle$ simplify considerably, namely to
\begin{eqnarray}
\langle \bV_{\q_i\g_j,k}(\tilde z_i,y_{ij,k})\rangle 
&=&
\CF\left\{\frac{2}{1-\tilde z_i+\tilde z_iy_{ij,k}}-(1+\tilde z_i)\right\}
\label{eq:FFPqq}\,,\\
\langle \bV_{\g_i\g_j,k}(\tilde z_i,y_{ij,k})\rangle 
&=&
2\CA\left\{\frac{1}{1-\tilde z_i+\tilde z_iy_{ij,k}}+
                \frac{1}{\tilde z_i+y_{ij,k}-\tilde z_iy_{ij,k}}-
                2+\tilde z_i\,(1-\tilde z_i)\right\}
\label{eq:FFPgg}\,,\\
\langle \bV_{\q_i\q_j,k}(\tilde z_i)\rangle
&=&
\TR\left\{1-2\tilde z_i\,(1-\tilde z_i)\right\}
\label{eq:FFPgq}\,.
\end{eqnarray}
When combining the factorised form of the $(m+1)$-parton phase space,
\bea
\d\Phi_{m+1} 
= 
\d\Phi_m \sum\limits_{ij}\sum\limits_{k\neq ij}\frac{2p_ip_j}{16\pi^2}\,
\frac{\d y_{ij,k}}{y_{ij,k}}\,d\tilde z_i\,\frac{d\phi}{2\pi}\,(1-y_{ij,k})\,
\Theta(\tilde z_i\,(1-\tilde z_i))\,\Theta(y_{ij,k}(1-y_{ij,k}))\,,
\eea
with the corresponding expression for the $(m+1)$-parton matrix element,
\bea
\left|{\cal M}_{m+1}\right|^2 
= 
\left|{\cal M}_{m}\right|^2\,\sum\limits_{ij}\sum\limits_{k\neq ij}
\frac{1}{2p_ip_j}\,\frac{1}{{\cal N}^{spec}_{ij}}\,8\pi\as\,
\langle \bV_{ij,k}(\tilde z_i,y_{ij,k})\rangle\,,
\eea
the fully factorised form of the $(m+1)$-parton differential cross section is 
recovered
\bea
\d\hat\sigma_{m+1} 
= 
\d\hat\sigma_m\,\sum\limits_{ij}\sum\limits_{k\neq ij}\,
\frac{\d y_{ij,k}}{y_{ij,k}}\,d\tilde z_i\,\frac{d\phi}{2\pi}\,
\frac{\as}{2\pi}\,\frac{1}{{\cal N}^{spec}_{ij}}\, J(y_{ij,k})
\langle \bV_{ij,k}(\tilde z_i,y_{ij,k}) \rangle\,. 
\eea
However, in this case, the Jacobian $J(y_{ij,k})$ simply is given by
\bea
J(y_{ij,k}) = 1-y_{ij,k}\,.
\eea
With the transverse momentum defined according to Eq.\ (\ref{eq:FFKT}) again the 
identity 
\bea
\frac{\d y_{ij,k}}{y_{ij,k}}=\frac{\d \kperpbfsq}{\kperpbfsq}\,,
\eea
is found.  Choosing $\kperpbfsq$ as the evolution variable with its lower cut-off 
given by $\kperpzero^2$ and the upper limit by $\kperpmax^2$ the $\tilde z_i$ 
integration range reduces to 
\bea
z_\mp(\kperpmax^2,\kperpzero^2) &=& 
\frac12\left(1\mp\sqrt{1-\frac{\kperpzero^2}{\kperpmax^2}}\right)\,.
\eea
Given a valid set of $\kperpbfsq$ and $\tilde z_i$ this can be solved for 
\bea
y_{ij,k} = \frac{\kperpbfsq}{Q^2\tilde z_i(1-\tilde z_i)}\,,
\eea
completing the determination of the splitting variables.  Making the necessary 
replacements when going from massive partons to massless the Sudakov form factor
given in Eq.~(\ref{eq:FFsudakov}) yields the corresponding non-branching probability.
The massless kinematics can be derived from Eqs.\ (\ref{eq:FFpiKIN})-(\ref{eq:FFpkKIN}) 
by setting $\mu_{ij}=\mu_i=\mu_j=0$, accordingly
\bea
p_i 
&=& 
\hphantom{(1-\tilde)}\tilde z_i \,\tilde p_{ij}+
\frac{\hphantom{(1}\kperpbfsq\hphantom{(())}}
     {\hphantom{(1-}\tilde z_i\,2\tilde p_{ij}\tilde p_k\hphantom{))}}\,
\tilde p_k + \kperp\,,
\label{eq:FFpiKINmassless}\\
p_j 
&=& 
(1-\tilde z_i) \,\tilde p_{ij}+
\frac{\kperpbfsq}{(1-\tilde z_i)\,2\tilde p_{ij}\tilde p_k}\,\tilde p_k - \kperp\,,
\label{eq:FFpjKINmassless}\\
p_k 
&=& 
\left(1-y_{ij,k}\right)\tilde p_k
\label{eq:FFpkKINmassless}\,.
\eea

\subsection{Final-state emitter and initial-state spectator}
\label{sec:kinematicsFI}

\begin{figure}
\centerline{
  \begin{picture}(160,120)(0,0)
  \Line(80,50)(110,70)
  \Line(110,70)(130,100)
  \Line(80,50)( 40, 20)
  \LongArrow( 52,18)( 67, 30)
  \LongArrow(110,85)(120, 100)
  \LongArrow(120,60)(135, 60)
  \Line(110,70)(143,70)
  \Vertex(110,70){2.5}
  \GCirc(80,50){10}{1}
  \put( 90, 70){$\widetilde{ij}$}
  \put(140, 100){$i$}
  \put(145, 60){$j$}  
  \put( 28, 18){$a$}
  \put( 40, 50){$\bV_{ij}^a$}
  \put( 65, 18){$p_a$}
  \put(100, 99){$p_i$}
  \put(120, 50){$p_j$}
  \end{picture} }
\caption{Sketch of the splitting of a final-state parton accompanied by an 
        initial-state spectator.  The blob denotes the $m$-parton matrix element.  
        The incoming and outgoing lines label the initial- and final-state partons, 
        respectively.}
\label{fig:split_FI}
\end{figure}
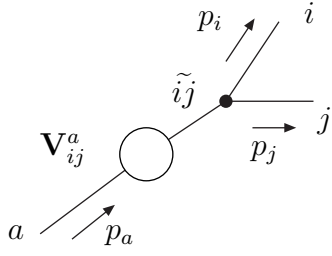

In this section, the case of a final-state emission with the spectator being an 
initial-state parton $a$ is worked out.  The splitting schematically reads 
$\{\widetilde{ij},\tilde a\}\to \{i,j,a\}$, for a pictorial representation of the 
configuration, cf.\ Fig.\ \ref{fig:split_FI}.  This configuration emerges for the 
first time when considering deep-inelastic lepton scattering (DIS), where one 
incoming line carries colour charge, or in configurations like vector boson fusion,
with no colour exchange between the two hadrons.  However, besides the singularity 
related to a final-state splitting, there is also a singular region for the splitting 
of the initial-state QCD parton, which needs to be included in such processes. 
This situation will be investigated in detail in Sec.\ \ref{sec:kinematicsIF}.  

\subsubsection{Massive case}
\label{sec:kinematicsFImassive}

The initial line is always assumed to be massless, however, all final-state 
particles can be massive.  Accordingly,
\bea
\tilde p^2_{ij}=m^2_{ij}\,\quad 
\tilde p^2_{a}=p^2_a=0\,\quad 
p^2_i=m^2_i\,,\quad p^2_j=m^2_j\,.
\eea
To avoid on-shell decays being described incorrectly, again 
$m^2_{ij}\leq m^2_i + m^2_j$ should hold true.  
\begin{itemize}
\item \underline{Kinematics:}\\
        Four-momentum conservation is incorporated through the condition
        \bea
        \tilde p_{ij}-\tilde p_a = p_i+p_j-p_a \equiv Q\,. 
        \eea
        Defining the Lorentz-invariants 
        \bea
        x_{ij,a}   
        &=& 
        \frac{p_ip_a+p_jp_a-p_ip_j + 
        \frac12(m^2_{ij}-m^2_i-m^2_j)}{p_ip_a+p_jp_a}\,,\\
        \tilde z_i 
        &=& 
        \frac{p_ip_a}{p_ip_a+p_jp_a}\,,\quad 
        \tilde z_j 
        = 
        \frac{p_jp_a}{p_ip_a+p_jp_a} = 1- \tilde z_i\,,\label{eq:FIzizj}
        \eea
        the relative transverse momentum of the new emerging final-state partons 
        is given by
        \bea\label{eq:FIKTmassive}
        \kperpbfsq 
        = 
        2\tilde p_a\tilde p_{ij}\,\frac{1-x_{ij,a}}{x_{ij,a}}\,
                \tilde z_i\,(1-\tilde z_i) - 
        (1-\tilde z_i)^2m^2_i - \tilde z^2_im^2_j\,.
        \eea
\item \underline{Phase space:}\\
        The factorised form of the three-parton phase space reads 
        \cite{Catani:2002hc}
        \bea
        \d\Phi(p_i,p_j;Q+p_a)
        =
        \int\limits_0^1\d x\d\Phi(\tilde p_{ij};Q+xp_a)\,
                [\d p_i(\tilde p_{ij};p_a,x)]\,\Theta(x_+-x)\,,
        \eea
        with the single-parton phase space factor
        \bea
        [\d p_i(\tilde p_{ij};p_a,x)]
        =
        \frac{2\tilde p_{ij}p_a}{16\pi^2}\,\frac{\d\phi}{2\pi}\,
        \d\tilde z_i\,\d x_{ij,a}\,\delta(x-x_{ij,a})\,,
        \eea
        and the integration boundaries 
        \bea\label{eq:FInaivelimits}
        x_- &=& 0\,,\quad 
        x_+ = 1+\mu^2_{ij}-(\mu_i+\mu_j)^2\,,\\
        z_\mp &=& \frac{1-x+\mu^2_{ij}+\mu^2_i-\mu^2_j\mp
                        \sqrt{(1-x+\mu^2_{ij}-\mu^2_i-\mu^2_j)^2-4\mu^2_i\mu^2_j}}
                        {2(1-x+\mu^2_{ij})}\,.
        \eea
        Here, again rescaled parton masses have been introduced, 
        \bea
        \mu_n = \frac{m_n}{\sqrt{2\tilde p_{ij}\tilde p_a / x_{ij,a}}} 
        \quad (n=i,j,ij)\,.
        \eea
\item \underline{Splitting kernels:}\\
        The polarisation-averaged QCD dipole splitting kernels 
        $\langle \bV^a_{ij}(\tilde z_i,x_{ij,a})\rangle$ read
        \begin{eqnarray}
        \langle \bV^a_{\Q_i\g_j}(\tilde z_i,x_{ij,a})\rangle 
        &=& 
        \CF\left\{\frac{2}{1-\tilde z_i+(1-x_{ij,a})}-(1+\tilde z_i)-
                \frac{m^2_i}{p_ip_j}\right\}
        \label{eq:FIPqqmassive}\,,\\&&\nn\\
        \langle \bV^a_{\g_i\g_j}(\tilde z_i,x_{ij,a})\rangle 
        &=& 
        2\CA\left\{\frac{1}{1-\tilde z_i+(1-x_{ij,a})}+
                \frac{1}{\tilde z_i+(1-x_{ij,a})}-
                2+\tilde z_i\,(1-\tilde z_i)\right\}
        \label{eq:FIPggmassive}\,,\nn\\&&\\
        \langle \bV^a_{\Q_i\Q_j}(\tilde z_i)\rangle 
        &=& 
        \TR\left\{1-2(z_+-\tilde z_i)(z_--\tilde z_i)\right\}
        \label{eq:FIPgqmassive}\,.
        \end{eqnarray}
        The scalar product of the {\it a priori} unknown momenta $p_i$ and $p_j$ in 
        Eq.\ (\ref{eq:FIPqqmassive}) can again be expressed according to Eq.\ 
        (\ref{eq:FFSplitpipj}).  
\item \underline{Matrix element:}\\
        Combining the $(m+1)$-parton phase space with the factorised form of the matrix 
        element,
        \bea\label{eq:FIMEmassive}
        \left|{\cal M}_{m+1}\right|^2 
        = 
        \left|{\cal M}_{m}\right|^2\,\sum\limits_{ij}\sum\limits_{a}
        \frac{1}{(p_i+p_j)^2-m^2_{ij}}\,\frac{1}{{\cal N}^{spec}_{ij}}\,\frac{1}{x_{ij,a}}\,
        8\pi\as\,\langle \bV^a_{ij}(\tilde z_i,x_{ij,a})\rangle\,,
        \eea
        one obtains the fully differential cross section for the emission of one additional 
        parton in that configuration
        \bea\label{eq:FISigmamassive}
        \d\hat\sigma_{m+1} 
        = 
        \d\hat\sigma_m\sum\limits_{ij}\sum\limits_{a}\,
        \frac{\d x_{ij,a}}{x_{ij,a}}\,\d\tilde z_i\,\frac{\d\phi}{2\pi}\,
        \frac{\as}{2\pi}\,\frac{1}{{\cal N}^{spec}_{ij}}\,\frac{1}{1-x_{ij,a}}\,
        \langle \bV^a_{ij}(\tilde z_i,x_{ij,a})\rangle\,,
        \eea
        where the sum covers all the possible colour-connected emitter-spectator pairings.  
        The Jacobian of the variable transformation in this case reads
        \bea
        J(x_{ij,a})=\frac{1}{1-x_{ij,a}}\,.
        \eea 
        Taking into account that the initial parton actually stems from a hadronic initial 
        state, a corresponding parton distribution function (PDF) emerges.  Absorbing it 
        into the Jacobian yields
        \bea
        \tilde J({x_{ij,a};\mu_F^2})
        =
        \frac{1}{1-x_{ij,a}}\,\frac{f_a(\eta_a/x_{ij,a},\mu_F^2)}{f_a(\eta_a,\mu_F^2)}\,. 
        \eea
        Here, $\eta_a$ is the momentum fraction of the spectator parton $a$ and 
        $f_a(\eta_a,\mu_F^2)$ the corresponding hadronic PDF evaluated at some scale 
        $\mu_F^2$.  In Sec.\ \ref{sec:CSscales} this scale has been set to $\mu_F = \kperpbf$.  
        The parton distribution function $f_a(\eta_a/x_{ij,a},\mu_F^2)$ corresponds to 
        the new incoming momentum and is also evaluated at scale $\mu_F^2$. 
\item \underline{Sudakov form factor:}\\
        Note that Eq.\ (\ref{eq:FIKTmassive}) implies that 
        \bea\label{eq:dkperpFImassive}
        \frac{\d x_{ij,a}}{x_{ij,a}}=(1-x_{ij,a})\,\frac{\d \kperpbfsq}{\kperpbfsq}\,.
        \eea
        With $\kperpbfsq$ taken as the evolution scale with an upper limit $\kperpmax^2$ and 
        the cut-off $\kperpzero^2$ the $\tilde z_i$ integration boundaries therefore are
        given by
        \bea\label{eq:FIPSlimits}
        z_-(\kperpmax^2,\kperpzero^2) 
        &=& 
        {\rm Max}\left(\frac12\left(1-\sqrt{1-\frac{\kperpzero^2}{\kperpmax^2}}\right)\,,\,
                        z_-\right)\,,\\
        z_+(\kperpmax^2,\kperpzero^2) 
        &=& 
        {\rm Min}\left(\frac12\left(1+\sqrt{1-\frac{\kperpzero^2}{\kperpmax^2}}\right)\,,\,
        z_+\right)
        \eea
        with $z_\pm$ given in Eq.\ (\ref{eq:FInaivelimits}).  Having determined $\kperpbfsq$ 
        and $\tilde z_i$ the variable $x_{ij,a}$ is calculated through
        \bea
        x_{ij,a} 
        = 
        1-\frac{\kperpbfsq+(1-\tilde z_i)^2m_i^2+\tilde z_i^2m_j^2-
                        \tilde z_i(1-\tilde z_i)(m_{ij}^2-m_i^2-m_j^2)}
                {\kperpbfsq+(1-\tilde z_i)^2m_i^2+\tilde z_i^2m_j^2+
                        \tilde z_i(1-\tilde z_i)(Q^2+2m_i^2+2m_j^2)}\,,
        \eea
        and has to fulfil the condition
        \bea 
        x_{ij,a} \in \left[\eta_a/\eta_{\rm max}\,,\,x_+\right]
        \eea
        to yield a valid branching.  Here, $\eta_{\rm max}$ corresponds to the maximal 
        allowed Bj{\o}rken-$x$ for the PDF.  Having at hand all ingredients, the Sudakov 
        form factor associated to the splitting of a final-state parton with an initial-state 
        spectator reads
        \bea\label{eq:FIsudakov}
        \lefteqn{\Delta_{\rm FI}(\kperpmax^2,\kperpzero^2)}\nn\\
        &=&
        \exp\left(-\sum\limits_{ij}\sum\limits_{a}\,
                \frac{1}{{\cal N}^{spec}_{ij}}\,
                \int\limits_{\kperpzero^2}^{\kperpmax^2}\frac{\d \kperpbfsq}{\kperpbfsq}\,
                \int\limits_{z_-}^{z_+}\d\tilde z_i\,\frac{\as(\kperpbfsq)}{2\pi}\,
                \frac{f_a(\eta_a/x_{ij,a},\kperpbfsq)}{f_a(\eta_a,\kperpbfsq)}\,
                \langle \bV^a_{ij}(\tilde z_i,x_{ij,a})\rangle\right)\,.\nn\\
        \eea
\item \underline{Physical kinematics:}\\
        The actual branching kinematics can be given in a Sudakov parametrisation. In the 
        Breit-frame of the emitter and spectator the two final-state momenta can be written 
        as  
        \bea
        p_i 
        &=& 
        \hphantom{(1-\tilde)}\tilde z_i\,\tilde p_{ij} + 
        \frac{\hphantom{1-}\kperpbfsq+m^2_i-\tilde z^2_i\,m^2_{ij}\hphantom{()),}}
             {\hphantom{(1-}\tilde z_i\, 2\tilde p_{ij}\tilde p_a\hphantom{))}}\,\tilde p_a + 
        \kperp\,,\\
        p_j 
        &=& 
        (1-\tilde z_i)\,\tilde p_{ij} + 
        \frac{\kperpbfsq+m^2_j-(1-\tilde z_i)^2\,m^2_{ij}}
             {(1-\tilde z_i)\, 2\tilde p_{ij}\tilde p_a}\,\tilde p_a - 
        \kperp\,,
        \eea
        with the spacelike-$\kperp$ being perpendicular to both the emitter and the 
        spectator momentum.  After the splitting the latter remains parallel to 
        $\tilde p_{a}$ but is rescaled according to
        \bea
        p_a &=& \frac{1}{x_{ij,a}}\tilde p_a\,.
        \eea
\end{itemize}

\subsubsection{Massless case}
\label{sec:kinematicsFImassless}

The modifications emerging in the massless limit are briefly discussed.  The splitting 
variable $x_{ij,a}$ simplifies to
\bea\label{eq:FIxmassless}
x_{ij,a}=\frac{p_ip_a+p_jp_a-p_ip_j}{p_ip_a+p_jp_a}\,,
\eea
whereas the momentum fractions $\tilde z_i$ and $\tilde z_j$ are still defined according to 
Eq.\ (\ref{eq:FIzizj}).  The invariant spacelike transverse momentum is simplified and reads
\bea\label{eq:FIKT}
\kperpbfsq = 
2\tilde p_a\tilde p_{ij}\,\frac{1-x_{ij,a}}{x_{ij,a}}\,\tilde z_i\,(1-\tilde z_i)\,.
\eea

While the $\g \to \g\g$ splitting function remains the same, the mass dependent terms drop 
out in the $\q \to \q\g$ and $\g \to \q\qbar$ kernels,
\begin{eqnarray}
\langle \bV^a_{\q_i\g_j}(\tilde z_i,x_{ij,a})\rangle 
&=& 
\CF\left\{\frac{2}{1-\tilde z_i+(1-x_{ij,a})}-(1+\tilde z_i)\right\}\label{eq:FIPqq}\,,\\
\langle \bV^a_{\q_i\q_j}(\tilde z_i)\rangle 
&=& 
\TR\left\{1-2\tilde z_i\,(1-\tilde z_i)\right\}\label{eq:FIPgq}\,.
\end{eqnarray}

Incorporating the factorisation of the $(m+1)$-parton matrix element 
and the corresponding phase space the fully differential $(m+1)$-parton cross section is 
still given by Eq.\ (\ref{eq:FISigmamassive}), with the appropriate Jacobian for hadronic 
initial states.  In the massless limit the phase-space boundaries are no longer constrained 
through finite mass terms, and therefore extend to 
\bea
x_{ij,a},\,\tilde z_i \in [0,1]\,. 
\eea

Eq.\ (\ref{eq:FIKT}) still implies that 
\bea\label{eq:dkperpFI}
\frac{\d x_{ij,a}}{x_{ij,a}}=(1-x_{ij,a})\,\frac{\d \kperpbfsq}{\kperpbfsq}\,.
\eea
When evolving in $\kperpbfsq$ from $\kperpmax^2$ and asking for a minimum separation 
$\kperpzero^2$ the allowed $\tilde z_i$ range is reduced to 
\bea
\tilde z_i \in \left[\frac12\left(1-\sqrt{1-\frac{\kperpzero^2}{\kperpmax^2}}\right)\,,\,
                     \frac12\left(1+\sqrt{1-\frac{\kperpzero^2}{\kperpmax^2}}\right)\right]
\eea
in the massless case.  The expression of the Sudakov from factor, Eq.\ (\ref{eq:FIsudakov}), 
of course remains unaltered.

The kinematics of the new final-state partons simplify to 
\bea
p_i 
&=& 
\hphantom{(1-\tilde)}\tilde z_i\,\tilde p_{ij} + 
\frac{\kperpbfsq}
        {\hphantom{(1-}\tilde z_i\, 2\tilde p_{ij}\tilde p_a\hphantom{))}}\,\tilde p_a + 
\kperp\,,\\
p_j 
&=& 
(1-\tilde z_i)\,\tilde p_{ij} + 
\frac{\kperpbfsq}{(1-\tilde z_i)\,2\tilde p_{ij}\tilde p_a}\,\tilde p_a - \kperp\,,
\eea
with $\kperp$ still being perpendicular to both the emitter and the spectator momentum. 
The new spectator momentum is still given by
\bea
p_a &=& \frac{1}{x_{ij,a}}\tilde p_a\,,
\eea
with $x_{ij,a}$ taken from Eq.\ (\ref{eq:FIxmassless}).

\subsection{Initial-state emitter and final-state spectator}
\label{sec:kinematicsIF}

\begin{figure}
\centerline{
  \begin{picture}(160,120)(0,0)
  \Line(80,50)( 20, 90)
  \Line(80,50)(120, 20)
  \LongArrow( 90,30)(105, 18)
  \LongArrow( 65,70)( 80, 80)
  \LongArrow( 20,80)( 35, 70)
  \Line( 50,70)( 80,90)
  \Vertex( 50,70){2.5}
  \GCirc(80,50){10}{1}
  \put( 49, 46){$\widetilde{ai}$}
  \put( 10, 88){$a$}
  \put( 87, 90){$i$}
  \put(125, 15){$k$}
  \put(100, 50){$\bV_k^{ai}$}
  \put( 90, 14){$p_k$}
  \put( 20, 63){$p_a$}
  \put( 76, 67){$p_i$}
  \end{picture} }
\caption{Splitting of an initial-state parton accompanied by a final-state spectator. 
         The blob denotes the $m$-parton matrix element.  The incoming and outgoing lines 
        label the initial- and final-state partons, respectively.}
\label{fig:split_IF}
\end{figure}
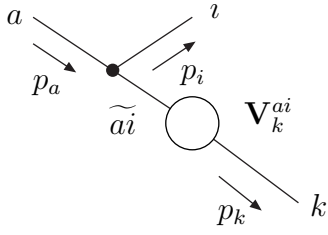

The case of an initial-state parton branching ($\widetilde {ai}$), accompanied by a 
final-state spectator ($\tilde k$) is sketched in Fig.\ \ref{fig:split_IF}.  This accounts 
for the situation where the emitter and the spectator parton studied in 
Sec.\ \ref{sec:kinematicsFI} exchange their r\^oles. 

\subsubsection{Massive case}
\label{sec:kinematicsIFmassive}

As stated above, treating initial-state particles as massless, final-state particles emitted 
from the initial state are assumed massless as well, the spectator mass, however, is 
arbitrary.  Accordingly, the momenta involved in the splitting 
$\{\widetilde{ai},\tilde k\} \to \{a,i,k\}$ have to fulfil the mass-shell relations 
\bea
\tilde p^2_{ai}= p^2_{i} = p^2_a = 0\,,\quad \tilde p^2_k = p^2_k = m^2_k\,.
\eea
and the momentum conservation condition
\bea
\tilde p_k - \tilde p_{ai} = p_i+p_k-p_a \equiv Q\,.
\eea 
\begin{itemize}
\item \underline{Kinematics:}\\
        The splitting can be specified by the variables
        \bea
        x_{ik,a}=\frac{p_ip_a+p_kp_a-p_ip_k}{p_ip_a+p_kp_a}\,,\quad 
        u_i = \frac{p_ip_a}{p_ip_a+p_kp_a}\,.
        \eea
        The transverse momentum squared parametrising the singular region where the emitted 
        parton $i$ becomes collinear with the initial-state parton $a$ then reads
        \bea\label{eq:IFKT}
        \kperpbfsq = 2\tilde p_{ai}\tilde p_k\,\frac{1-x_{ik,a}}{x_{ik,a}}\,u_i(1-u_i)\,.
        \eea
        To allow for a more compact notation, the rescaled spectator mass 
        \bea
        \mu_k = \frac{m_k}{\sqrt{2\tilde p_{ai}\tilde p_k / x_{ik,a}}}
        \eea
        is introduced.
\item \underline{Splitting kernels:}\\
        The QCD splitting kernels, taking into account possible non-zero spectator masses, 
        read
        \begin{eqnarray}
        \langle \bV_k^{\q_a\g_i}(x_{ik,a},u_i)\rangle 
        &=& 
        \CF\left\{\frac{2}{1-x_{ik,a}+u_i}-(1+ x_{ik,a})\right\}
        \label{eq:IFPqqmassive}\,,\\
        \langle \bV_k^{\q_a\q_i}(x_{ik,a})\rangle 
        &=& 
        \CF\left\{x_{ik,a}+2\frac{1-x_{ik,a}}{x_{ik,a}}-
                \frac{2\mu^2_k}{x_{ik,a}}\,\frac{u_i}{1-u_i}\right\}
        \label{eq:IFPqgmassive}\,,\\
        \langle \bV_k^{\g_a\g_i}(x_{ik,a},u_i)\rangle 
        &=& 
        2\CA\left\{\frac{1}{1-x_{ik,a}+u_i}+
                \frac{1-x_{ik,a}}{x_{ik,a}}-1 \right.\nn\\
        && \hspace*{15mm}\left.
                + x_{ik,a}(1 - x_{ik,a}) - 
                \frac{\mu^2_k}{x_{ik,a}}\,\frac{u_i}{1-u_i}\right\}\,,
                \label{eq:IFPggmassive}\\
        \langle \bV_k^{\q_a\q_i}(x_{ik,a})\rangle 
        &=& 
        \TR\left\{1-2 x_{ik,a}(1- x_{ik,a})\right\}\label{eq:IFPgqmassive}\,.
        \end{eqnarray}
\item \underline{Phase space:}\\
        The three-parton phase space is again obtained by a convolution of a two-parton 
        piece and a single-parton part,
        \bea\label{eq:IF3PS}
        \d\Phi(p_i,p_k;Q+p_a)=
        \int\limits_0^1\d x\d\Phi(\tilde p_k;Q+xp_a)\,[\d p_i(\tilde p_k;p_a,x)]\,,
        \eea
        where
        \bea
        [\d p_i(\tilde p_k;p_a,x)]=
        \frac{\d^4p_i}{2\pi}\delta(p^2_i)\,\Theta(x)\,
        \Theta(1-x)\,\delta(x-x_{ik,a})\,\frac{1}{1-u_i}\,,
        \eea
        or, more conveniently,
        \bea\label{eq:IF1PS}
        [\d p_i(\tilde p_k;p_a,x)]=
        \frac{2\tilde p_kp_a}{16\pi^2}\,\frac{\d\phi}{2\pi}\,\d x_{ik,a}\,\d u_i\,
        \Theta(u_i(1-u_i))\,\Theta(x(1-x))\,\delta(x-x_{ik,a})\,.
        \eea
        The upper limit for the $u_i$-integration contains a dependence on the spectator mass,
        \bea
        u_+ = \frac{1-x_{ik,a}}{1-x_{ik,a}+\mu^2_k}\,.
        \eea
\item \underline{Matrix element:}\\
        Using the factorisation property of the $(m+1)$-parton matrix element
        \bea\label{eq:IFME}
        \left|{\cal M}_{m+1}\right|^2 
        = 
        \left|{\cal M}_{m}\right|^2\,\sum\limits_{ai}\sum\limits_{k}
        \frac{1}{2p_ap_i}\,\frac{1}{{\cal N}^{spec}_{ai}}\,\frac{1}{x_{ik,a}}\,
        8\pi\as\,\langle \bV_k^{ai}(x_{ik,a},u_i)\rangle
        \eea
        in the soft and collinear limits and the relation
        \bea
        \frac{2\tilde p_kp_a}{2p_ap_i} = \frac{1}{u_i}
        \eea
        the $(m+1)$-parton fully differential cross section reads
        \bea\label{eq:dsigmaIF}
        \d\hat\sigma_{m+1} 
        = 
        \d\hat\sigma_m\sum\limits_{ai}\sum\limits_{k}\,
        \frac{\d u_i}{u_i}\,\d x_{ik,a}\,\frac{\d\phi}{2\pi}\,
        \frac{\as}{2\pi}\,\frac{1}{{\cal N}^{spec}_{ai}}\,\frac{1}{x_{ik,a}}\,
        \langle \bV_k^{ai}(x_{ik,a},u_i)\rangle\,.
        \eea
        The integration range of the variables $u_i$ and $x_{ik,a}$ is $[0,u_+]$ and $[0,1]$, 
        respectively, and $[0,2\pi]$ for $\phi$.  The Jacobian 
        \bea
        J(x_{ik,a})=\frac{1}{x_{ik,a}}
        \eea
        for the parton matrix element again is changed in hadronic interactions to
        include the effect of the PDFs, such that
        \bea
        \tilde J(x_{ik,a};\mu_F^2) = \frac{1}{x_{ik,a}}\,
        \frac{f_a(\eta_{ai}/x_{ik,a},\mu_F^2)}{f_{ai}(\eta_{ai},\mu_F^2)}\,,
        \eea
        where again, in the implementation here the choice for the factorisation scale is
        $\mu_F = \kperpbf$, cf.\ Sec.\ \ref{sec:CSscales}.  Note that the Jacobian takes into 
        account not only a change in Bj{\o}rken-$x$ but also a possible flavour change in 
        the process' initial state.  
\item \underline{Sudakov form factor:}\\
        The integration over $u_i$ in Eq.~(\ref{eq:dsigmaIF}) can be replaced by an 
        integration over $\kperpbfsq$ according to
        \bea\label{eq:dkperpIF}
        \frac{\d u_i}{u_i}=\frac{1-u_i}{1-2u_i}\frac{\d \kperpbfsq}{\kperpbfsq}\,.
        \eea
        The arising Jacobian is combined with the function $\tilde J(x_{ik,a};\mu_F^2)$ 
        to 
        \bea
        \tilde J(x_{ik,a},u_i;\mu_F^2) = 
        \frac{1-u_i}{1-2u_i}\,\frac{1}{x_{ik,a}}\,
        \frac{f_a(\eta_{ai}/x_{ik,a},\mu_F^2)}{f_{ai}(\eta_{ai},\mu_F^2)}\,.
        \eea
        With $\kperpbfsq > 0$ as the evolution variable and its cut-off being 
        $\kperpzero^2$ the $x_{ik,a}$ phase-space boundaries are
        \bea
        x_{ik,a} \in \left[\frac{\eta_{ai}}{\eta_{\rm max}}\,,
          \frac{Q^2}{Q^2+4\kperpzero^2}\right]\,,
        \eea
        with $\eta_{\rm max}$ the maximal allowed Bj{\o}rken-$x$ of the PDF.  With 
        $\kperpbfsq$ and $x_{ik,a}$ given, $u_i$ can be calculated and yields
        \bea
        u_i = \frac12\,\left(1 - \sqrt{1-\frac{4\kperp^2x_{ik,a}}{Q^2(1-x_{ik,a})}}\right)\,.
        \eea
        When $u_i \le u_+$ an allowed branching is found.  Thus the Sudakov form factor 
        for having no emission from an initial-state parton accompanied by a final-state 
        spectator between scales $\kperpmax^2$ and $\kperpzero^2$ can be written down,
        \bea
        \lefteqn{\Delta_{\rm IF}(\kperpmax^2,\kperpzero^2)}\nn\\
        &=& 
        \exp\left(-\sum\limits_{ai}\sum\limits_{k}\,\frac{1}{{\cal N}^{spec}_{ai}}\,
        \int\limits_{\kperpzero^2}^{\kperpmax^2}\frac{\d \kperpbfsq}{\kperpbfsq}\,
        \int\limits_{x_-}^{x_+}\d x_{ik,a}\,\frac{\as(\kperpbfsq/4)}{2\pi}\,
        \tilde J(x_{ik,a},u_i;\kperpbfsq)\,
        \langle \bV_k^{ai}(x_{ik,a},u_i)\rangle\right)\,.\nn\\
        \eea
\item \underline{Physical kinematics:}\\
        The new initial-state particle $a$ remains parallel to the original initial-state 
        parton, and is just rescaled by the splitting variable $x_{ik,a}$ such that
        \bea
        p_a = \frac{1}{x_{ik,a}}\,\tilde p_{ai}\,.
        \eea
        The two final-state momenta are most conveniently evaluated in the rest-frame of 
        $Q+p_a$ with $p_a$ pointing along the positive $z$-axis.  The corresponding light-cone 
        momenta read
        \bea
        Q+p_a = (M,M,\vec{0})\;\;\;\mbox{\rm and}\;\;\;
        p_a   = (2E_a,0,\vec{0})\,.
        \eea
        Note that the massless vector $p_a$ only has a light-cone $+$-component, given 
        by twice the energy of the parton. For $p_i$ and $p_k$ the ansatz
        \bea
        p_i = (l_\perp\,e^y,l_\perp\,e^{-y},\vec{l}_\perp)\,,\;\;\;
        p_k = (m_{k,\perp}\,e^z,m_{k,\perp}\,e^{-z},-\vec{l}_\perp)\,,
        \eea
        is used, with $m_\perp$ being the transverse mass.  Besides the energy- and 
        momentum-\-conservation requirement the momenta are constrained by the splitting 
        variables, 
        \bea
        u_i &=& \frac{p_ip_a}{(p_i+p_k)p_a} = \frac{l_\perp e^{-y}}{M}\,.
        \eea
        yielding 
        \bea
        \vec{l}_\perp^2 = (M^2-m^2_k)u_i - M^2u^2_i\,, 
        \eea
        for the transverse momentum squared.  This equals the physical transverse 
        momentum squared of parton $i$, $\kperpbfsq$.  Employing the relations
        \bea
        \cosh y &=& \frac{M^2-m^2_k}{2Ml_\perp}\,,\quad 
        \sinh y = \frac12\left(\frac{l_\perp}{Mu_i}-\frac{Mu_i}{l_\perp}\right)\,,\\ 
        \cosh z &=& \frac{M^2+m^2_k}{2Mm_{k,\perp}}\,,\quad 
        \sinh z = \frac12\left(\frac{m_{k,\perp}}{M(1-u_i)}-
                        \frac{M(1-u_i)}{m_{k,\perp}}\right)\,, 
        \eea
        the four-momenta of the final-state partons, in the frame specified above, read
        \bea
        p_i &=& (l_\perp\cosh y,l_{\perp}\cos\phi,l_{\perp}\sin\phi,l_\perp\sinh y)\,,\\
        p_k &=& (m_{k,\perp}\cosh z,-l_{\perp}\cos\phi,-l_{\perp}\sin\phi,m_{k,\perp}\sinh z)\,.
        \eea
        Again, $\phi$ has been uniformly distributed in the transverse plane.  The kinematics 
        is completed by rotating and boosting the momenta $p_a$, $p_i$ and $p_k$ back in the 
        laboratory frame.
\end{itemize}

\subsubsection{Massless case}
\label{sec:kinematicsIFmassless}

The massless limit of the scenario above, initial-state splittings accompanied by final-state 
spectators, $\{\widetilde{ai},\tilde k\}\to \{a,i,k\}$, corresponds to neglecting the 
spectator mass, $\tilde p^2_k=p^2_k=0$.  Apart from that, the splitting variables remain 
unchanged and the dependence on $m_k$, of course, disappears in the corresponding phase space 
boundaries.

Dropping the explicit mass terms present in $\langle \bV_k^{\q_a\q_i}(x_{ik,a})\rangle$ and 
$\langle \bV_k^{\g_a\g_i}(x_{ik,a},u_i)\rangle$ given in Eqs.\ (\ref{eq:IFPqgmassive}) and 
(\ref{eq:IFPggmassive}), respectively, the factorised form of the fully differential cross 
section can completely be taken over.

Neglecting the finite spectator masses the splitting kinematics is significantly simplified. 
In the emitter--spectator Breit-frame 
\bea
p_a &=& \frac{1}{x_{ik,a}}\,\tilde p_{ai}\,,\\
p_i &=& (1-u_i)\,\frac{1-x_{ik,a}}{x_{ik,a}}\,\tilde p_{ai}+
                \hphantom{(1-))}u_i\,\tilde p_k+\kperp\,,\\
p_k &=& \hphantom{(1-))}u_i\,\frac{1-x_{ik,a}}{x_{ik,a}}\,\tilde p_{ai}+
                (1-u_i)\,\tilde p_k-\kperp\,,
\eea
with $\kperp$ perpendicular to both the emitter and the spectator.

\subsection{Initial-state emitter and initial-state spectator}
\label{sec:kinematicsII}

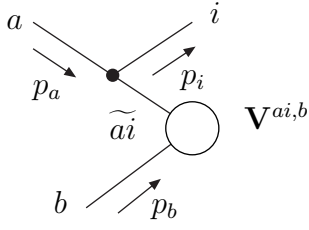
\begin{figure}
\centerline{
  \begin{picture}(160,120)(0,0)
  \Line(80,50)( 20, 90)
  \Line(80,50)( 40, 20)
  \LongArrow( 52,18)( 67, 30)
  \LongArrow( 65,70)( 80, 80)
  \LongArrow( 20,80)( 35, 70)
  \Line( 50,70)( 80,90)
  \Vertex( 50,70){2.5}
  \GCirc(80,50){10}{1}
  \put( 49, 46){$\widetilde{ai}$}
  \put( 10, 88){$a$}
  \put( 87, 90){$i$}
  \put( 28, 18){$b$}
  \put(100, 50){$\bV^{ai,b}$}
  \put( 65, 18){$p_b$}
  \put( 20, 63){$p_a$}
  \put( 76, 67){$p_i$}
  \end{picture} }
\caption{Schematical view of the splitting of an initial-state parton with an initial-state 
        parton as spectator.  The blob denotes the $m$-parton matrix element.  Incoming and 
        outgoing lines label the initial- and final-state partons, respectively.}
\label{fig:split_II}
\end{figure}

The last scenario to be studied is the splitting of an initial-state particle 
$\widetilde {ai}$, with the spectator $b$ being an initial-state parton as well, 
cf.\ Fig.\ \ref{fig:split_II}.  This type of branching occurs when considering hadron-hadron 
collisions, where both the initial-state particles are colour charged and therefore can be 
colour connected.  The simplest example for this configuration is the lowest order Drell-Yan 
process, where both the incoming quark and anti-quark can serve as emitter and spectator.

In contrast to all other cases discussed before, it turns out to be convenient to preserve 
the spectator momentum $p_b$ in this branching.  Since also the emitter momentum remains 
parallel to $p_a$,
\bea
\tilde p_{ai} = x_{i,ab}\,p_a\,,\quad{\rm{with}}\quad 
x_{i,ab}=\frac{p_ap_b-p_ip_a-p_ip_b}{p_ap_b}\,,
\eea
the transverse momentum of the emitted parton, $p_i$, has to be balanced by {\it all} other 
final-state momenta $k_j$. This does not only include the QCD partons, but all non-QCD 
particles, \eg leptons, as well.

\begin{itemize}
\item \underline{Kinematics:}\\
        Defining the variable
        \bea
        \tilde v_i = \frac{p_ip_a}{p_ap_b}
        \eea
        the transverse momentum squared of parton $i$ is given by 
        \bea\label{eq:IIKT}
        \kperpbfsq = 2\tilde p_{ai}p_b\,\tilde v_i\,\frac{1-x_{i,ab}-\tilde v_i}{x_{i,ab}}\,.
        \eea
        The four-momenta of the $m$-parton ensemble fulfil 
        \bea\label{eq:IIfourmom1}
        \tilde p_{ai}+ p_b - \sum\limits_{j=1}^m \tilde k_j  = 0\,,
        \eea 
        correspondingly the full set of $m+1$ particles has to satisfy
        \bea\label{eq:IIfourmom2}
        p_a+p_b - \sum\limits_{j=1}^m k_j -p_i = 0\,.
        \eea
\item \underline{Splitting kernels:}\\
        The polarisation-averaged splitting kernels $\langle \bV^{ai,b}\rangle$ depend on 
        $x_{i,ab}$ only and read
        \begin{eqnarray}
        \langle \bV^{\q_a\g_i,b}(x_{i,ab})\rangle 
        &=& 
        \CF\left\{\frac{2}{1-x_{i,ab}}-(1+ x_{i,ab})\right\}\label{eq:IIPqq}\,,\\
        \langle \bV^{\q_a\q_i,b}(x_{i,ab})\rangle 
        &=& 
        \CF\left\{x_{i,ab}+2\frac{1-x_{i,ab}}{x_{i,ab}}\right\}\label{eq:IIPqg}\,,\\
        \langle \bV^{\g_a\g_i,b}(x_{i,ab})\rangle 
        &=& 
        2\CA\left\{\frac{1}{1-x_{i,ab}}+\frac{1-x_{i,ab}}{x_{i,ab}}-
                1+ x_{i,ab}(1 - x_{i,ab})\right\}\label{eq:IIPgg}\,,\\
        \langle \bV^{\q_a\q_i,b}(x_{i,ab})\rangle
        &=&
        \TR\left\{1-2x_{i,ab}(1-x_{i,ab})\right\}\label{eq:IIPgq}\,.
        \end{eqnarray}
\item \underline{Phase space:}\\
        The final-state phase space can be written as follows \cite{Catani:2002hc}
        \bea
        \d\Phi(p_i,k_1,\dots;p_a+p_b)=
        \int\limits_0^1\d x\,\d\Phi(\tilde k_1,\dots;xp_a+p_b)\,[\d p_i(p_a,p_b,x)]\,,
        \eea
        with
        \bea
        [\d p_i(p_a,p_b,x)]=
        \frac{2p_ap_b}{16\pi^2}\,\frac{\d\phi}{2\pi}\,\d x_{i,ab}\,\d\tilde v_i\,
        \Theta(x(1-x))\,\Theta(\tilde v_i)\,\Theta\left(1-\frac{\tilde v_i}{1-x}\right)\,
        \delta(x-x_{i,ab})\,,
        \eea
        where $\phi$ is the polar angle in the plane perpendicular to $p_a$ and $p_b$.
\item \underline{Matrix element:}\\
        Combining this with the expression for the $(m+1)$-parton matrix element
        \bea
        \left|{\cal M}_{m+1}\right|^2 
        = 
        \left|{\cal M}_{m}\right|^2\,\sum\limits_{ai}\sum\limits_{b\neq ai}
        \frac{1}{2p_ap_i}\,\frac{1}{{\cal N}^{spec}_{ai}}\,\frac{1}{x_{i,ab}}\,
        8\pi\as\,\langle \bV^{ai,b}(x_{i,ab})\rangle
        \eea
        the differential cross section becomes
        \bea
        \d\hat\sigma_{m+1} = 
        \d\hat\sigma_m\sum\limits_{ai}\sum\limits_{b\neq ai}\,
        \frac{\d\tilde v_i}{\tilde v_i}\,\d x_{i,ab}\,\frac{\d\phi}{2\pi}\,
        \frac{\as}{2\pi}\,\frac{1}{{\cal N}^{spec}_{ai}}\,\frac{1}{x_{i,ab}}\,
        \langle \bV^{ai,b}(x_{i,ab})\rangle\,,
        \eea
        where $1-x_{i,ab}-\tilde v_i>0$ has to hold. The Jacobian can be read off as
        \bea
        J(x_{i,ab}) = \frac{1}{x_{i,ab}}\,,
        \eea
        or, including again the PDFs,  
        \bea
        \tilde J(x_{i,ab};\mu_F^2) = 
        \frac{1}{x_{i,ab}}\,\frac{f_a(\eta_{ai}/x_{i,ab},\mu_F^2)}
                        {f_{ai}(\eta_{ai},\mu_F^2)}\,.
        \eea
\item \underline{Sudakov form factor:}\\
        Regarding the transverse momentum given by Eq.~(\ref{eq:IIKT}) the identity
        \bea\label{eq:dkperpII}
        \frac{\d\tilde v_i}{\tilde v_i}=
        \frac{1-x_{i,ab}-\tilde v_i}{1-x_{i,ab}-2\tilde v_i}\frac{\d \kperpbfsq}{\kperpbfsq}\,,
        \eea
        can be employed to replace the $\tilde v_i$ integration with a $\kperpbfsq$-integral.  
        The resulting Jacobian, combined with $\tilde J(x_{i,ab};\mu_F^2)$, amounts to
        \bea
        \tilde J(x_{i,ab},\tilde v_i;\mu_F^2) = 
        \frac{1-x_{i,ab}-\tilde v_i}{1-x_{i,ab}-2\tilde v_i}\,\frac{1}{x_{i,ab}}\,
        \frac{f_a(\eta_{ai}/x_{i,ab},\mu_F^2)}{f_{ai}(\eta_{ai},\mu_F^2)}\,.
        \eea
        When evolving in $\kperpbfsq$ the dependence of the $x_{i,ab}$-integration 
        boundaries on the cut-off $\kperpzero^2$ read
        \bea
        x_{i,ab} \in \left[\frac{\eta_{ai}}{\eta_{\rm max}}\,,\frac{2\tilde p_ap_b}{2\tilde p_ap_b+4\kperpzero^2}
        \right]\,.
        \eea
        $\tilde v_i$ can be calculated from $\kperpbfsq$ and $x_{i,ab}$,
        \bea
        \tilde v_i = \frac{1-x_{i,ab}}{2}\, 
        \left(1-\sqrt{1-\frac{2\kperpbfsq x_{i,ab}}{\tilde p_ap_b(1-x_{i,ab})^2}}\right)\,.
        \eea
        The Sudakov form factor then reads
        \bea
        \lefteqn{\Delta_{\rm II}(\kperpmax^2,\kperpzero^2)}\nn\\
        &=&
        \exp\left(-\sum\limits_{ai}\sum\limits_{b\neq ai}\,\frac{1}{{\cal N}^{spec}_{ai}}\,
        \int\limits_{\kperpzero^2}^{\kperpmax^2}\frac{\d\kperpbfsq}{\kperpbfsq}\,
        \int\limits_{x_-}^{x_+}\d x_{i,ab}\,\frac{\as(\kperpbfsq/4)}{2\pi}\,
        \tilde J(x_{i,ab},\tilde v_i;\kperpbfsq)\,\langle 
        \bV^{ai,b}(x_{i,ab})\rangle\right)\,.\nn\\
        \eea
\item \underline{Physical kinematics:}\\
        The momenta of the $(m+1)$-parton ensemble, expressed through the emitter and 
        spectator momentum and the momenta of all other final-state particles of the 
        $m$-parton process, read
        \bea
        p_a &=& \frac{1}{x_{i,ab}}\,\tilde p_{ai}\,,\\
        p_i &=& \frac{1-x_{i,ab}-\tilde v_i}{x_{i,ab}}\,\tilde p_{ai}+
                        \tilde v_i\,p_b+\kperp\,,\\
        k_j &=& \Lambda(\tilde p_{ai}+p_b,p_a+p_b-p_i)\,\tilde k_j\,,
        \eea
        with $\kperp/\sqrt{\kperpbfsq}$ uniformly distributed in the transverse plane and 
        $\Lambda(\tilde p_{ai}+p_b,p_a+p_b-p_i)=\Lambda(\tilde K,K)$ being a proper 
        Lorentz transformation given by
        \bea\label{eq:IILorentztrafo}
        \Lambda^\mu_{\,\,\,\nu}(\tilde K,K) = 
        g^\mu_{\,\,\,\nu} - 
        \frac{2\,(\tilde K+K)^\mu\,(\tilde K+K)_\nu}{(\tilde K+K)^2} + 
        \frac{2\,K^\mu \tilde K_\nu}{\tilde K^2}\,.
        \eea
        Accordingly, the full set of final-state momenta compensates for the transverse 
        momentum of $p_i$, although they do not participate in the splitting.
\end{itemize}

\subsection{SUSY QCD splitting functions}
\label{sec:kinematicsSUSY}

In the minimal supersymmetric extension of the Standard Model the sector of strongly 
interacting particles is extended by the superpartners of the ordinary quark- and 
gluon-fields \cite{Haber:1984rc}.  The new particles participating in the strong interaction 
are the scalar-quarks, called squarks and the gluino. While the former are colour-triplets 
the gluino is a Majorana fermion in the adjoint representation, a colour-octet.

In order to be consistent with todays experimental (non-)observations the assumed SUSY 
particles have to be rather heavy. This renders the massless limit for these fields not 
applicable when describing their QCD interactions at the energies of the forthcoming 
colliders.  Based on that argument it is beyond the present scope to describe possible 
branchings like $g \to \tilde q \tilde q^*$, $g \to \tilde g \tilde g$ in a
quasi-collinear limit.  Rather, they are appropriately described using exact 
matrix element methods, as discussed \eg in \cite{Hagiwara:2005wg,Berdine:2007uv}. 

\begin{figure}
\bc
\includegraphics{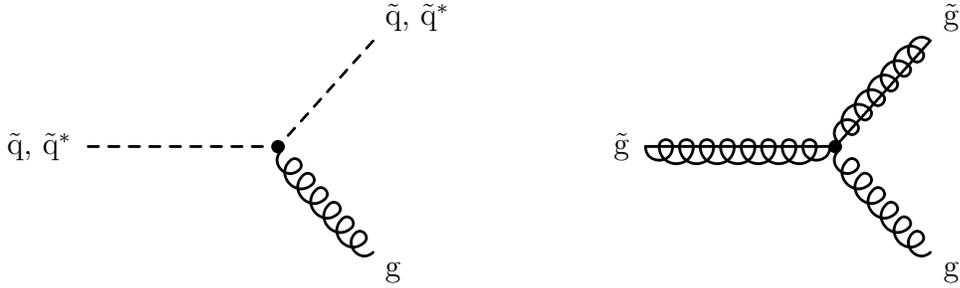}
\caption{The SUSY QCD vertices corresponding to gluon emission off (anti-)squarks and 
        gluinos.\label{fig:SUSYsplit}}
\ec
\end{figure}

Since the spin and the flavour of the spectator parton do not enter the splitting functions, 
the branchings of the Standard Model particles are not altered in supersymmetric extensions. 
The only SUSY QCD splittings that appear to be relevant in the context of a parton shower 
formulation are related to the emission of a gluon off a squark or anti-squark and off a 
gluino, cf.\ Fig.\ \ref{fig:SUSYsplit}.  Further, assuming that supersymmetric particles do 
not appear as partonic initial states those are solely final-state splittings.  The 
associated spectator, however, can be either in the final state or in the initial state.  

Due to its fermionic nature the splitting functions involving gluinos are equal to the 
corresponding splittings of massive quarks, cf.\ Eq.\ (\ref{eq:FFPqqmassive}) and 
Eq.\ (\ref{eq:FIPqqmassive}), only the colour factors have to be adopted from $C_F$ to 
$C_A$.

The kernel of the branching $\tilde q \to \tilde q g$ with the spectator also in the final 
state reads
\bea
\langle \bV_{\tilde \q_i\g_j,k}(\tilde z_i,y_{ij,k})\rangle
=
\CF\left\{\frac{2}{1-\tilde z_i+\tilde z_iy_{ij,k}}
-\frac{\tilde v_{ij,k}}{v_{ij,k}}\left(2 + \frac{m^2_i}{p_ip_j}\right)\right\}\,,
\eea
where all the variables have been defined in Sec.~\ref{sec:kinematicsFFmassive}.  If the 
spectator is in the initial state this becomes
\bea
\langle \bV^a_{\tilde \q_i\g_j}(\tilde z_i,x_{ij,a})\rangle=
\CF\left\{\frac{2}{1-\tilde z_i+(1-x_{ij,a})}- 2 - \frac{m^2_i}{p_ip_j}\right\}\,,
\eea
for the definitions of the variables used see Sec.~\ref{sec:kinematicsFImassive}.

Apart from the splitting kernels all the results derived in the corresponding sections 
describing the branchings of massive final-state partons with spectators in the final- 
or initial state can be taken over without any alteration.  This includes the exact 
phase-space factorisation as well as the parton kinematics defined there. 

%% file: cschecks.tex
\section{Comparing the hardest emission with matrix elements}
\label{sec:MEPScomparison}

In the following, the predictions for the hardest (first) emission of the parton 
shower algorithm will be worked out for different processes and compared with 
corresponding exact tree-level matrix element calculations.  The set of processes 
to be considered covers three-jet production in $e^+e^-$ collisions, 
cf.\ Sec.\ \ref{sec:MEPSee}, the first order real correction process to DIS, 
cf.\ Sec.\ \ref{sec:MEPSDIS}, and the production of a weak gauge boson accompanied 
by a light jet at hadron colliders, cf.\ Sec.\ \ref{sec:MEPSDY}.  These three 
examples constitute a full set of generic processes to reliably test the first 
emission of the proposed parton shower approach.

\subsection{Three-jet production at lepton-colliders}
\label{sec:MEPSee}

\begin{figure}
\bc
\includegraphics{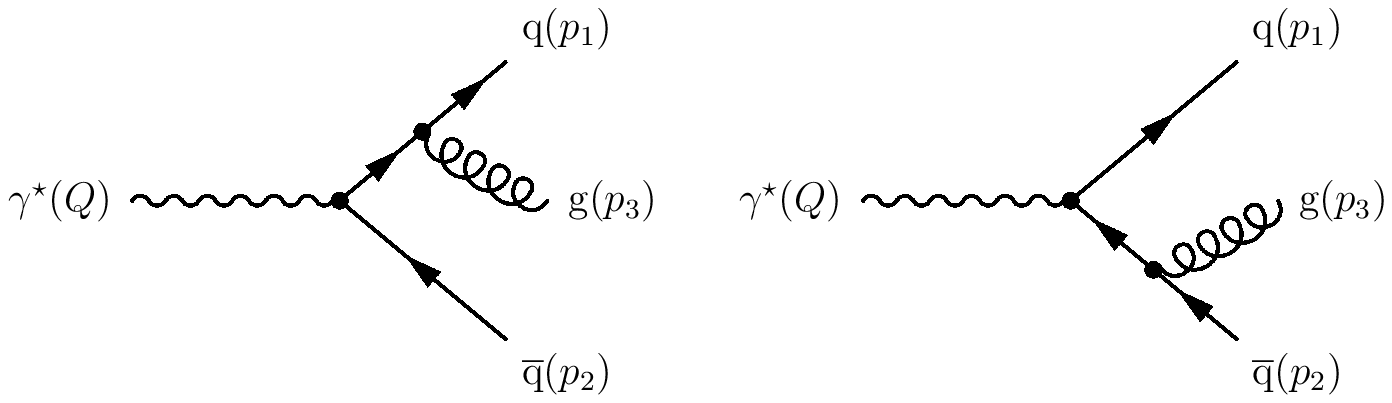}
\caption{The two first order $\as$ Feynman diagrams contributing to 
        $\gamma^\star\to \q\qbar\g$.\label{fig:eeqqg}}
\ec
\end{figure}

In this example the production of three jets at a lepton-collider is investigated.  Jet 
production proceeds via the $s$-channel exchange of a colour-singlet particle, namely 
a $\gamma^\star$ or $\Z^0$-boson.  The latter will be ignored in the discussion here.  
At first perturbative order in $\as$, two Feynman diagrams contribute to the 
matrix element $\gamma^\star\to \q\qbar\g$, corresponding to the emission of a gluon 
from either the final-state quark or the anti-quark, cf.\ Fig.\ \ref{fig:eeqqg}. \\

For convenience, the centre-of mass energy 
\bea
E_{\rm c.m.} \equiv\sqrt{Q^2}\,,
\eea
and the momentum fractions 
\bea
x_i \equiv \frac{2p_iQ}{Q^2}\,.
\eea
are introduced.  Neglecting the masses of the final-state particles the 
Lorentz-invariant Mandelstam variables for the $1\to 3$ process become
\bea
\shat &\equiv& (p_1+p_3)^2 = 2p_1p_3=Q^2(1-x_2)\,,\\
\that &\equiv& (p_2+p_3)^2 = 2p_2p_3=Q^2(1-x_1)\,,\\
\uhat &\equiv& (p_1+p_2)^2 = 2p_1p_2=Q^2(1-x_3)\,.
\eea
Energy-momentum conservation implies that 
\bea
x_1+x_2+x_3 =2\;\;\;\mbox{\rm and}\;\;\;
\shat+\that+\uhat =Q^2\,. 
\eea
The partonic differential cross section with respect to the quark and anti-quark
momentum fractions $x_{1,2}$ reads
\bea
\left.\frac{\d\hat\sigma}{\d x_1\d x_2}\right|_{\mathrm{ME}} = 
\hat\sigma_0\,\frac{\as}{2\pi}\,\CF\left[\frac{x_1^2+x_2^2}{(1-x_1)(1-x_2)}\right]\,, 
\eea 
where $\hat\sigma_0$ denotes the total cross section of the two-jet process 
$\gamma^\star \to \q\qbar$,
\bea
\hat\sigma_0 = 2\alpha_{\mathrm{qed}} e_{\q}^2E_{\rm c.m.}\,,
\eea
see for instance \cite{Field:1989uq}.

In the parton shower approach, two contributions occur as well.  They correspond to 
the timelike splitting of either the quark or the anti-quark, and the total result 
is just the incoherent sum of the two pieces.  To work this out, consider the case of 
the quark splitting with the anti-quark being the spectator parton.  Then, the 
shower variables are, cf.\ Sec.\ \ref{sec:kinematicsFFmassless},
\bea
y_{13,2}&=& 
\frac{p_1p_3}{p_1p_3+p_1p_2+p_2p_3}=\frac{\shat}{\shat+\uhat+\that}=
\frac{\shat}{Q^2}\,,\\
\tilde z_1 &=& \frac{p_1p_2}{p_1p_2+p_3p_2}=\frac{\uhat}{\uhat+\that}\,,
\eea
which, expressed in terms of the $x_{i}$, translate into
\bea
y_{13,2} = 1-x_2\;\;\;\mbox{\rm and}\;\;\;
\tilde z_1 = \frac{1-x_3}{x_2}=1-\frac{1-x_1}{x_2}\,.  
\eea
Accordingly, the cross section for the emission off the quark can be cast into the 
form
\bea\label{eq:PSeeq}
\left.\frac{\d\hat\sigma}{\d x_1\d x_2}\right|_{\mathrm{PS\q}}
=
\hat\sigma_0\,\frac{\as}{2\pi}\,\CF\left[\frac{1}{1-x_2}
\left(\frac{2}{2-x_1-x_2}-(1+x_1)\right) + \frac{1-x_1}{x_2}\right]\,.
\eea
The result for the emission of a gluon off the anti-quark can be obtained from 
Eq.\ (\ref{eq:PSeeq}) by $1 \leftrightarrow 2$.  Taken together, the total parton 
shower cross section yields
\bea
\lefteqn{\left.\frac{\d\hat\sigma}{\d x_1\d x_2}\right|_{\mathrm{PS}} =
\left.\frac{\d\hat\sigma}{\d x_1\d x_2}\right|_{\mathrm{PSq}} + 
\left.\frac{\d\hat\sigma}{\d x_1\d x_2}\right|_{\mathrm{PS\qbar}}}\nn\\
&=&
\hat\sigma_0\,\frac{\as}{2\pi}\,\CF
\left[\frac{x_1^2+x_2^2}{(1-x_1)(1-x_2)}+ \frac{1-x_1}{x_2}+\frac{1-x_2}{x_1}\right]\,.
\eea
Obviously, the parton shower cross section reproduces the matrix element calculation 
in both the soft and the collinear limit.  The only difference between the two results 
are two non-singular terms in the parton shower result that vanish as $x_{1,2} \to 1$.

\subsection{Real corrections to leading order DIS}
\label{sec:MEPSDIS}

The simplest physical process involving initial-state hadrons is deep-inelastic 
lepton-nucleon scattering (DIS), \ie $e^\pm p\to e^\pm+X$.  At leading order, two 
partonic processes contribute, namely $e^\pm\q \to e^\pm\q$ and 
$e^\pm\qbar \to e^\pm\qbar$, both of which must be convoluted with the initial 
hadron's PDF to obtain the hadronic cross section.  The interaction is mediated by 
virtual-photon and $\Z^0$-boson exchange.  In the following, however, only the 
$\gamma^\star$ channel is taken into account, for which the two partonic cross 
sections are equal. 

\begin{figure}
\bc
\includegraphics{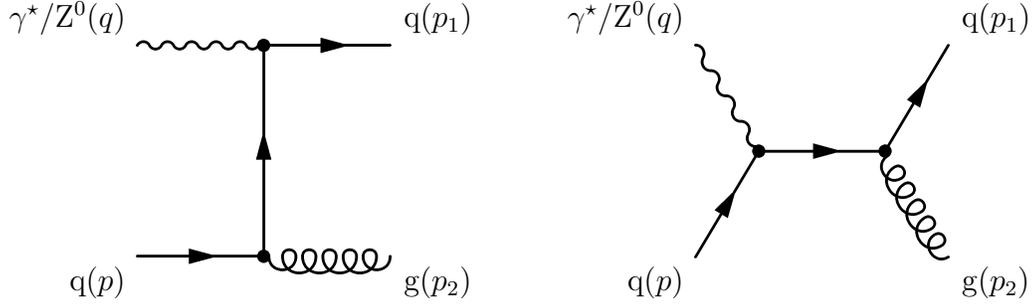}
\caption{The two leading order Feynman diagrams contributing to $\gamma^\star \q \to \q\g$.\label{fig:DISqg}}
\ec
\end{figure}

\begin{figure}
\bc
\includegraphics{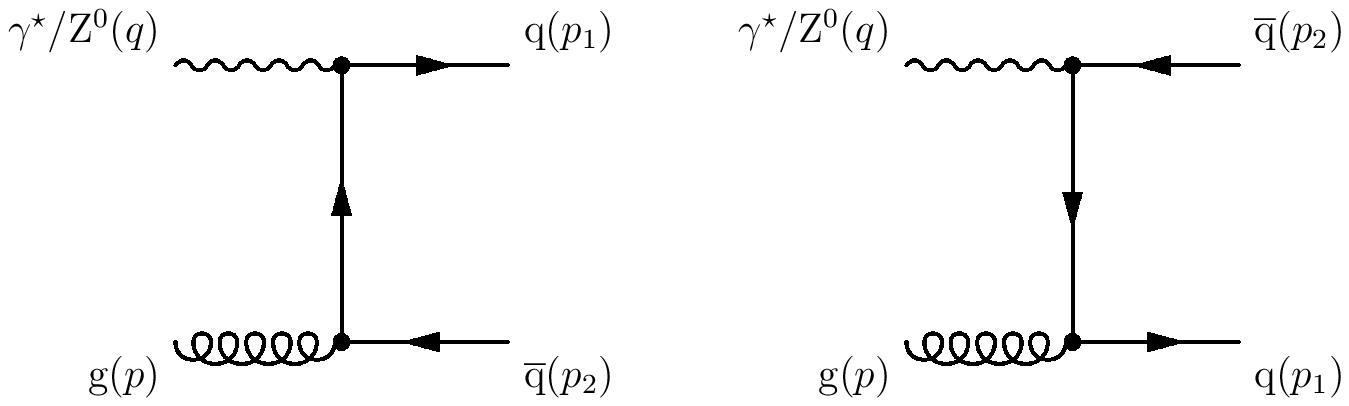}
\caption{The two possible Feynman diagrams for $\gamma^\star \g \to \q\qbar$.\label{fig:DISqq}}
\ec
\end{figure}

At next-to-leading order the quark can radiate a gluon before or after its interaction 
with the virtual photon, cf.\ Fig.\ \ref{fig:DISqg}.  Beyond this, at NLO the incoming 
quark may originate from a gluon in the initial hadron that produces a 
quark--anti-quark pair which the $\gamma^\star$ then couples to, cf.\ 
Fig.\ \ref{fig:DISqq}. The real emission matrix elements can be expressed through 
the kinematic variables 
\bea
Q^2 = -q^2\,,\quad x=\frac{Q^2}{2pq},\,\quad z_i=\frac{p_ip}{pq}\,, 
\eea 
where $q$ denotes the four-momentum of the off-shell photon, $p$ the incoming parton 
momentum and the $p_i$ label the momenta of the final-state partons.  The Mandelstam 
variables for the $2\to 2$ processes $\gamma^\star(q)\q(p) \to \q(p_1)\g(p_2)$ and  
$\gamma^\star(q)\g(p) \to \q(p_1)\qbar(p_2)$ are
\bea
\shat &\equiv& (q+p)^2\hphantom{_1}=\hphantom{-}2pq-Q^2\hphantom{_1}=
\hphantom{-}Q^2\,\frac{1-x}{x}\,,\\
\that &\equiv& (p_1-q)^2=-2p_1q-Q^2=-Q^2\,\frac{x+z_1}{x}\,,\\
\uhat &\equiv& (p_2-q)^2=-2p_2q-Q^2=-Q^2\frac{x+z_2}{x}\,.
\eea
Momentum conservation implies that $q+p = p_1+p_2$ and
\bea
\shat+\that+\uhat+Q^2=0\,.
\eea
In the following, the two real emission processes will be discussed in detail. 
 
\subsubsection{The gluon emission process}

The matrix element of the gluon emission channel $\gamma^\star(q)\q(p) \to \q(p_1)\g(p_2)$ 
reads \cite{Catani:1996vz,Field:1989uq}
\bea\label{eq:DISMEqg}
\left|{\cal{M}}_{2,\q}(p_1,p_2;p)\right|_{\mathrm{ME}}^2 
&=& 
\frac{8\pi\as}{Q^2}\CF \left[\frac{x^2+z_1^2}{(1-x)(1-z_1)}+2(1-3xz_1)\right]
\cdot\left|{\cal{M}}_{1,\q}(q+xp;xp)\right|^2\,,\nn\\
\eea
with ${\cal{M}}_{1,\q}(q+p;p)$ the matrix element of the lowest order process.

In the parton shower approach two contributions to this final state emerge.  First, 
the emission of the gluon from the initial-state quark with the final-state parton 
serving as spectator (IF) has to be considered.  Second, the initial-state parton acts 
as the spectator of the final-state splitting $\q \to \q\g$ (FI).
\begin{itemize}
\item IF:\\
        The ``parton shower''-matrix element of the initial-state splitting with 
        final-state spectator is obtained from Eq.\ (\ref{eq:IFME}) and is given by
        \bea
        \lefteqn{\left|{\cal{M}}_{2,\q}(p_1,p_2;p)\right|_{\mathrm{PSif}}^2}\nn\\ 
        &=& 
        \frac{1}{2pp_2}\,\frac{1}{x_{21,p}}\,
        8\pi\as\,\CF\left[\frac{2}{1-x_{21,p}+u_2}-(1+x_{21,p})\right]
        \cdot\left|{\cal{M}}_{1,\q}(q+xp;xp)\right|^2\,,
        \eea
        where the appropriate splitting function, Eq.\ (\ref{eq:IFPqqmassive}) with 
        $\mu^2_k=0$, has been inserted.  Employing the identities
        \bea
        x_{21,p} &=& 
        \frac{p_1p+p_2p-p_2p_1}{p_1p+p_2p}=\frac{\uhat+\that+\shat}{\uhat+\that}=
        \frac{Q^2}{\shat+Q^2}=x\,,\\
        u_2 &=& \frac{p_2p}{p_2p+p_1p} = \frac{\that}{\uhat+\that}=z_2=1-z_1\,, \\ 
        \frac{1}{2pp_2x}&=&\frac{1}{Q^2pp_2/pq}=\frac{1}{Q^2(1-z_1)}\,,
        \eea
        the expression above becomes
        \bea
        \left|{\cal{M}}_{2,\q}(p_1,p_2;p)\right|_{\mathrm{PSif}}^2
        = 
        \frac{8\pi\as}{Q^2(1-z_1)}\,\CF\left[\frac{2}{2-x-z_1}-(1+x)\right]
        \cdot\left|{\cal{M}}_{1,\q}(q+xp;xp)\right|^2\,.
        \eea
\item FI:\\
        In full analogy the shower expression for the final-state emission process 
        yields 
        \bea
        \lefteqn{\left|{\cal{M}}_{2,\q}(p_1,p_2;p)\right|_{\mathrm{PSfi}}^2}\nn\\ 
        &=& 
        \frac{1}{2p_1p_2}\,\frac{1}{x_{12,p}}\,
        8\pi\as\,\CF\left[\frac{2}{1-\tilde z_1+(1-x_{12,p})}-(1+\tilde z_1)\right]
        \cdot\left|{\cal{M}}_{1,\q}(q+xp;xp)\right|^2\,.\nn
        \eea
        With 
        \bea
        x_{12,p} = \frac{p_1p+p_2p-p_1p_2}{p_1p+p_2p}=x\;\;\;\mbox{\rm and}\;\;\;
        \tilde z_1 = \frac{p_1p}{p_1p+p_2p} = z_1\,,
        \eea
        this can be cast into the form
        \bea
        \left|{\cal{M}}_{2,\q}(p_1,p_2;p)\right|_{\mathrm{PSfi}}^2
        = 
        \frac{8\pi\as}{Q^2(1-x)}\,\CF\left[\frac{2}{2-x-z_1}-(1+x)\right]
        \cdot\left|{\cal{M}}_{1,\q}(q+xp;xp)\right|^2\,,
        \eea
        where in addition 
        \bea
        2p_1p_2 = Q^2\frac{1-x}{x}
        \eea
        has been employed. 
\end{itemize}
Combining the two parton shower contributions yields the final result, namely
\bea
\lefteqn{\left|{\cal{M}}_{2,\q}(p_1,p_2;p)\right|_{\mathrm{PS}}^2 
        = \left|{\cal{M}}_{2,\q}(p_1,p_2;p)\right|_{\mathrm{PSif}}^2 
        + \left|{\cal{M}}_{2,\q}(p_1,p_2;p)\right|_{\mathrm{PSfi}}^2}\nn\\
&=& \frac{8\pi\as}{Q^2}\,\CF\left[\frac{x^2+z_1^2}{(1-x)(1-z_1)}\right]\cdot 
        \left|{\cal{M}}_{1,\q}(q+xp;xp)\right|^2\,.
\eea
When comparing this with the exact perturbative result of Eq.\ (\ref{eq:DISMEqg}), it 
can be inferred that the parton shower exactly reproduces the soft and collinear 
singular structure of the matrix element as $z_1\to 1$ or $x\to 1$.  The only 
difference is an additional finite non-singular term present in the exact result.

\subsubsection{The initial-state gluon channel}

Expressed in terms of the leading order matrix element the exact real emission 
next-to-leading order result for the process 
$\gamma^\star(q)\g(p) \to \q(p_1)\qbar(p_2)$ reads \cite{Catani:1996vz,Field:1989uq}
\bea\label{eq:DISMEqq}
\lefteqn{\left|{\cal{M}}_{2,\g}(p_1,p_2;p)\right|^2_{\mathrm{ME}}}\nn\\
&=& \frac{8\pi\as}{Q^2}\,\TR 
        \left[\frac{(z_1^2+(1-z_1)^2)(x^2+(1-x)^2)}{z_1(1-z_1)}+8x(1-x)\right]
        \cdot\left|{\cal{M}}_{1,\q}(q+xp;xp)\right|^2\,.
\eea
Starting from the leading order matrix element $\gamma^\star(q)\q(p) \to \q(p_1)$ 
there is only one possibility in the parton shower to reach the $2\to 2$ process,
the splitting of an initial-state gluon into $\q\qbar$ and the $\q$ interacting with 
the off-shell photon.  The second matrix element diagram, corresponding to the 
interaction of the anti-quark with the $\gamma^\star$, here has no parton shower 
counterpart. However, when starting the shower from the charge conjugated leading 
order process, namely $\gamma^\star(q)\qbar(p) \to \qbar(p_1)$, this contribution 
will occur while the $\gamma^\star\q$ interaction will be missing instead. 
The two terms are evaluated separately and then added incoherently.
\begin{itemize}
\item Emission off the quark:\\
        The case of an internal quark propagator is discussed first.  According to 
        Eqs.\ (\ref{eq:IFME}) and (\ref{eq:IFPgqmassive}) the parton shower 
        approximation to the matrix element reads
        \bea
        \lefteqn{\left|{\cal{M}}_{2,\g}(p_1,p_2;p)\right|_{\mathrm{PS\q}}^2}\nn\\ 
        &=& 
        \frac{1}{2pp_2}\,\frac{1}{x_{21,p}}\,8\pi\as\,\TR
        \left[1-2x_{21,p}(1-x_{21,p})\right]
        \cdot\left|{\cal{M}}_{1,\q}(q+xp;xp)\right|^2\nn\\
        &=& 
        \frac{8\pi\as}{Q^2(1-z_1)}\,\TR\left[1-2x(1-x)\right]
        \cdot\left|{\cal{M}}_{1,\q}(q+xp;xp)\right|^2\,.
        \eea
\item Emission off the anti-quark:\\
        Starting instead the shower from the $\qbar$ initiated process, and emitting 
        the quark into the final state yields, correspondingly,
        \bea
        \lefteqn{\left|{\cal{M}}_{2,\g}(p_1,p_2;p)\right|_{\mathrm{PS\qbar}}^2}\nn\\ 
        &=& \frac{1}{2pp_1}\,\frac{1}{x_{12,p}}\,8\pi\as\,\TR
                \left[1-2x_{12,p}(1-x_{12,p})\right]
                \cdot\left|{\cal{M}}_{1,\qbar}(q+xp;xp)\right|^2\nn\\
        &=&\frac{8\pi\as}{Q^2z_1}\,\TR\left[1-2x(1-x)\right]
                \cdot\left|{\cal{M}}_{1,\qbar}(q+xp;xp)\right|^2\,.
        \eea
\end{itemize}
Due to the charge conjugation invariance of the leading order matrix element,
\bea
\left|{\cal{M}}_{1,\qbar}(q+xp;xp)\right|^2 = \left|{\cal{M}}_{1,\q}(q+xp;xp)\right|^2\,,
\eea
the two parton shower contributions can directly be combined and yield
\bea
\left|{\cal{M}}_{2,\g}(p_1,p_2;p)\right|_{\mathrm{PS}}^2 &=&
\left|{\cal{M}}_{2,\g}(p_1,p_2;p)\right|_{\mathrm{PS\q}}^2 +
\left|{\cal{M}}_{2,\g}(p_1,p_2;p)\right|_{\mathrm{PS\qbar}}^2\nn\\
&=& 
\frac{8\pi\as}{Q^2}\,\TR\left[\frac{x^2+(1-x)^2}{z_1(1-z_1)}\right]
\cdot\left|{\cal{M}}_{1,\q}(q+xp;xp)\right|^2\,.
\eea
Again the parton shower matches the soft and collinear behaviour of the matrix element 
given in Eq.\ (\ref{eq:DISMEqq}) and reproduces the exact result up to non-singular 
terms.

\subsection{Associated production of a weak gauge boson and a light parton}
\label{sec:MEPSDY}

The lowest order production process of weak gauge bosons 
($\W^\pm$, $\Z^0$, $\gamma^\star$) at a hadron collider proceeds via the $s$-channel 
fusion of two initial-state quarks.  Without loosing generality $\W^\pm$ boson production 
will be investigated in the following.  The leading order process then simply reads 
$\q \qbar' \to \W^\pm$. At order $\as$ there are three processes emerging: 
$\q \qbar' \to \W^\pm \g$, $\g\qbar' \to \W^\pm\qbar$ and $\q \g \to \W^\pm \q'$. 
Considering on-shell $\W^\pm$ bosons for simplicity\footnote{
        This corresponds to neglecting the off-shell gauge boson 
  decays which, however, do not affect the QCD dynamics of the processes under 
  consideration.  The decay products of the gauge boson can be introduced 
  into the process using the narrow-width-approximation, or by incorporating 
  the full off-shell $\W^\pm$ propagator.}, 
only $2 \to 2$ processes have to be discussed, which can be described using the 
Mandelstam variables
\bea
\shat &\equiv& (p_1 + p_2)^2 = \hphantom{-}2p_1p_2\,,\\
\that &\equiv& (p_1 - p_3)^2 = -2p_1p_3\,,\\
\uhat &\equiv& (p_2 - p_3)^2 = -2p_2p_3\,.
\eea
Momentum conservation then implies that 
\bea
\shat + \that + \uhat = \mW^2\,,
\eea
where $\mW$ denotes the $\W^\pm$-boson mass. 

\subsubsection{The gluon emission channel}
\begin{figure}
\bc
\includegraphics{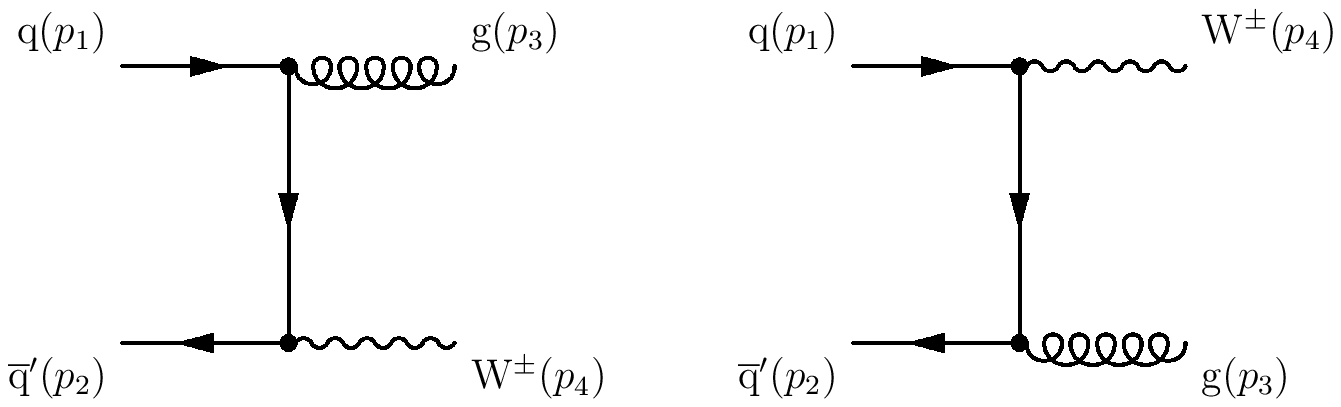}
\caption{The leading order Feynman diagrams contributing to the process 
        $\q\qbar' \to \W^\pm\g$.\label{fig:Wg}}
\ec
\end{figure}
The first channel to be discussed is the gluon emission process 
$\q\qbar' \to \W^\pm \g$.  At tree-level, there are two Feynman diagrams contributing 
to the matrix element, cf.\ Fig.\ \ref{fig:Wg}.  The partonic differential cross 
section can be written as \cite{Field:1989uq}  
\bea 
\left.\frac{\d\hat\sigma}{\d\that}\right|_{\mathrm ME} = 
\frac{\hat\sigma_0}{\shat}\,
\frac{\as}{2\pi}\,\CF\left[\frac{\that^2+\uhat^2+2\mW^2\shat}{\that\uhat}\right]\,, 
\eea
with $\hat\sigma_0$ the cross section of the leading order process $\q\qbar' \to \W^\pm$
\bea
\hat\sigma_0 = \frac13\frac{4\pi}{\shat}\frac{g_{\mathrm{W}}^2}{2\pi}\,.
\eea
In the parton shower approach there are two ways to produce the final-state gluon,
which have to be added  incoherently: either the gluon can be emitted from the 
initial-state quark or from the anti-quark.
\begin{itemize}
\item Emission off the quark:\\
        As a first step, the kinematical variables used in the parton shower 
        approximation should be related to the Mandelstam variables.  In the first
        case, the initial quark as emitter and the initial anti-quark as spectator, 
        the parton shower variables become, cf.\ Sec.\ \ref{sec:kinematicsII}, 
        \bea
        \tilde v_3 = \frac{p_3p_1}{p_1p_2} = -\frac{\that}{\shat}
        \;\;\;\mbox{\rm and}\;\;\;
        x_{3,12} = \frac{p_1p_2-p_3p_1-p_3p_2}{p_1p_2} = 
        \frac{\shat+\that+\uhat}{\shat} = \frac{\mW^2}{\shat}\,.
        \eea
        Using the appropriate splitting function of Eq.\ (\ref{eq:IIPqq}), the parton 
        shower differential cross section 
        \bea
        \left.\frac{\d\hat\sigma}{\d \tilde v_3}\right|_{\mathrm{PSq}} = 
        \hat\sigma_0\,\frac{\as}{2\pi}\,\frac{1}{\tilde v_3}\,
        \CF\left[ \frac{2}{1-x_{3,12}}-(1-x_{3,12}) \right] 
        \eea
        can be cast into
        \bea
        \left.\frac{\d \hat\sigma}{\d \that}\right|_{\mathrm{PSq}} = 
        \hat\sigma_0\,\frac{\as}{2\pi} \CF \frac{1}{-\that}\,
        \left[ \frac{2}{1-x_{3,12}}-(1-x_{3,12}) \right]\,.
        \eea
        Using the relation
        \bea
        (1-x_{3,12}) = -\frac{\that + \uhat}{\shat}
        \eea
        and multiplying with $\shat/\shat$ yields
        \bea
        \left.\frac{\d\hat\sigma}{\d\that}\right|_{\mathrm{PSq}} = 
        \frac{\hat\sigma_0}{\shat}\,\frac{\as}{2\pi}\,\CF
        \left[\frac{\shat^2+\mW^4}{\that(\that+\uhat)}\right] \,.
        \eea
\item  Emission off the anti-quark:\\
        Swapping the r\^ole of the emitter and the spectator parton amounts to only 
        interchanging $\that$ and $\uhat$ in the results above.  Accordingly,
        the differential cross section in this case is given by
        \bea
        \left.\frac{\d\hat\sigma}{\d\that}\right|_{\mathrm{PS\qbar}} = 
        \frac{\hat\sigma_0}{\shat}\,\frac{\as}{2\pi}\,\CF
        \left[\frac{\shat^2+\mW^4}{\uhat(\that+\uhat)}\right] \,.
        \eea
\end{itemize}
The full parton shower result is the sum of the two contributions and reads
\bea
\left.\frac{\d \hat\sigma}{\d\that}\right|_{\mathrm{PS}} = 
\left.\frac{\d \hat\sigma}{\d\that}\right|_{\mathrm{PSq}}+ 
\left.\frac{\d \hat\sigma}{\d\that}\right|_{\mathrm{PS\qbar}} =
 \frac{\hat\sigma_0}{\shat}\,\frac{\as}{2\pi}\,\CF
\left[\frac{\shat^2+\mW^4}{\that\uhat}\right] \,.
\eea
Again, the parton shower approach provides the correct description for soft and 
collinear phase-space configurations but misses non-singular terms.  The difference 
of the parton shower and the exact result can be quantified by the ratio 
\bea
\frac{\left.\d\hat\sigma/\d\that\right|_{\mathrm{ME}}}
        {\left.\d\hat\sigma/\d\that\right|_{\mathrm{PS}}}=
\frac{\that^2+\uhat^2+2\mW^2\shat}{\shat^2+\mW^4} = 1-\frac{2\that\uhat}{\shat^2+\mW^4}\,,
\eea
which can take values between $0.5$ and $1$ in full agreement with the result of the 
parton shower algorithm implemented in {\sc Pythia} \cite{Miu:1998ju}.  This indicates 
that the parton shower approximation tends to overestimate the matrix element - a feature 
already present, \eg, in $e^+e^-\to q\bar qg$.

\subsubsection{The initial-state gluon case}

\begin{figure}
\bc
\includegraphics{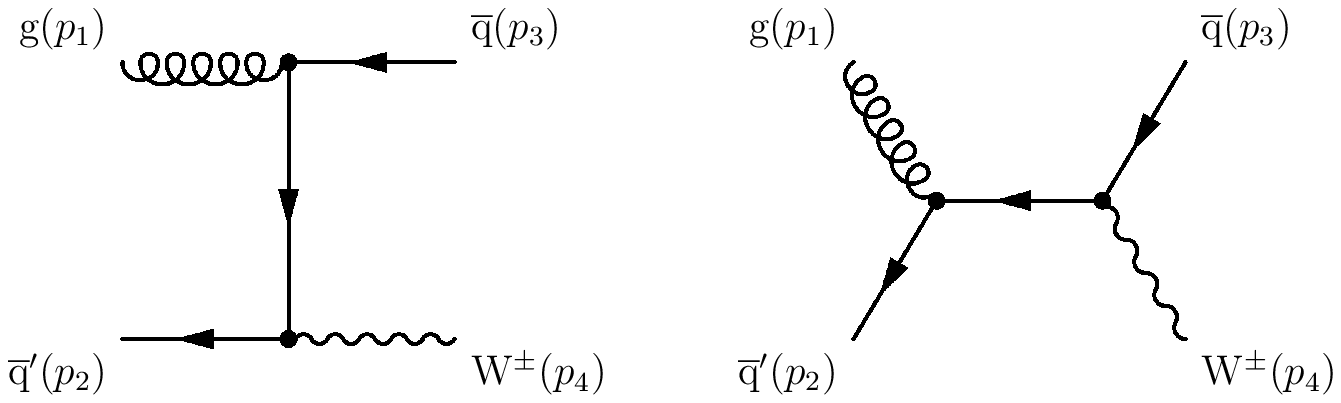}
\caption{The leading order Feynman diagrams contributing to the process 
        $\g\qbar'\to \W^\pm\qbar$.\label{fig:Wq}}
\ec
\end{figure}
There are two Feynman diagrams, cf.\ Fig.\ \ref{fig:Wq}, contributing to the channel 
with an initial-state gluon, \ie to the process $\g\qbar' \to \W^\pm \qbar$.  The 
result of the full matrix element calculation reads \cite{Field:1989uq}
\bea
\left.\frac{\d\hat\sigma}{\d\that}\right|_{\mathrm{ME}} &=& 
\frac{\hat\sigma_0}{\shat}\,
\frac{\as}{2\pi}\,\TR\left[\frac{\shat^2+\uhat^2+2\mW^2\that}{-\shat\that}\right]\,. 
\eea
In the parton shower approach only one emission process contributes to this channel, 
corresponding to the $t$-channel diagram. The $s$-channel contribution is not 
realised in the shower ansatz.  Using the definitions of the shower variables as 
given above and the corresponding splitting function, cf.\ Eq.\ (\ref{eq:IIPgq}), 
the parton shower cross section reads
\bea
\left.\frac{\d\hat\sigma}{\d\that}\right|_{\mathrm{PS}} 
&=& 
\frac{\hat\sigma_0}{\shat}\,
\frac{\as}{2\pi}\,\TR\frac{\shat}{-\that}\left[1-2x_{3,12}(1-x_{3,12})\right] \nn\\
&=& 
\frac{\hat\sigma_0}{\shat}\,
\frac{\as}{2\pi}\,\TR\left[\frac{\shat^2+2\mW^2(\that+\uhat)}{-\shat\that}\right]\,,
\eea
where
\bea
(1-x_{3,12}) = -\frac{\that + \uhat}{\shat}
\;\;\;\mbox{\rm and}\;\;\;x_{3,12}=\mW^2/\shat
\eea
has been used.  The ratio of the parton shower and the matrix element result is
\bea
\frac{\left.\d\hat\sigma/\d\that\right|_{\mathrm{ME}}}
        {\left.\d\hat\sigma/\d\that\right|_{\mathrm{PS}}}=
\frac{\shat^2+\uhat^2+2\mW^2\that}{\shat^2+\mW^4(\that+\uhat)} = 
1+\frac{\uhat(\uhat-2\mW^2)}{(\shat-\mW^2)^2+\mW^4}\,,
\eea
varying between $1$ and $3$ \cite{Miu:1998ju}.  Accordingly, the parton shower ansatz 
tends to undershoot the exact matrix element. However, the shower is constructed to 
give the correct answer in the logarithmically enhanced phase-space regions and thus 
has the correct limiting behaviour in the soft and collinear limits.  The differences
identified here are a result of differences in the non-singular terms, contributing only
in hard regions of phase space.  The process $\q\g\to\W^\pm \q'$ closely follows the 
above example solely $\that$ and $\uhat$ have to be exchanged.  This leads to the same 
qualitative results and the same conclusions.

%% file: csapplications.tex
\section{Applications}
\label{sec:applications}

In this section, the abilities of the newly developed parton shower formulation in 
describing QCD dynamics will be highlighted by comparing its results for various physics 
processes with experimental data and other calculations: In Sec.\ \ref{sec:applee}, the 
predictions for hadron production in $e^+e^-$ collisions as measured at LEP will be 
studied and some results related to a future machine operated at $\sqrt{s}= 500$ GeV 
will be discussed.  In Sec.\ \ref{sec:applpp}, emphasis is put on the capabilities of 
the shower to describe particle production at hadron colliders such as the 
Tevatron or the upcoming LHC.

\subsection{Jet production at $e^+e^-$ colliders}
\label{sec:applee}

Measurements of hadronic final states produced in $e^+e^-$ collisions provide a very 
precise probe of QCD dynamics in the final state and an excellent means to deduce
its fundamental parameters such as the value of $\alpha_S(m_Z)$, see for instance 
\cite{Abbiendi:2001qn}, and the colour charges $C_F$ and $C_A$ in three- and four-jet 
events as discussed \eg in \cite{Abbiendi:2001qn}-\cite{Barate:1997ha}.  Therefore it 
is not surprising that in the past years calculations for relevant three-jet observables, 
such as thrust, have become available at NNLO \cite{GehrmannDe Ridder:2007bj} and that 
full parton-level Monte Carlo codes for four-jet final states at NLO have been constructed 
\cite{Signer:1997gs,Nagy:1998bb}.  Obviously such observables also provide a critical
test of the corresponding final-state radiation piece of a parton shower model.  However, 
due to the fragmentation of partons into hadrons, which at the moment can be simulated 
with phenomenological models only, the parton shower predictions can not directly be 
compared with experimental data but rather have to be supplemented with a hadronisation 
model. The new parton shower presented here therefore has been interfaced to the Lund 
string fragmentation routines of \Pythia version 6.2 \cite{Sjostrand:2001yu} in the 
framework of the \Sherpa event generator.  The large number of very precisely measured 
observables at various energies allows tuning the intrinsic parameters of the parton 
shower in conjunction with the phenomenological parameters of the fragmentation model.  
Such a procedure has been performed, for instance, for the new parton shower and 
fragmentation code in \Herwigpp \cite{Gieseke:2003hm}.  In principle, such a tuning 
is a very time-consuming and delicate procedure, see for instance \cite{Hamacher:1995df}, 
deserving a publication in its own right.  Recent developments to automatise the task 
of generator tuning and validation to a large extend are reported in \cite{Buckley:2007hi}.
Here, only a very limited tuning based on few parameters and observables only has been 
performed.  The results of this tuning are presented in Sec.~\ref{sec:appllep1}.  
In Sec.~\ref{sec:applhQ} the focus is on heavy-quark production at LEP1 and ILC energies 
to validate the treatment of finite parton masses in the shower model. 

\subsubsection{Comparison with LEP1 data}\label{sec:appllep1}

The most extensive data set available to validate QCD Monte Carlo predictions are 
LEP measurements at the $Z^0$ pole.  A selection of event shape variables, multiplicity 
distributions, differential jet rates, four-jet angle measurements and various
particle momentum distributions have been used to select values for the unconstrained,
phenomenological parameters of the simulation, namely the value of the strong coupling 
constant at $m_Z$, the infrared shower cut-off $\kperpzero$ and the three Lund string 
hadronisation parameters $a$ ({\tt PARP(41)}), $b$ ({\tt PARP(42)}) and $\sigma_q$ 
({\tt PARP(21)}).  For the results presented in the following, they have been fixed to 
$\as(m_Z)=0.125$, $\kperpzero=0.63$ GeV, $a=0.33$, $b=0.75$ $\GeV^{-2}$, 
and $\sigma_q=0.358$ GeV, respectively.  This yields a  mean charged multiplicity per 
event of $\langle N_{ch}\rangle=20.87$ at $\sqrt{s}=m_Z$, in good agreement with 
the experimentally found value of $\langle N_{ch}\rangle=20.92 \pm 0.24$ 
\cite{Abreu:1996na}.  

\begin{figure*}[t!]
\begin{picture}(500,500)
\put(0,250){\includegraphics[width=230pt]{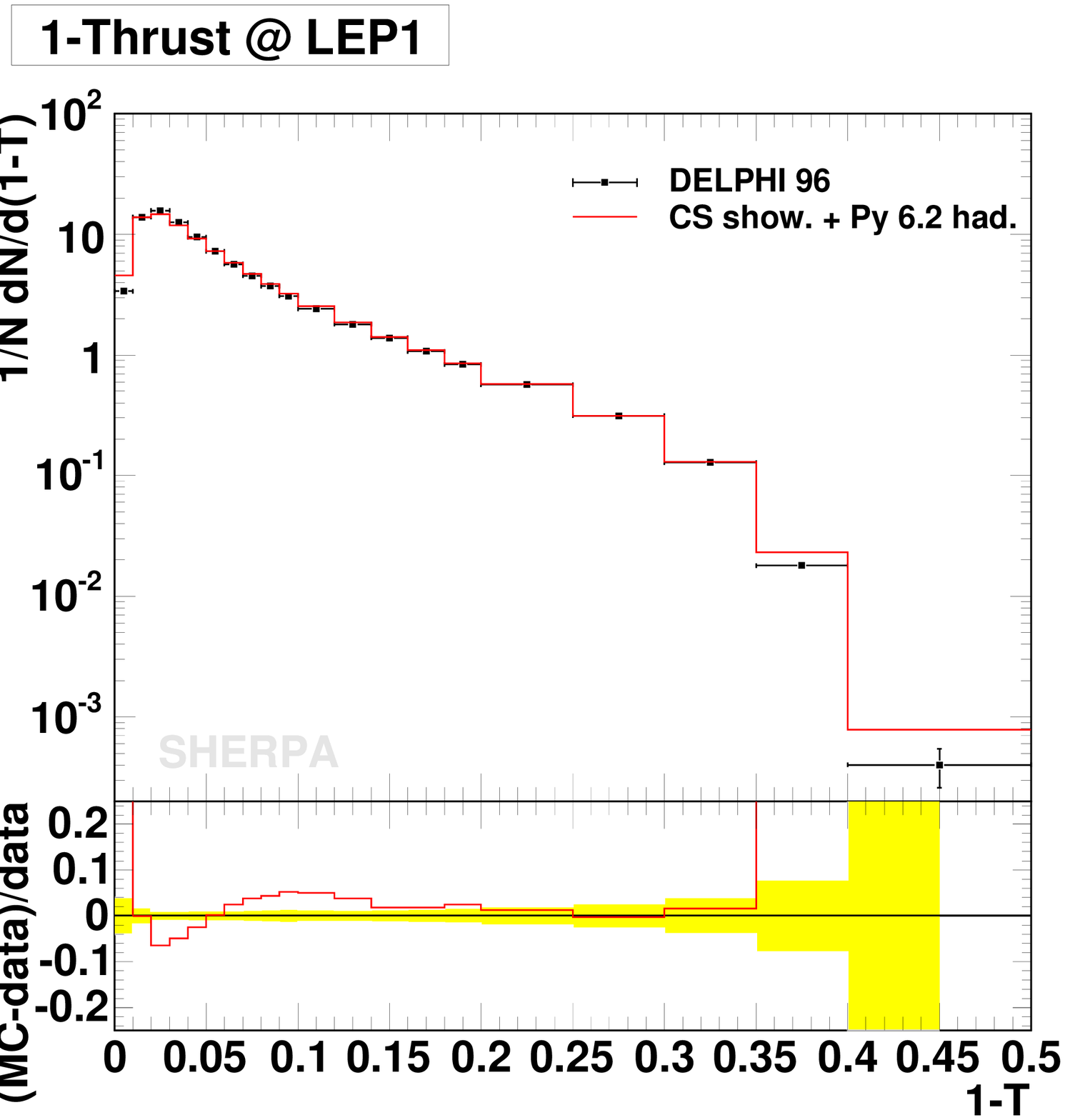}}
\put(250,250){\includegraphics[width=230pt]{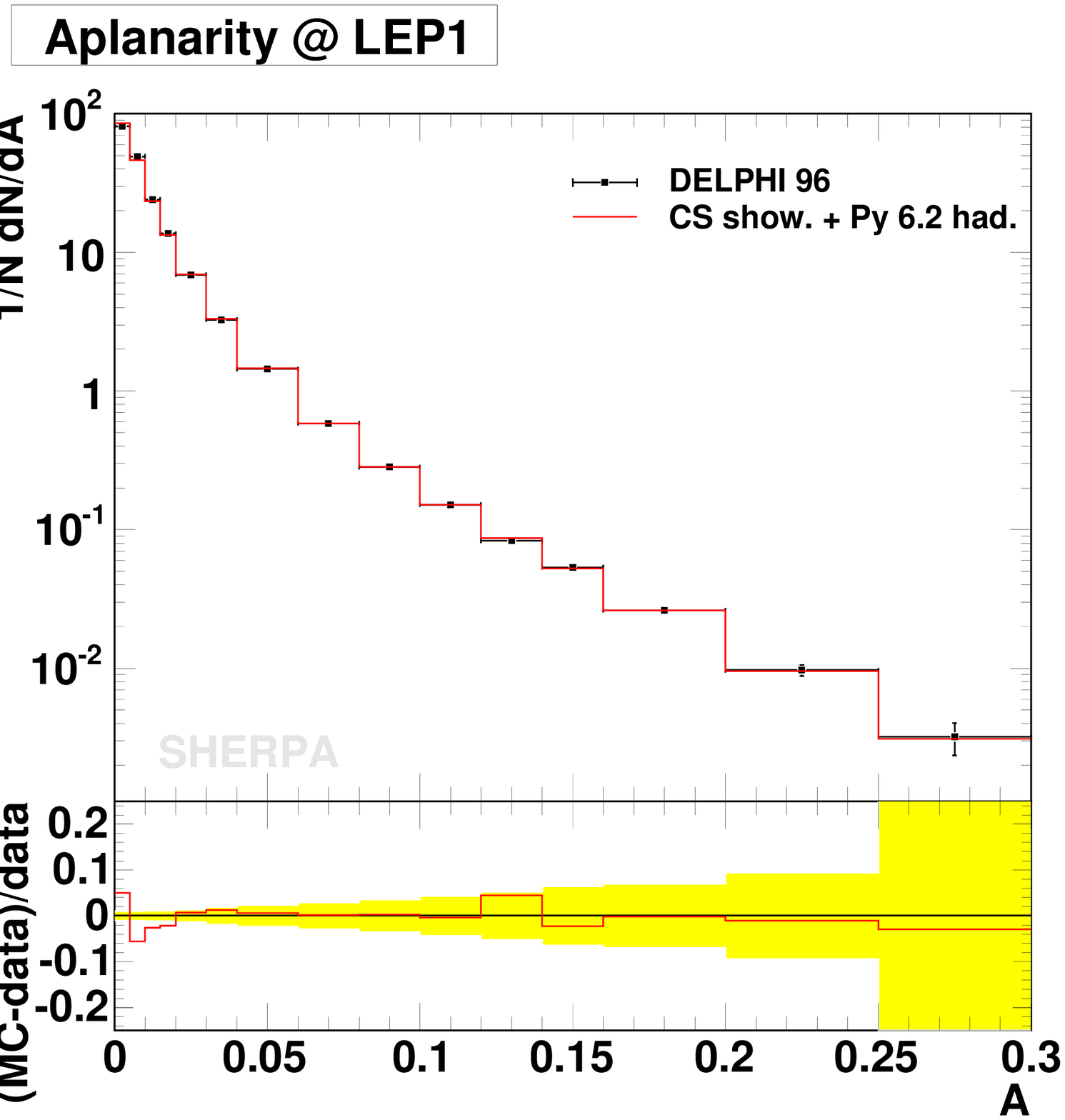}}
\put(0,0){\includegraphics[width=230pt]{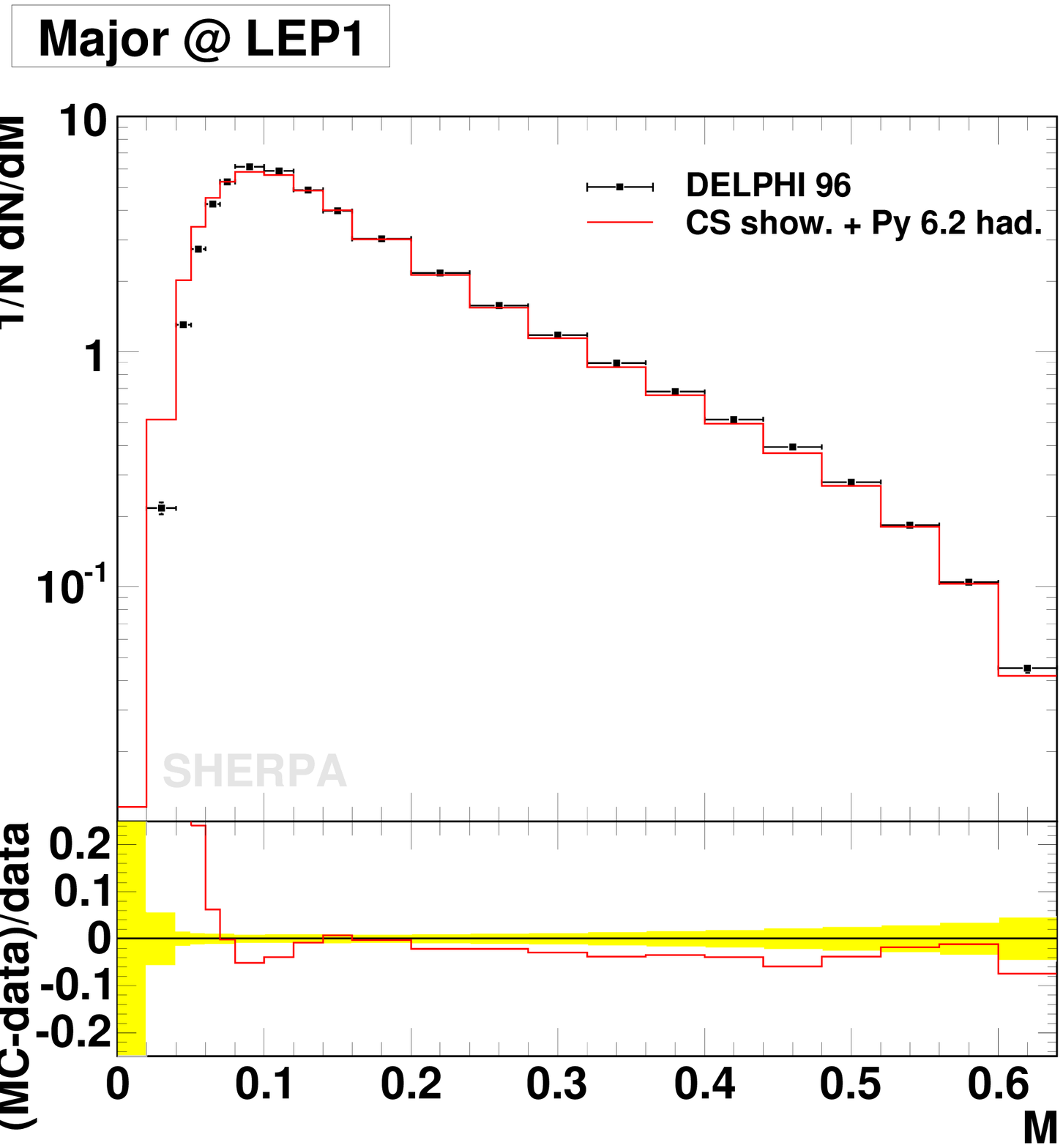}}
\put(250,0){\includegraphics[width=230pt]{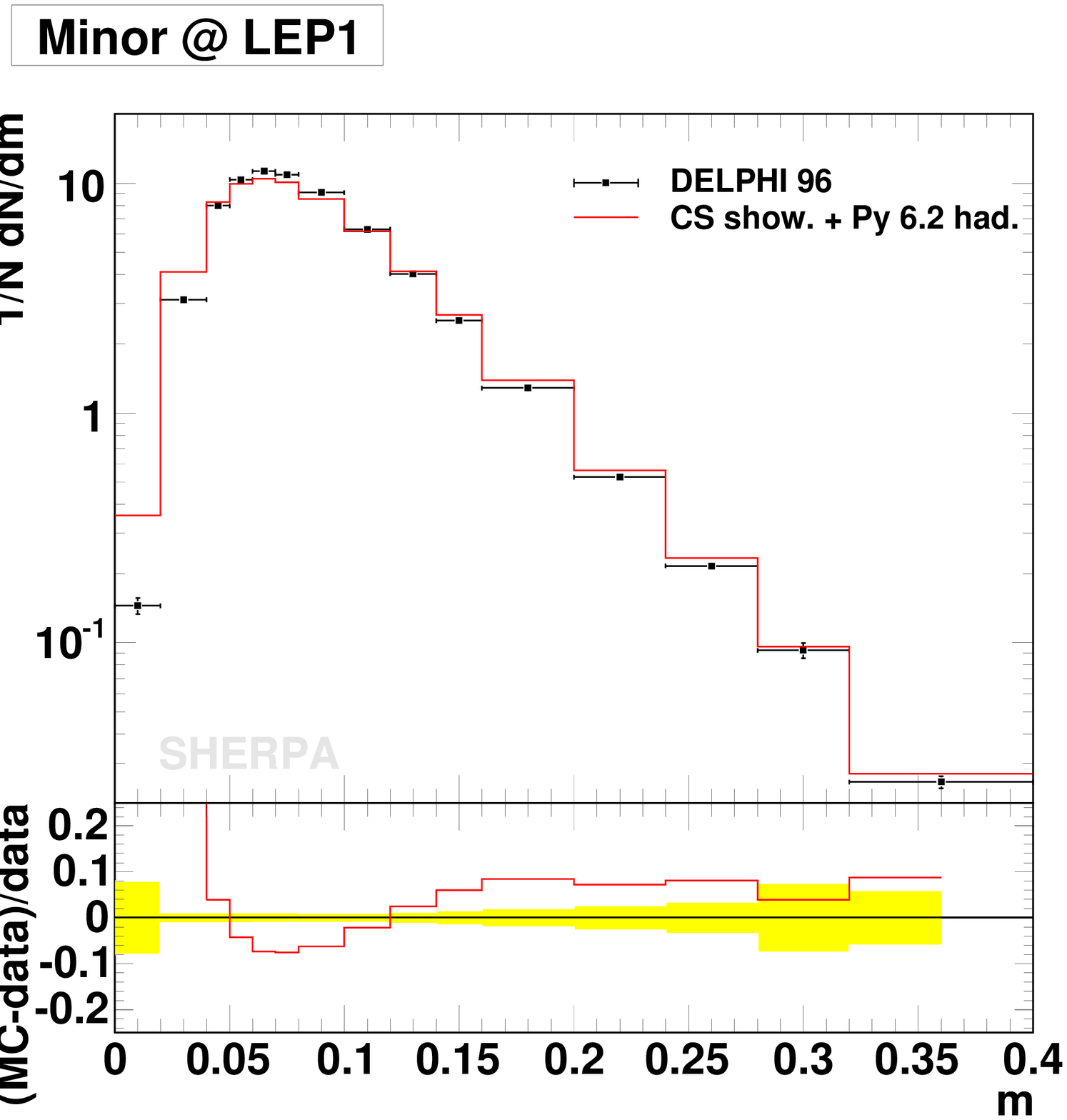}}
\end{picture}
\caption{\label{fig:LEP1eventshapes}
        The event-shape variables $1-$Thrust ($1-T$), 
        Aplanarity ($A$), Major ($M$) and Minor ($m$) in comparison with \protect\Delphi\ data 
        \cite{Abreu:1996na}.}
\end{figure*}

\begin{figure*}[t!]
\begin{picture}(250,250)
\put(0,0){\includegraphics[width=230pt]{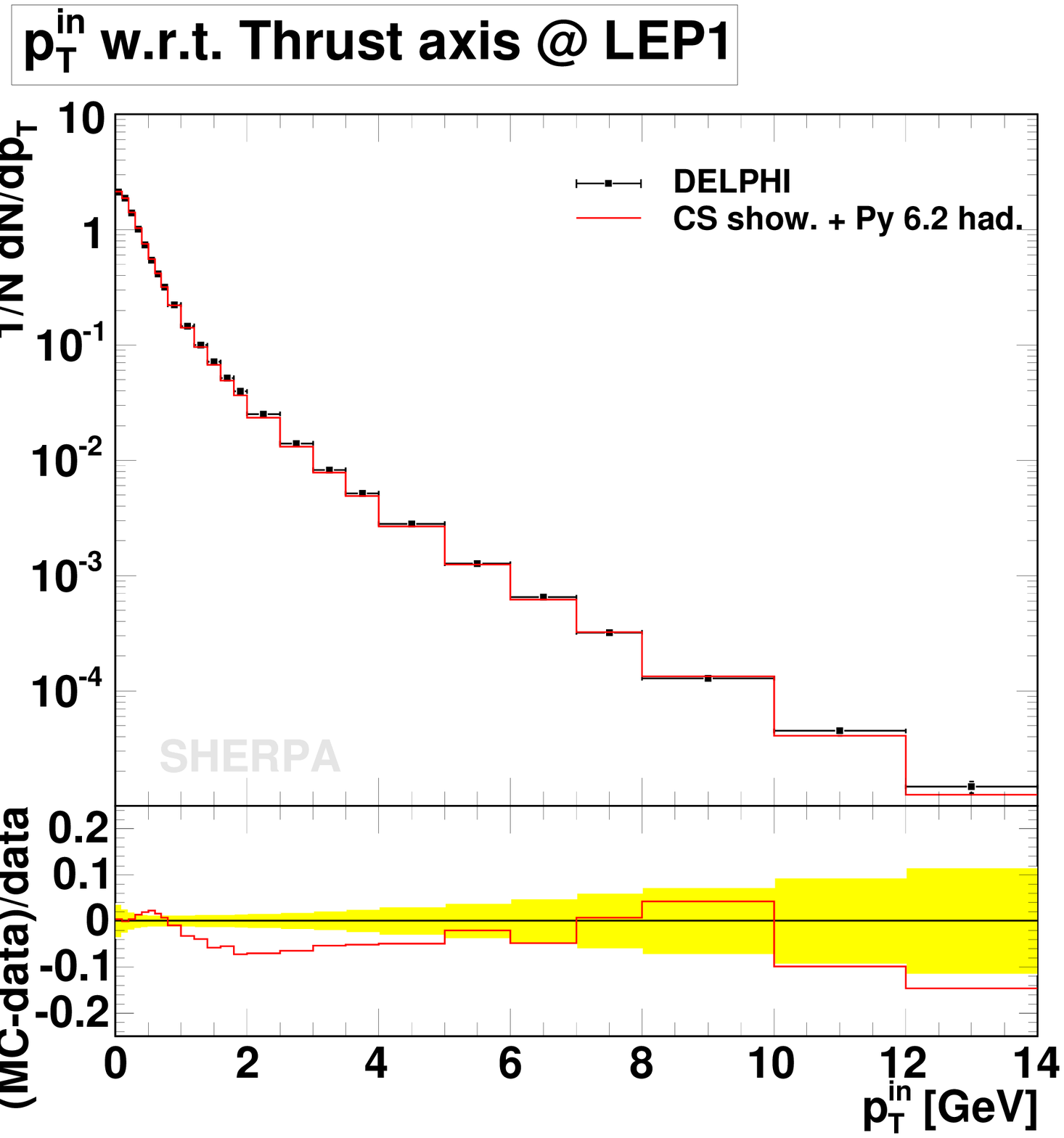}}
\put(250,0){\includegraphics[width=230pt]{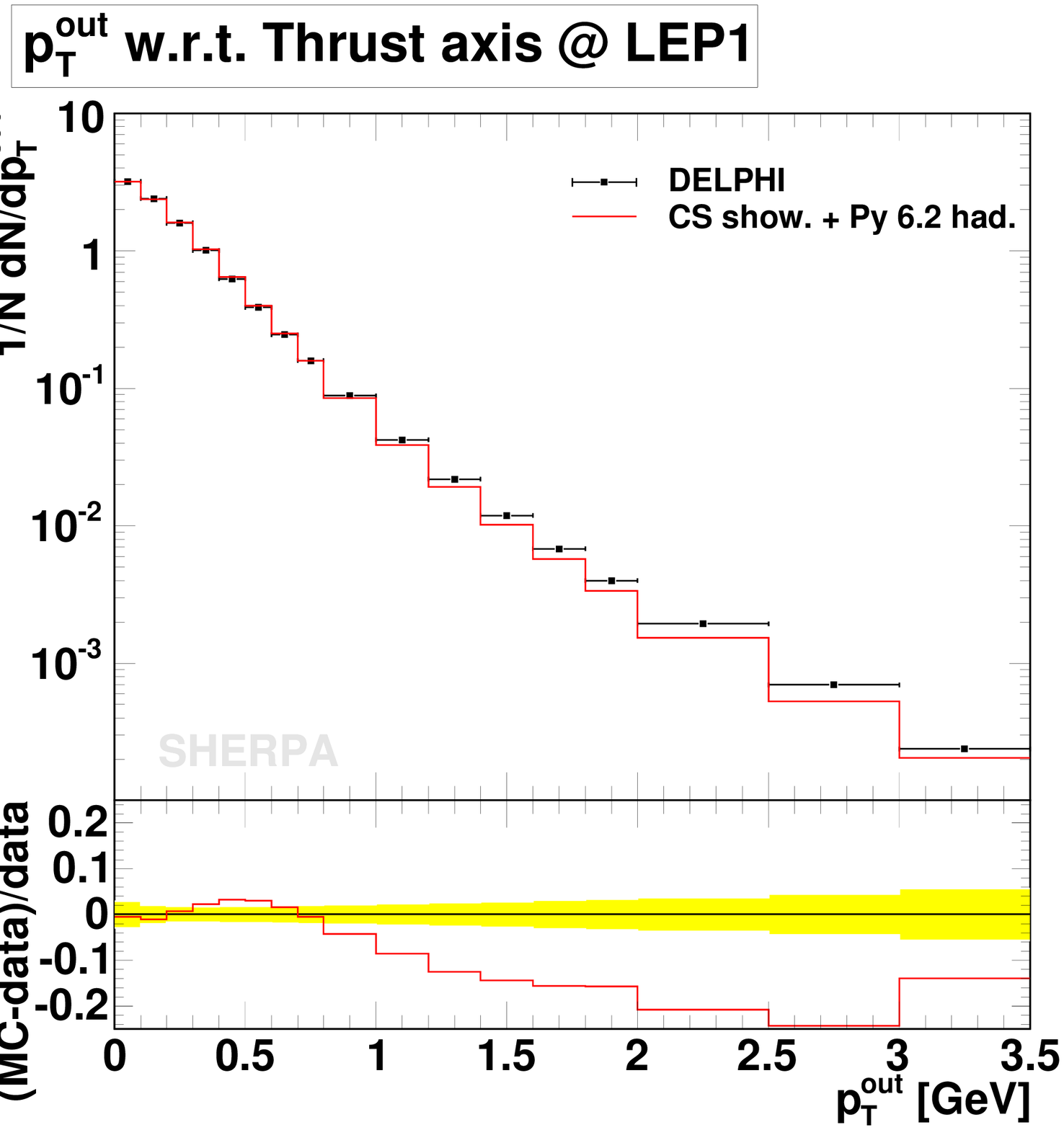}}
\end{picture}
\caption{\label{fig:LEP1PTinoutT}
        The $p^{\rm in}_T$ and $p^{\rm out}_T$ observables measured with 
        respect to the thrust axis compared to a \protect\Delphi\ measurement 
        \cite{Abreu:1996na}. }
\end{figure*}

\begin{figure*}[t!]
\begin{picture}(500,500)
\put(0,250){\includegraphics[width=230pt]{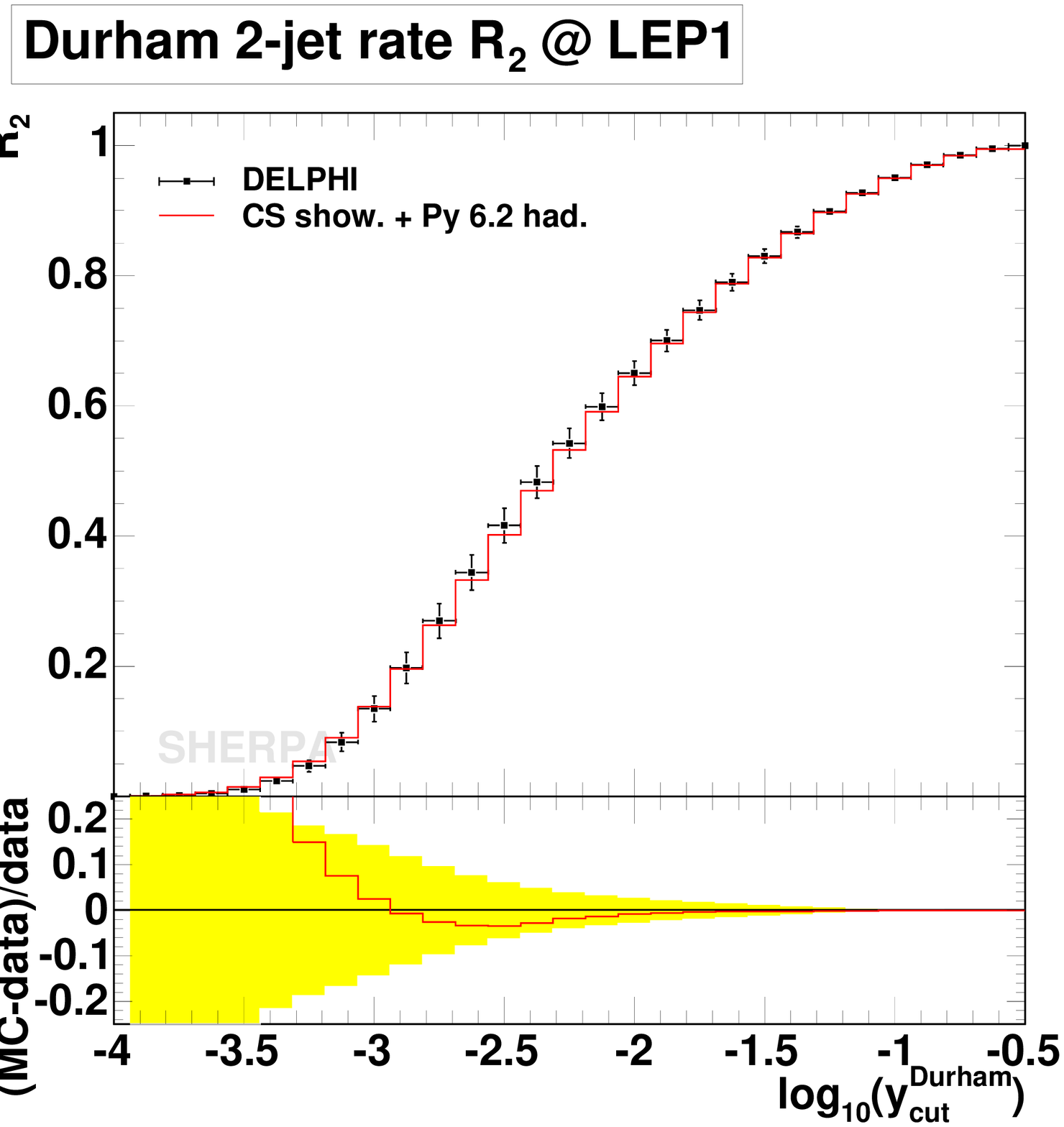}}
\put(250,250){\includegraphics[width=230pt]{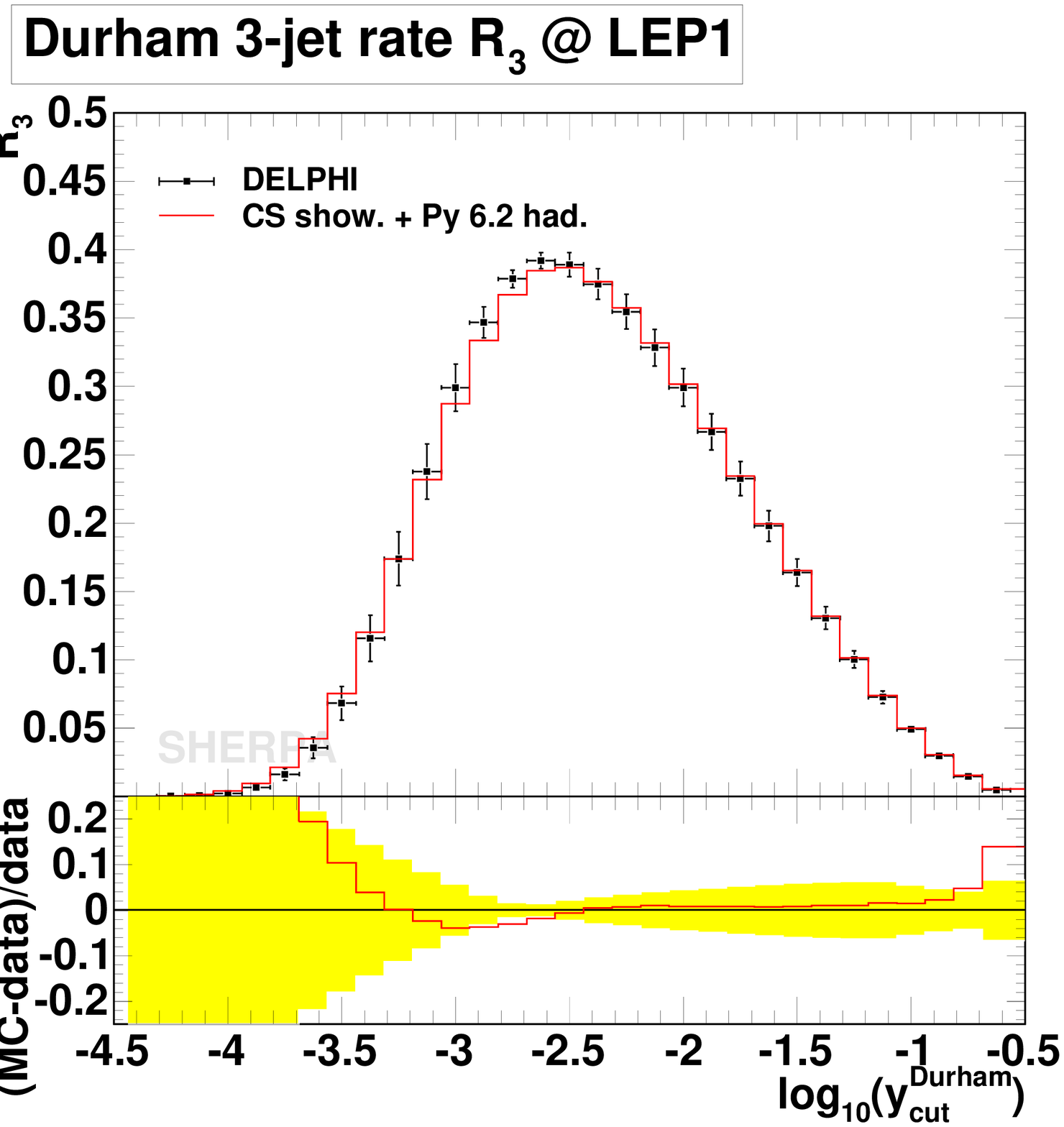}}
\put(0,0){\includegraphics[width=230pt]{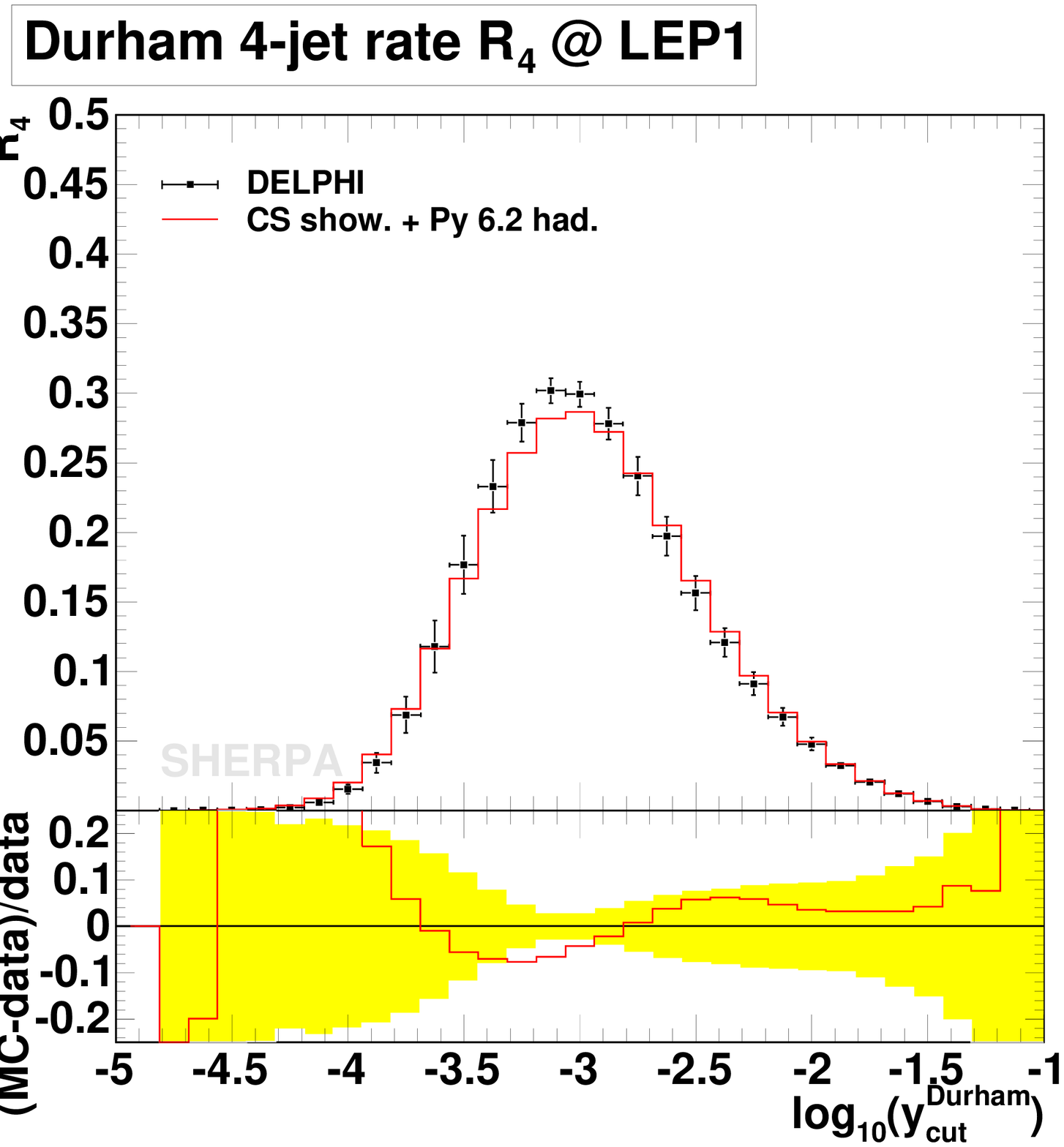}}
\put(250,0){\includegraphics[width=230pt]{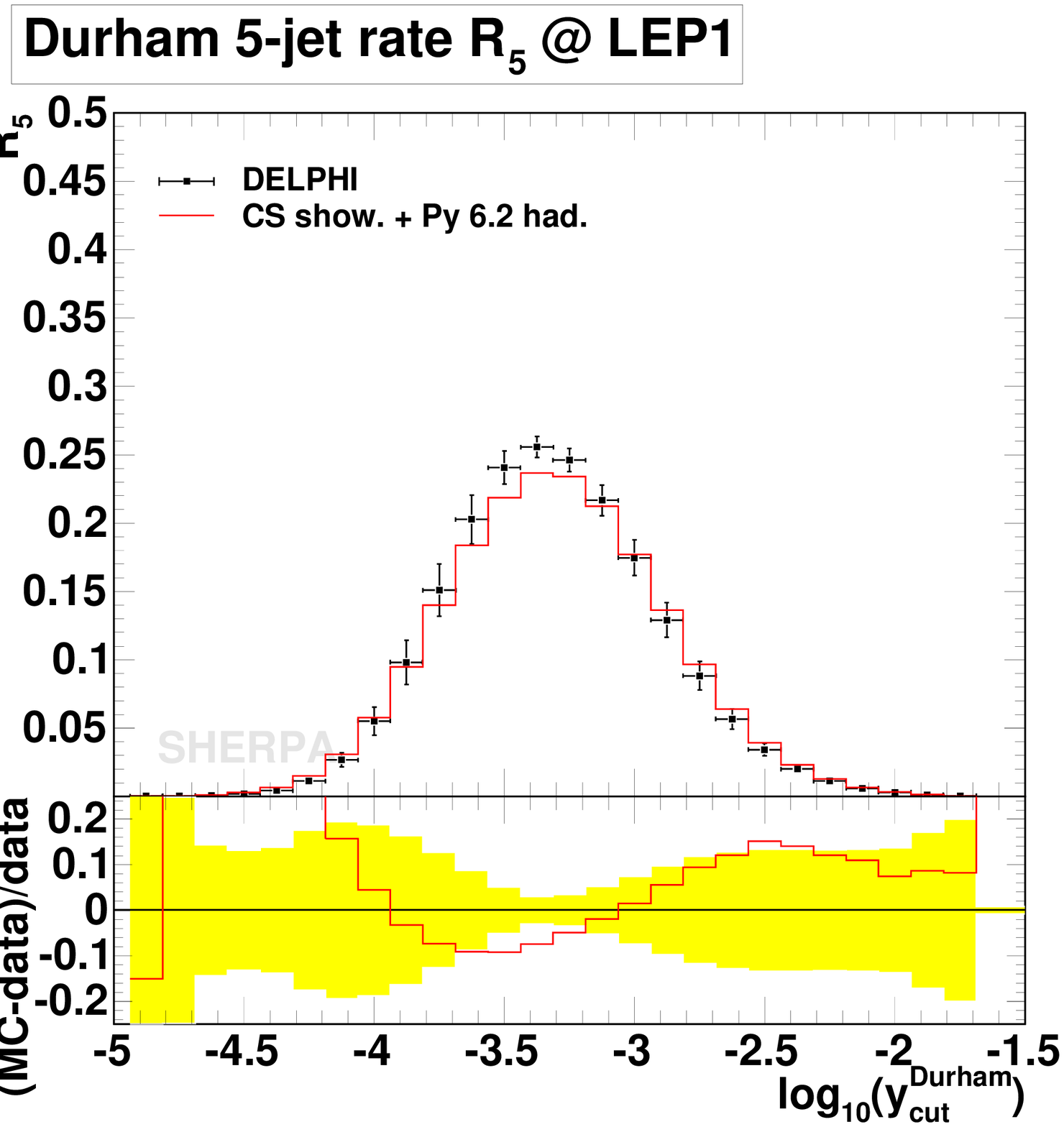}}
\end{picture}
\caption{\label{fig:LEP1jetrates}
        The $n$-jet rates $R_n$ for the Durham jet algorithm as a function
        of the jet-resolution parameter $y^{\rm Durham}_{\rm cut}$. Data taken from
        \cite{Hendrik:2003}. }
\end{figure*}

\begin{figure*}[t!]
\begin{picture}(250,250)
\put(0,0){\includegraphics[width=230pt]{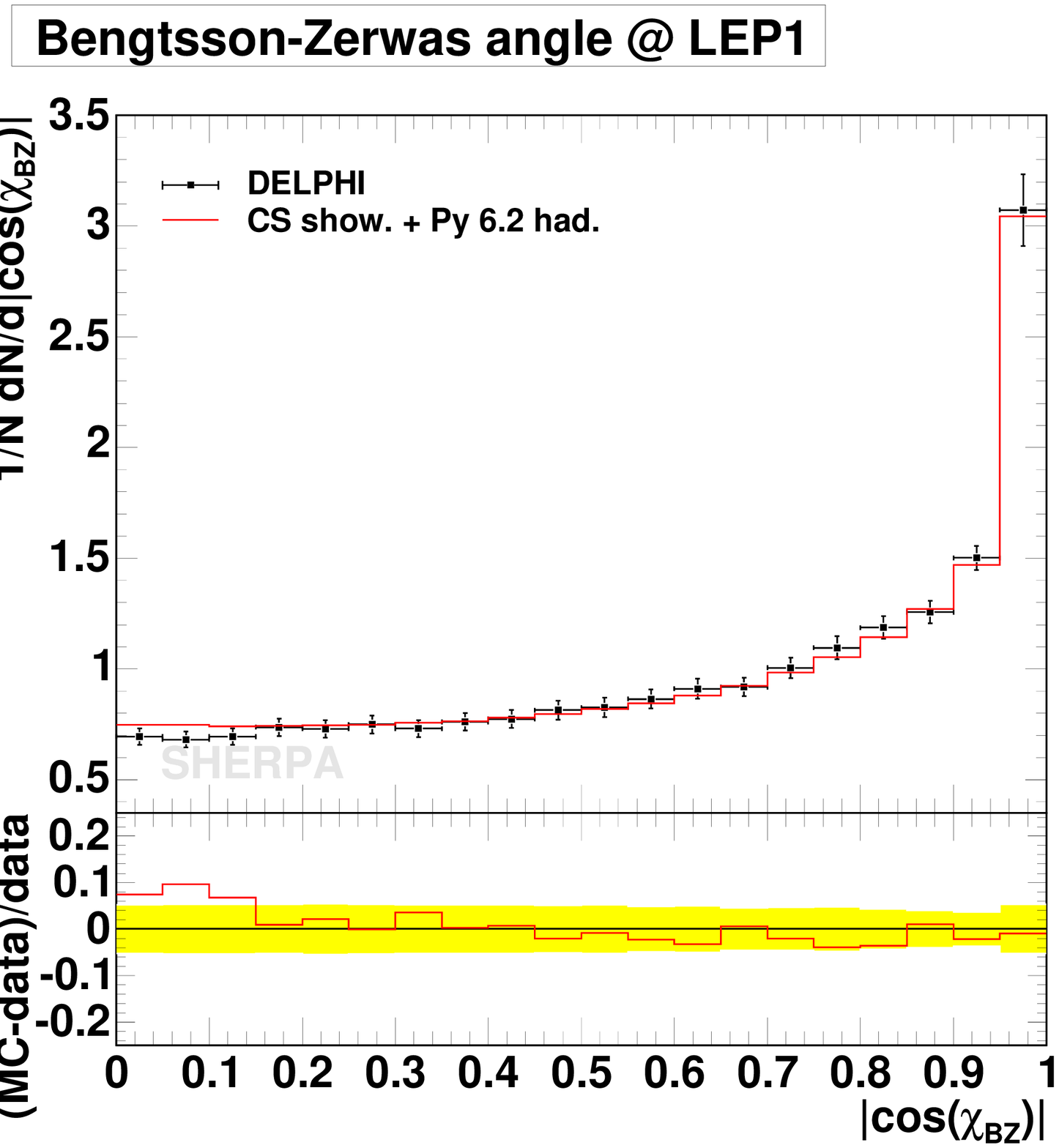}}
\put(250,0){\includegraphics[width=230pt]{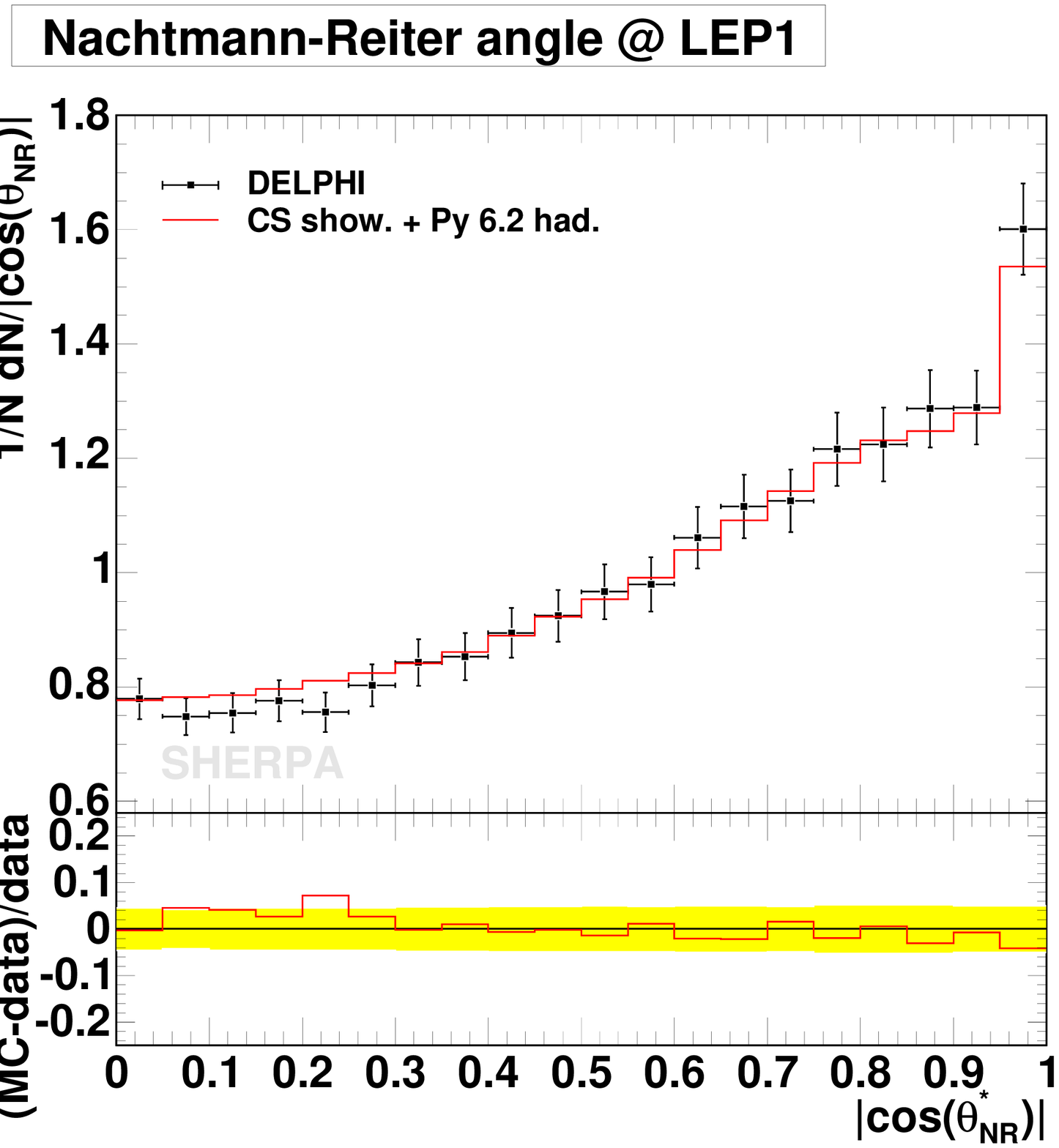}}
\end{picture}
\caption{\label{fig:LEP14jangles}
        The Bengtsson-Zerwas and Nachtmann-Reiter 
        four-jet angles compared with \protect\Delphi\ data 
        \cite{Hendrik:2003}. }
\end{figure*}

Figures~\ref{fig:LEP1eventshapes} to \ref{fig:LEP14jangles} show some exemplary 
results obtained with the new shower implementation compared to \Delphi LEP1 
data at $\sqrt{s}=91.2$ GeV \cite{Abreu:1996na}.  
 
In Fig.\ \ref{fig:LEP1eventshapes} the new algorithm, denoted as ``CS shower'' in the 
following, is compared with some event-shape measurements by \Delphi \cite{Abreu:1996na}. 
The distributions of thrust, thrust-major, thrust-minor and aplanarity are displayed.  
The lower panel of each plot contains the bin-wise ratio (MC-data)/data, and the yellow 
bands show the statistical plus systematic error of the respective measurements. All the 
observables are sensitive to the pattern of QCD radiation probing both soft and hard 
emissions off the shower initiating $q\bar q$ pair. The Monte Carlo predictions agree 
very well with the event-shape data.  There is some slight excess at very low $1-T$ 
corresponding to two-jet like events. This region of phase space however is very sensitive 
to hadronisation corrections and therefore dominated by non-perturbative physics.  The same 
reasoning holds for the major and minor distributions at low $M$ or $m$.  

The transverse-momentum distribution within and out of the event plane defined by the 
thrust and thrust-major axes, ($p^{\rm in}_T$) and ($p^{\rm out}_T$), respectively, are 
presented in Fig.\ \ref{fig:LEP1PTinoutT}.  While $p^{\rm in}_T$ is quite well modeled by 
the Catani-Seymour shower, $p^{\rm out}_T$ is significantly underestimated for values above 
$1$ GeV.  This tendency, however, is observed in other QCD Monte Carlo simulations as well 
\cite{Abreu:1996na}. 

In Fig.\ \ref{fig:LEP1jetrates} the predictions for the exclusive two-, three-, four- 
and five-jet rates in the Durham algorithm \cite{Catani:1991hj} as a function of the 
jet resolution $y^{\rm Durham}_{\rm cut}$ are compared with data taken by the \Delphi 
experiment \cite{Hendrik:2003}.  They all exhibit a sufficient agreement with data 
within the experimental uncertainty bands.  For the four- and five-jet rate the shower 
seems to underestimate the region of $y^{\rm Durham}_{\rm cut} \approx 0.001$, however, 
this region is also affected by hadronisation effects and a more sophisticated tuning 
may provide an even better agreement with data here.  The dependence on the choice of 
hadronisation parameters is even more pronounced for jet resolutions smaller than 
$0.001$ where the results for the new shower preferably lie on the upper side of 
the experimental uncertainty band.   

The last observables to be considered are jet angular distributions in events with four 
jets.  These observables can not be expected to be too well described by a pure 
parton shower as they should probe spin correlations of the produced partons.  Such 
correlations, however, are not taken into account in conventional showers but require 
full matrix element calculations (eventually combined with a parton shower) to be 
completely taken into account \cite{Catani:2001cc,Nagy:1997fk}.  
In Fig.\ \ref{fig:LEP14jangles}, the predictions for the Bengtsson-Zerwas 
\cite{Bengtsson:1988qg} and the Nachtmann-Reiter \cite{Nachtmann:1982xr} angle are 
compared with \Delphi data \cite{Hendrik:2003} for events with four jets at a 
jet resolution $y^{\rm Durham}_{\rm cut}=0.008$.  Both results agree surprisingly 
well with data. A similar level of agreement is observed for the other two 
prominent four-jet angles, $\alpha_{34}$ and the K\"orner-Schierholz-Willrodt 
angle, that are not shown here.

\subsubsection{Jet rates in heavy-quark production}\label{sec:applhQ}

\begin{figure*}[t!]
\begin{picture}(250,250)
\put(0,0){\includegraphics[width=230pt]{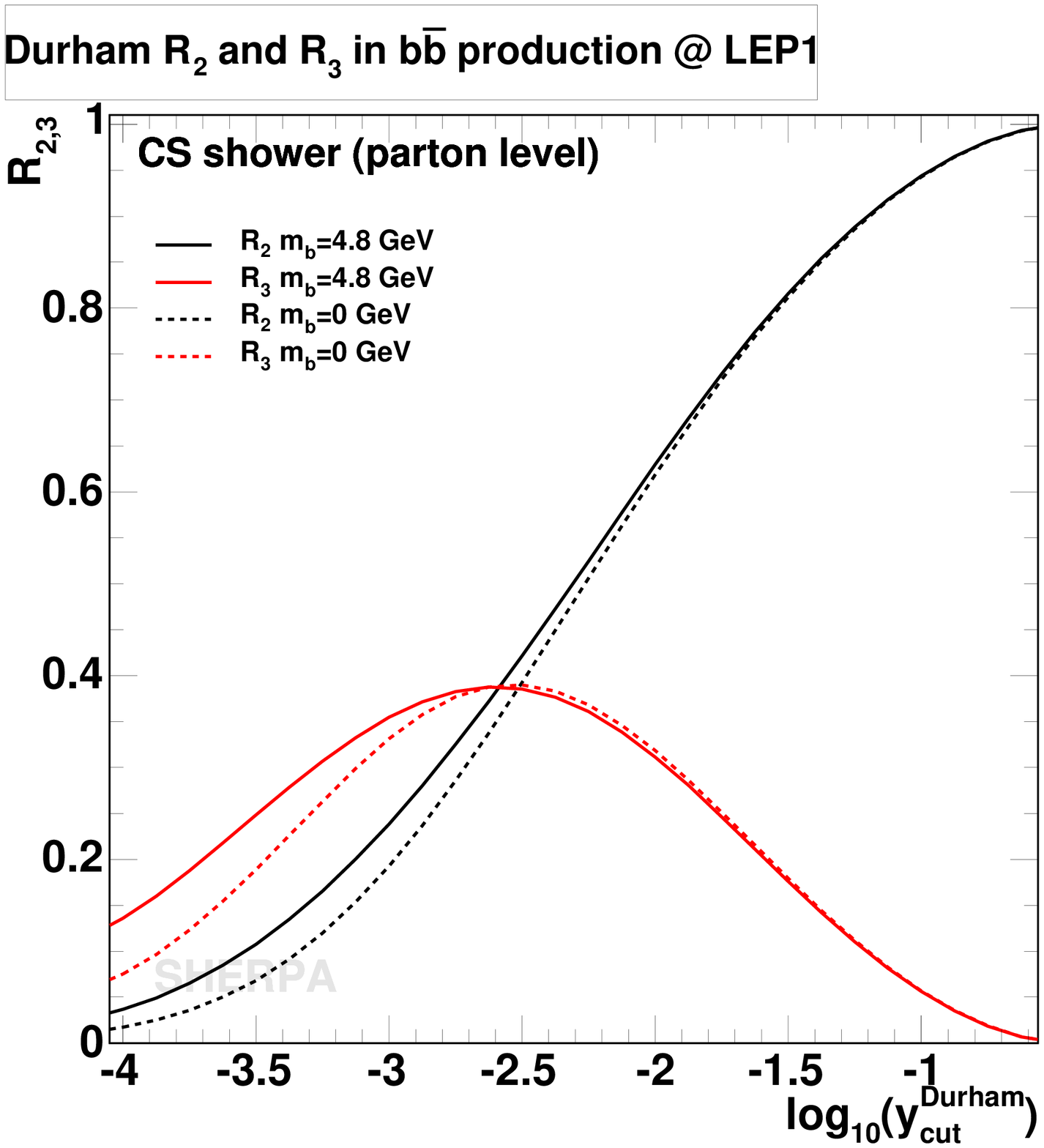}}
\put(250,0){\includegraphics[width=230pt]{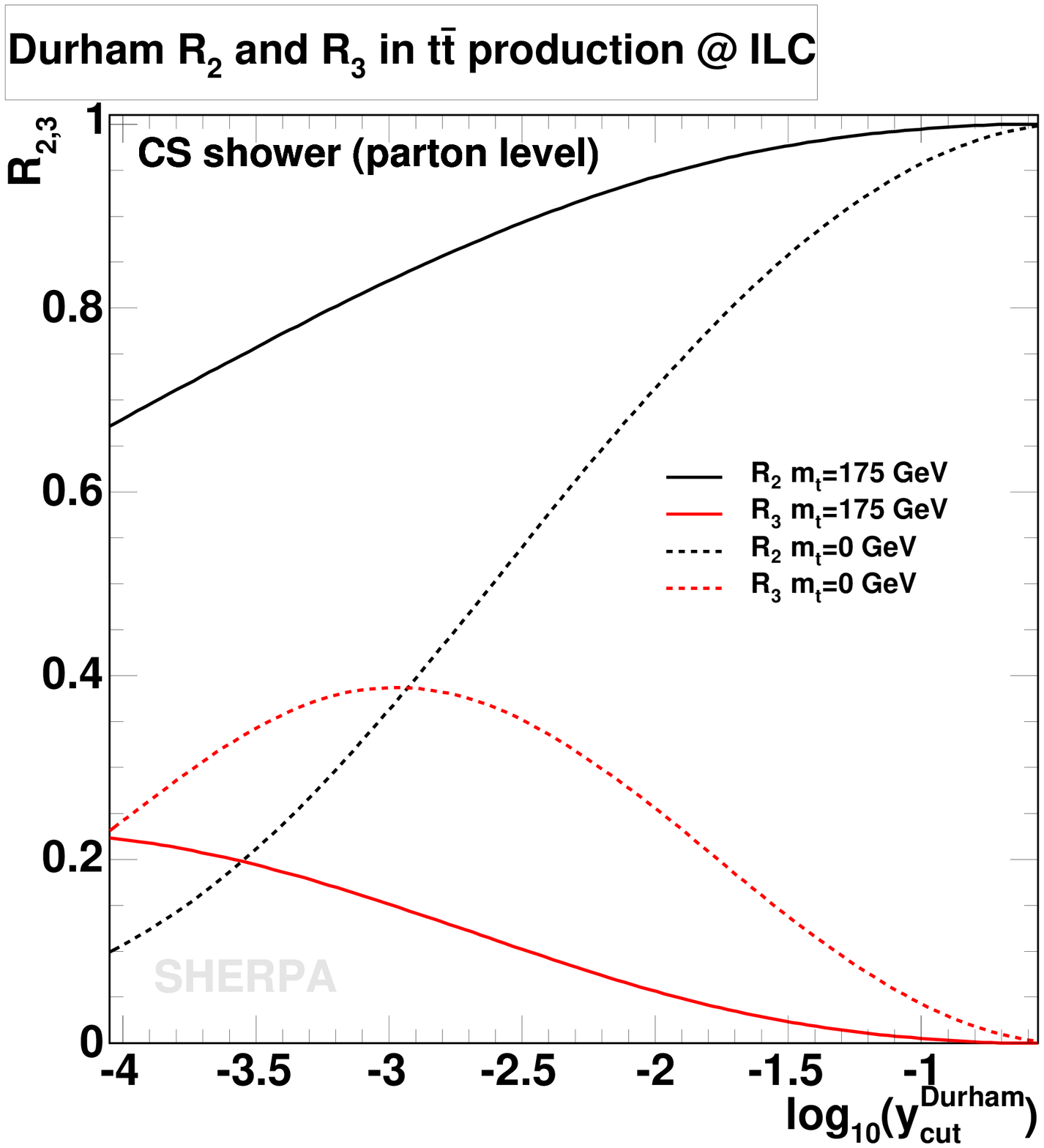}}
\end{picture}
\caption{\label{fig:hQR23}
  The exclusive Durham two- and three-jet rates in inclusive 
  $b\bar b$ production at LEP1 (left panel) and $t\bar t$ production at a 
  $500$ GeV linear collider (right panel). The solid curves correspond to 
  fully taking into account the quark masses in the parton shower simulation
  while for the dashed predictions the finite masses have been neglected. 
}
\end{figure*}

The leading order of heavy-quark production at lepton colliders also proceeds 
through an intermediate $\gamma^*$ or $Z^0$ in the $s$-channel.  Since pair 
production of top-quarks was outside the kinematical reach of LEP, only the production of 
bottom-quarks is available at these energies to discuss the treatment of heavy quarks 
in the new parton shower algorithm.  At a future international linear collider (ILC), 
operating at or around $\sqrt{s}=500$ GeV, pair production of top-quarks will play a 
key-r{\^o}le in the physics programme.  This is also true for the LHC where top-quarks 
will copiously be produced and constitute a major background in nearly all searches 
for new physics.  Therefore, a correct description of the radiation pattern of heavy 
quarks will be of enormous importance.  As already hinted at in Sec.\ 
\ref{sec:kinematicsFFmassive}, radiation off massive quarks is suppressed with 
respect to the case $m_Q=0$, also known as ``dead-cone''-effect 
\cite{Ellis:1991qj}.  The impact is however rather small when considering $b$-quark 
masses of $4.8$ GeV at collider energies that are much larger.  To illustrate the
impact of the finite $b$-quark mass in the shower approach the Durham two- and 
three-jet rates for $b\bar b$-production at LEP1 are presented in the left panel of 
Fig.\ \ref{fig:hQR23}.  There, results are shown for the fully massive case (\ie the 
mass has fully been taken into account in the splitting kernels, the phase-space 
boundaries and the splitting kinematics) and for the massless case are depicted.  As 
expected, in the massive case both $R_2$ and $R_3$ are slightly enhanced at low values 
of $y^{\rm Durham}_{\rm cut}$, corresponding to the suppressions of additional 
radiation that turns a two-jet event into three-jet and a three-jet into a four-jet 
event at the scale of the emission.
 
In the right panel of Fig.\ \ref{fig:hQR23} the same observables are presented, but this
time for the pair-production of $175$ GeV top-quarks at a $500$ GeV ILC. Obviously, the 
finite mass has to be taken into account in the description of QCD radiation off 
top-quarks, since the differences with respect to the massless case can exceed an 
order of magnitude for the two-jet rate.

\subsection{Particle production in hadron collisions}
\label{sec:applpp}

With the advent of the LHC era, the description and simulation of particle production 
processes at hadron colliders gained even more relevance.  Due to the colour-charged 
partonic initial states, {\it every} hard process at hadron colliders is accompanied by 
initial- and subsequent final-state radiation.  In the following, only two examples 
shall be considered to highlight the performance of the new parton shower model in
such situations.  First, the inclusive production of Drell-Yan lepton pairs, the simple 
most process that features initial-state emitter -- initial-state spectator dipoles, 
and, second, QCD jet production are discussed.  For the latter, besides looking at 
some inclusive two-jet distributions, three-jet observables sensitive to the inclusion 
of QCD colour coherence are considered and qualitatively compared with data. 

For all the predictions presented below, the CTEQ6L set of PDFs \cite{Pumplin:2002vw} has
been used, the strong coupling constant has been fixed to $\as(m_Z)=0.118$ with its 
running taken at two-loop level, in accordance with the choice in the PDF, and the 
infrared cut-off of the shower is chosen to be $\kperpzero = 2$ GeV.  Hadronisation of 
the partonic shower final states is again accomplished by an interface to the Lund 
string routines of \Pythia 6.2 \cite{Sjostrand:2001yu}.

\subsubsection{Inclusive gauge boson production}
\label{sec:applDY}

The production of electroweak gauge bosons, \eg $W^\pm$ and $Z^0$ bosons, and their
subsequent decay into leptonic final states, is one of the most prominent processes at 
hadron colliders due to their clean signature.  Although very interesting in their own
right, their inclusive production, \ie their production together with additional QCD 
jets, represents a serious background to many other interesting processes, like, \eg 
the production and decay of top-quarks or SUSY particles.  Therefore, many theoretical 
efforts have been undertaken to predict gauge boson production as precisely as possible, 
both at fixed order in the strong coupling, see for instance 
\cite{Altarelli:1979ub}-\cite{Anastasiou:2003ds}, or focusing on the analytical 
resummation of large logarithms from soft gluon emissions, see for example 
\cite{Gonsalves:1989ar}-\cite{Gonsalves:2005ng}.  An important ingredient in all cases 
have been parametrisations of the PDFs and a good perturbative control over their 
scaling behaviour, which by now is known at the three-loop level \cite{Vogt:2004mw}.
In addition, in the past few years, Drell-Yan production formed the testbed for approaches 
aiming at the combination of tree-level matrix elements with parton shower Monte Carlos 
\cite{Krauss:2004bs,Lavesson:2005xu,Alwall:2007fs,Mrenna:2003if,Krauss:2005nu}.
Parton shower Monte Carlos thereby have to deliver the correct description for the bulk 
of the events where the bosons are accompanied by rather soft emissions only. 

In the following, Drell-Yan production of $\gamma^*/Z^0$ at Tevatron Run I energies is 
considered with the bosons decaying into $e^+e^-$-pairs.  They are constrained to fall 
into a mass-window of $66\;{\rm GeV} < M_{e^+e^-} < 116\;{\rm GeV}$.  The predictions of 
the new shower algorithm will directly be compared to results obtained with the matrix 
element-parton shower merging approach as implemented in \Sherpa.  To this end, an 
inclusive sample combining matrix elements for no extra emission and one extra 
final-state QCD parton has been generated with \Sherpa version 1.0.10.  In the
figures this sample will be denoted by ``\Sherpa 1.0.10 CKKW (0+1 jet ME)''.  

\begin{figure*}[t!]
\begin{picture}(250,250)
\put(0,0){\includegraphics[width=230pt]{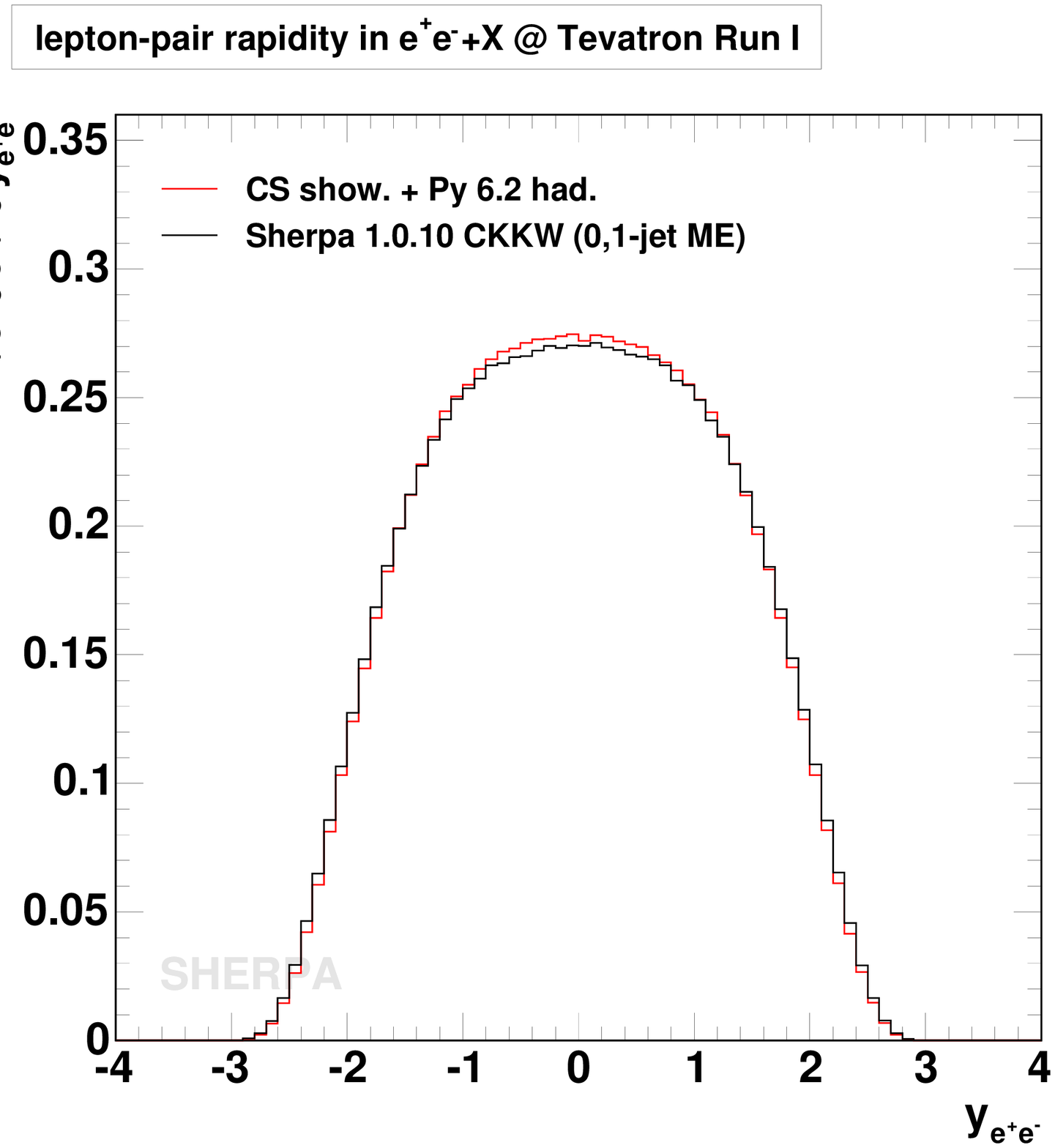}}
\put(250,0){\includegraphics[width=230pt]{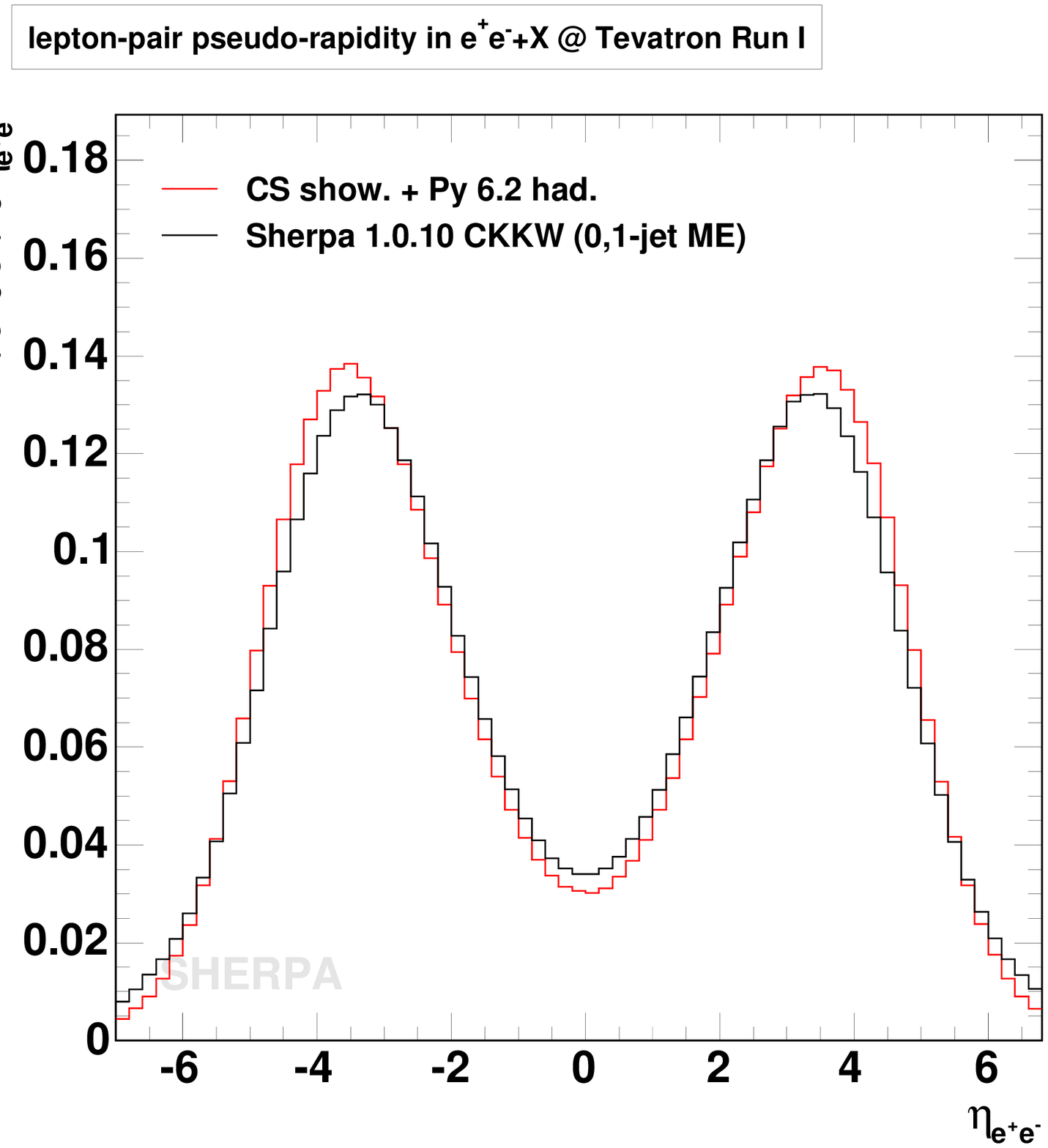}}
\end{picture}
\caption{\label{fig:DYTeVrapidities}
        The rapidity- (left panel) and pseudo-rapidity (right panel) 
        distribution of $e^+e^-$ Drell-Yan pairs produced in $p\bar p$ 
        collisions at $\sqrt{s}=1800$ GeV.}
\end{figure*}

The discussion of the results starts with the rapidity- and pseudo-rapidity distributions 
of the produced lepton-pair, see Fig.\ \ref{fig:DYTeVrapidities}.  As the shape of the 
former is already described well at the leading order, \ie without any radiation,
there is hardly any difference visible for the two results.  The gauge boson 
pseudo-rapidity distribution however, only emerges when there is some additional QCD 
radiation.  The radiation pattern, and especially the hardest emission, determines this 
leptonic observable.  The pure shower result is in excellent agreement with the merged 
result, which contains the exact tree-level matrix element for the first hard emission.  
However, the shower distribution is somewhat lower at central pseudo-rapidity and 
slightly exceeds the merged \Sherpa result for the two maxima around 
$\eta_{e^+e^-}\approx \pm4$.  These differences can be traced back to the lack of 
sufficiently hard radiation in the shower, which is constrained from above though the 
default shower start scale for this process, namely the invariant mass squared of the 
initial dipoles, $M^2_{e^+e^-}$.  Below that scale, however, the parton shower can be 
expected to deliver reliable results, and in order to fill the phase space above that
scale, matrix element--parton shower merging techniques should be added.  

\begin{figure*}[t!]
\begin{picture}(250,250)
\put(0,0){\includegraphics[width=230pt]{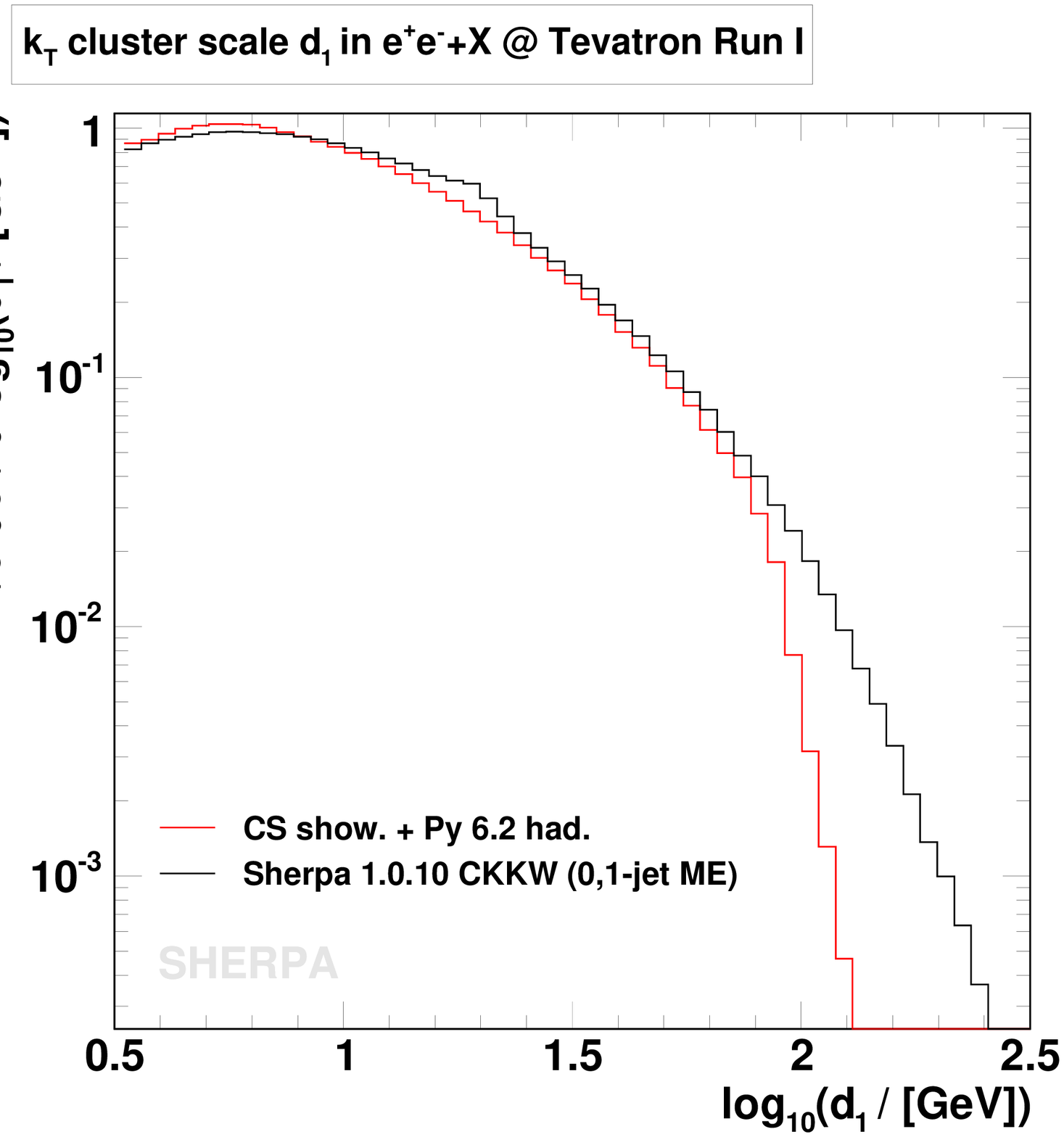}}
\put(250,0){\includegraphics[width=230pt]{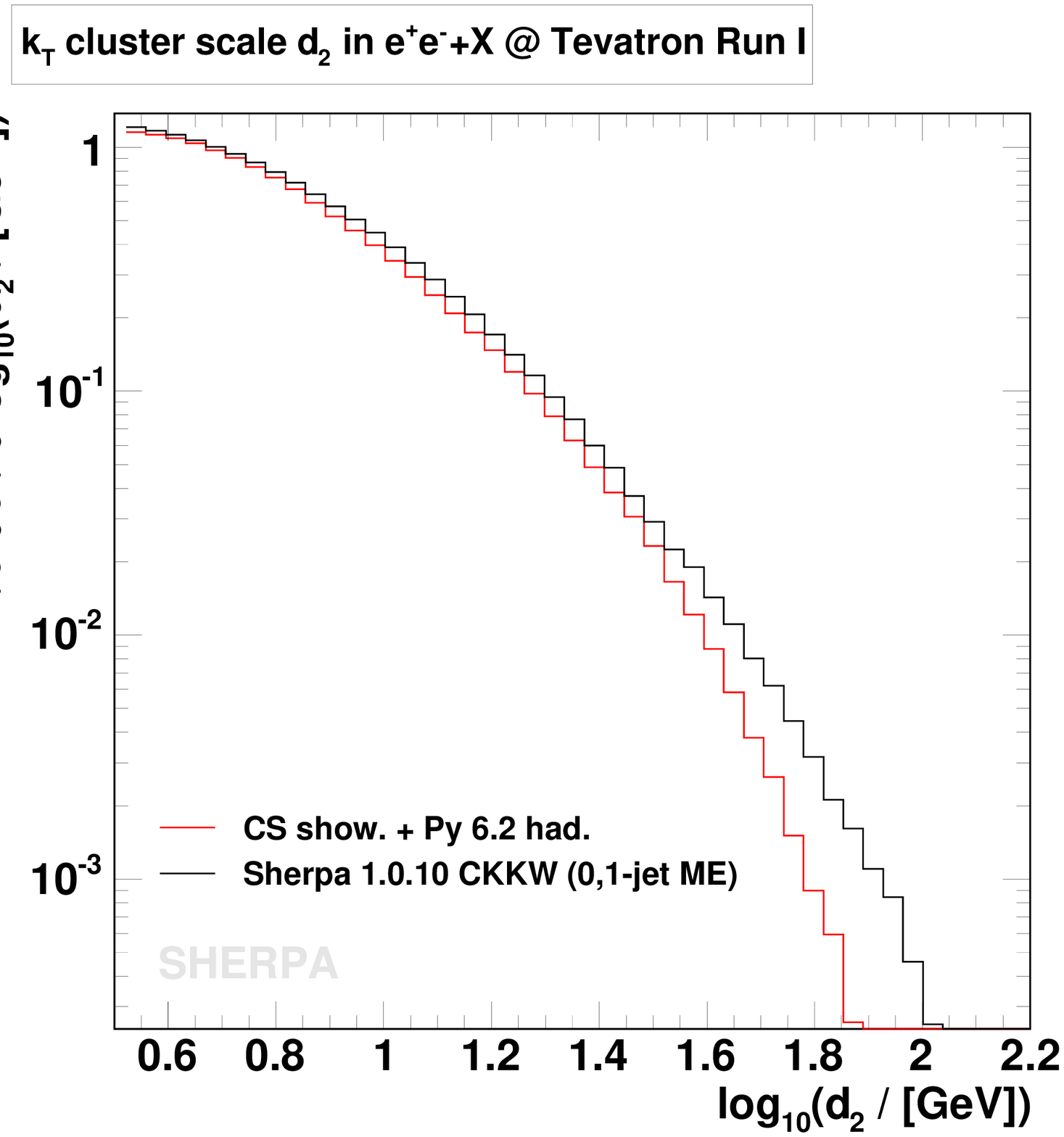}}
\end{picture}
\caption{\label{fig:DYTeVdiffrates} 
        The $k_\perp$ differential jet rates $d_1$ and $d_2$ in $e^+e^-+X$ production
        at Tevatron Run I.}
\end{figure*}

The smaller amount of hard radiation can be further quantified by looking at the 
differential jet rates $d_1$ and $d_2$ for the $\kperp$-jet algorithm \cite{Catani:1993hr},
displayed in Fig.~\ref{fig:DYTeVdiffrates}.  These observables determine the scales
where the first ($d_1$) and second ($d_2$) additional parton gets resolved as a jet 
from the core process.  The results for the Catani-Seymour based shower and the merged 
\Sherpa sample agree well for small cluster scales but, as can be expected, the shower 
is significantly lower for values of $d_i>m_Z$.\\

\begin{figure*}[t!]
\begin{center}
\includegraphics[width=230pt]{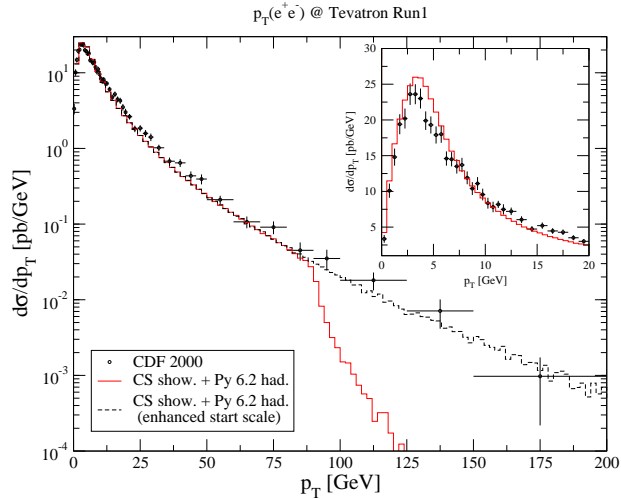}
\end{center}
\caption{\label{fig:DYTeVptZ} 
        The $p_\perp$ distribution of $e^+e^-$ Drell-Yan pairs in comparison 
        with data from CDF at the Tevatron, Run I \cite{Affolder:1999jh}.}
\end{figure*}

The last observable to be considered is the transverse momentum distribution of the
lepton-pair.  This distribution has been measured with high precision by the Tevatron 
experiments.  Like the Drell-Yan pseudo-rapidity it is very sensitive to both 
soft and hard radiation accompanying the produced boson.  Fig.\ \ref{fig:DYTeVptZ}
contains a comparison of the prediction from the new shower model with a CDF 
measurement \cite{Affolder:1999jh} 
\footnote{A comparison of the merged \Sherpa prediction with this data has already 
        been presented in \cite{Krauss:2004bs}.}. 
The agreement between data and simulation is quite good up to $p_T$'s of approximately
$80$ GeV.  The upper-right part of Fig.~\ref{fig:DYTeVptZ} contains a blow-up of the 
low transverse-momentum region of $p_T < 20$ GeV, this time, however, on a linear scale.  
There, the parton shower describes the turn-on of the distribution quite nice, the 
actual peak, however, is slightly higher and a bit broader than seen in data.  To 
describe the very low transverse-momentum region a Gaussian-smeared intrinsic $\kperp$ 
was introduced, with a mean of $0.52$ GeV and a width of $0.8$ GeV.  A more detailed 
tuning of these values combined with the shower cut-off $\kperpzero$ may yield an even 
better description of the distribution's peak. Above $80$ GeV the parton shower dies 
off very rapidly due to its phase space being constrained by the choice of the 
starting scale, $\kperpmax^2 = M^2_{e^+e^-}$.  For illustrative purposes a prediction 
has been added where the start scale has been enhanced to $4M^2_{e^+e^-}$.  While the 
results at low $p_T$ do not change significantly, the distribution continues in the 
tail, thereby following the experimental data.  But, of course, with this choice of 
parton shower starting scale, there is a similar drop-off of the distribution at 
scales of around $4M^2_{e^+e^-}$.  However, since there is no 
guarantee that the parton shower kernels do perform well enough at large scales, \ie 
outside the soft- and collinear phase-space regions, it seems to be overly optimistic to
stretch its predictions to such high scales.  Instead, the parton shower description 
should consistently be improved by incorporating exact higher-order corrections.

\subsubsection{Inclusive jet production}
\label{sec:appljets}

The most obvious QCD production process to look for at hadron colliders is inclusive
jet production.  However, from a theoretical point of view this is quite a complicated 
process.  Besides tree-level calculations for up-to six final-state jets, so far, 
there merely exist full next-to-leading order results up-to three-jet production 
\cite{Nagy:2001fj}, \cite{Kunszt:1992tn}-\cite{Trocsanyi:1996by}.  Despite of 
strong efforts, culminating in evaluating the complete set of necessary matrix elements 
\cite{Bern:2000dn}-\cite{Bern:2002tk} and in developing methods to isolate the 
infrared divergences in the real correction part \cite{Daleo:2006xa} a full NNLO 
calculation for inclusive jet production has not been finished yet.  Also, from the 
point of view of the parton shower presented here, jet production at hadron colliders 
is rather involved.  This is because the $2\to 2$ hard process will contain all possible 
colour connections between initial-state and final-state partons.  Hence, QCD 
jet production constitutes a severe test of the entire shower algorithm.  The input 
parameters for the simulations have been chosen as specified above.  The 
starting scale of the shower, however, is related to the transverse momentum of 
the $2 \to 2$ core process' outgoing partons, namely $\kperpmax^2=p^2_{\perp,j}$.\\

\begin{figure*}[t!]
\begin{center}
\includegraphics[width=230pt]{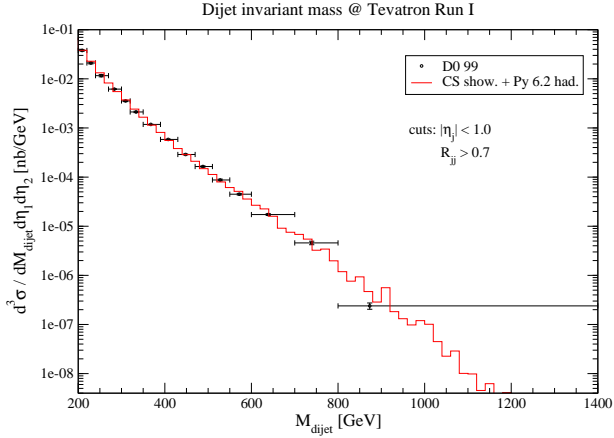}
\end{center}
\caption{\label{fig:dijetmass}Dijet mass $M_{dijet}$ measured by \Dzero at Tevatron Run I
        \cite{Abbott:1998wh}.}
\end{figure*}

The first thing to be looked at is a very inclusive quantity, the dijet invariant 
mass.  This has been measured by \Dzero during Run I \cite{Abbott:1998wh}.  The jets 
considered there have been reconstructed using a jet-cone algorithm with a cone opening 
angle of $R=0.7$ in the $\eta-\phi$ space and with jet transverse energies above $30$ 
GeV.  Dijet candidates have then been subjected to the requirement that both jets
satisfy $|\eta_j| < 1.0$.  Fig.\ \ref{fig:dijetmass} exhibits the resulting dijet-mass 
distribution starting at $M_{\rm dijet} > 200$ GeV.  It is a very steeply falling 
spectrum spanning six orders of magnitude in the mass range under consideration.  To 
compare with data the result of the (leading order) simulation has been normalised to 
the cross section observed in experiment.  In fact, the prediction of the proposed 
shower algorithm then is in very good agreement with the data and almost everywhere 
exactly hits the weighted bin centers.\\

\begin{figure*}[t!]
\begin{center}
\includegraphics[width=230pt]{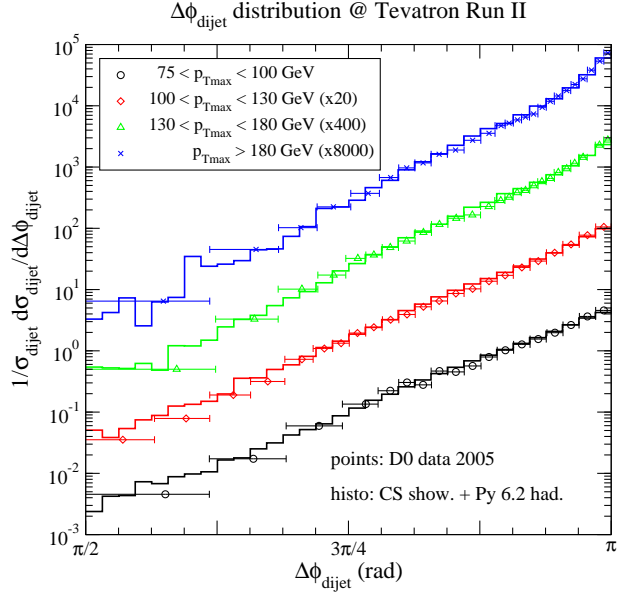}
\end{center}
\caption{\label{fig:dijetphi}Azimuthal decorrelation in dijet events measured by 
        \Dzero at Tevatron Run II \cite{Abazov:2004hm}.}
\end{figure*}

Another interesting observable when studying dijet events is the azimuthal angle between 
the two highest-$p_T$ jets.  If there is no additional QCD radiation the two jets have 
equal transverse momenta and they are oriented back-to-back.  Thus, in this case, their 
azimuthal separation $\Delta \phi_{\rm dijet} = |\phi_1 - \phi_2|$ equals $\pi$.  In the 
presence of merely soft radiation the azimuthal angles remain strongly correlated,
the strength of the decorrelation rises with the presence of additional hard radiation.  
Therefore, the dijet decorrelation provides a testbed for soft- and hard QCD emissions 
without the necessity to reconstruct further jets.  Fig.\ \ref{fig:dijetphi} contains 
the results of a recent \Dzero measurement for cone jets found for $R=0.7$ 
\cite{Abazov:2004hm}.  The data fall into different ranges of the leading-jet 
transverse momentum and are then multiplied with different constant prefactors in order 
to display them in one plot.  In all cases, the second-leading jet was required to have 
a transverse momentum $p_T > 40$ GeV and both jets are constrained to the 
central-rapidity region, $|y_j| < 0.5$.  The data are overlayed with the respective 
predictions of the Catani-Seymour dipole shower approach.  The simulation agrees very 
well with the data over the whole interval of $\Delta \phi_{\rm dijet}$ spanned by the 
experimental measurements. This is a very satisfying result as it proves that the 
proposed shower formulation not only correctly accounts for phase space regions related 
to soft and collinear radiation but also yields qualitatively and quantitatively correct 
estimates for rather hard emissions as well.  Furthermore, since this observable is 
quite sensitive to model-intrinsic scale choices such as the shower start scale and 
scales entering the running coupling constant and parton density functions, this 
agreement proves that the defaults have been chosen correctly.\\

The last item to be discussed are observables in QCD jet production at hadron
colliders that are known to be sensitive to the correct treatment of QCD soft
colour coherence in the parton shower simulation. Colour-coherence effects
have been widely studied for $e^+e^-$ collisions, for an early review see e.g.\
\cite{Khoze:1996dn}.  They manifest themselves in the fact that soft emissions 
are forbidden outside a certain angular cone around the emitting particle's 
direction, known as angular ordering \cite{Ellis:1991qj,Marchesini:1987cf}. 
To account for this in shower Monte Carlos the phase space for allowed emissions 
has to be properly constrained. Within the \Herwig Monte Carlo for instance this 
is realised by evolving the shower in terms of cone-opening angles. While the 
situation for pure final-state showers is quite clear, in hadronic collisions the 
situation is slightly more complicated due to the presence of more colour flows,
among them those that connect initial- and final-state partons. As colour-coherence 
here already influences the first emission from the initial- and final-state 
partons QCD three-jet events are the best place to look for the pattern of these 
phenomena at hadron colliders.

In one of the pioneering studies \cite{Abe:1994nj} three-jet events that 
feature a hard leading jet and a rather soft third jet have been considered.  
Observables potentially sensitive to colour coherence are spatial correlations 
between the third jet and the leading ones. In \cite{Abe:1994nj} such 
discriminating variables have been found and by comparison with Monte Carlo 
simulations evidence for the observation of colour coherence in hadron collisions
has been provided.  This ultimately has led to a refinement of the \Pythia shower 
algorithm in order to appropriately model colour coherence in the spirit of 
\cite{Bengtsson:1986et}.  In the CDF study \cite{Abe:1994nj} jets have been 
defined through a cone algorithm with a cone radius of $R=0.7$ and the following 
event selection criteria have been applied:
\begin{itemize}
\item For the two leading jets the pseudo-rapidity is constrained to 
$|\eta_1| < 0.7$ and $|\eta_2| < 0.7$;  
\item they have to be back-to-back within $20$ degrees in 
  the transverse plane, corresponding to $|\phi_1-\phi_2| > 2.79$ radian; 
\item and the transverse energy of the leading jet, $E_{T1}$, has to 
  exceed $110$ GeV, the third jet is required to have $E_{T3} > 10$ GeV.
\item Only for the study of the $\alpha$ variable defined below the additional 
 cut $1.1 <\Delta R_{23} <\pi$, where 
 $\Delta R_{23} = \sqrt{(\eta_2-\eta_3)^2+(\phi_2-\phi_3)^2}$, is imposed.
\end{itemize}
A number of observables has been considered, the two most convenient and 
discriminating ones have been the pseudo-rapidity distribution of the third jet, 
$\eta_3$, and the polar angle in the space parametrised by 
$\Delta\phi = \phi_3 - \phi_2$ and  
$\Delta H = \mbox{\rm sign}(\eta_2)(\eta_3-\eta_2)$, 
namely $\alpha = \arctan(\Delta H/|\Delta\phi|)\;$\footnote{
	A further observable considered in the CDF study is the spatial 
        separation of the second- and third jet in the $\eta-\phi$ space, 
	$\Delta R_{23}$.  This
	observable, however, seems to be less discriminatory between 
	theoretical models.  In addition, and more importantly, detector effects
	seem to have a larger impact on its discriminating power. Therefore 
	it is not taken into account here.}. 
\begin{figure*}[t!]
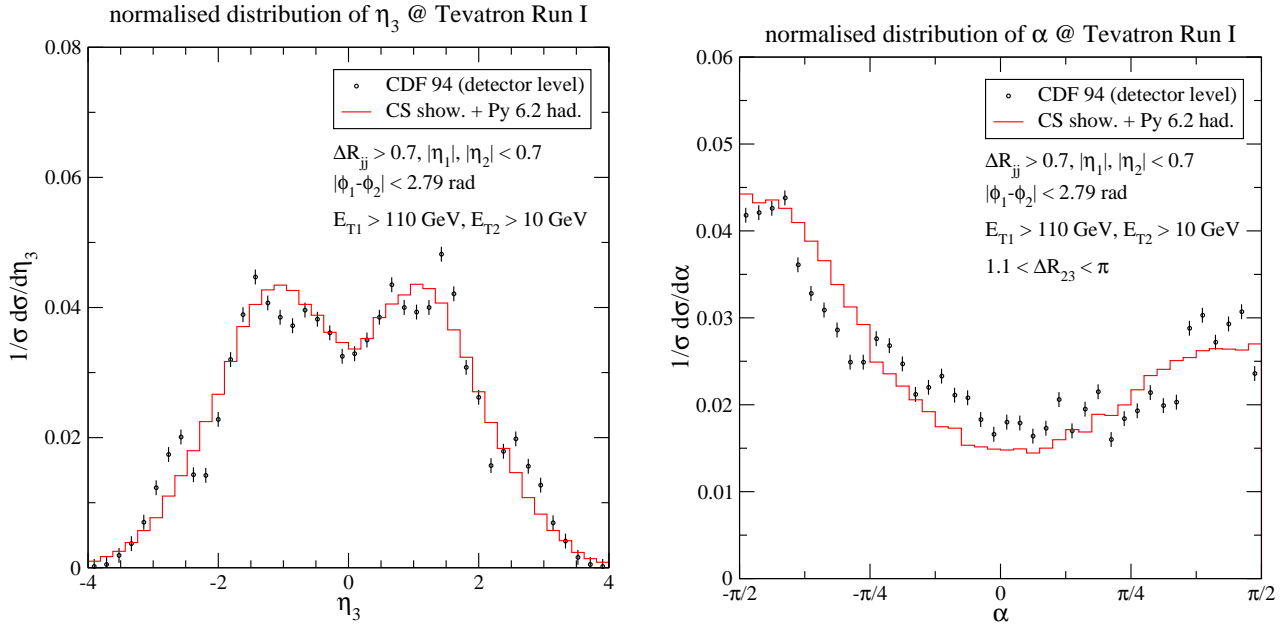

\begin{picture}(250,250)
\put(0,0){\includegraphics[width=230pt]{figures/eta3_ch_CDF.eps}}
\put(250,0){\includegraphics[width=230pt]{figures/alpha_ch_CDF.eps}}
\end{picture}
\caption{\label{fig:QCDjets_ch} The pseudo-rapidity distribution of the 
        third-hardest jet (left panel) and the distribution of the angle 
        $\alpha$ (right panel) in inclusive QCD three-jet production in 
        comparison with CDF data taking during Tevatron Run I. Experimental
        errors are statistical only. Histograms are normalised to one.}
\end{figure*}
It should be stressed that the published results, used for the comparison, 
are not corrected for detector effects, such as finite resolution and 
uninstrumented regions, and therefore can only qualitatively be compared with 
theoretical calculations.  In contrast, the results of the Monte Carlo simulations 
exhibited in \cite{Abe:1994nj} have passed the full chain of the CDF detector 
simulation.  In Fig.~\ref{fig:QCDjets_ch} the measurements are compared with 
simulated events at the hadron level.  In the left panel the $\eta_3$ distribution 
is shown and the right panel contains the comparison of the $\alpha$ distribution.  
Both predictions agree well with the data. The $\eta_3$ distribution tends to be 
broader in models that take into account colour-coherence effects and only then 
theoretical calculations show the significant dip around $\eta_3 \approx 0$ seen 
in data \footnote{
	The \Herwig Monte Carlo, incorporating colour coherence through explicit 
	angular ordering, describes the data very well.  Switching on the approximate 
	version of angular ordering in \Pythia, realised by a veto on rising opening 
	angles during shower evolution, significantly improves \Pythia's agreement 
	with data.}. 
The $\alpha$ variable is also very sensitive to the inclusion of colour coherence. 
It decreases from $\alpha = -\pi/2$ to $\alpha=0$ but then the slope changes and the 
distribution rises as $\alpha \to \pi/2$. This trend is clearly seen for the 
simulation with the new shower algorithm.  Models not taking into account coherence 
fail to describe the distribution's raise towards $\alpha \to \pi/2$ and have a 
clear excess of events at small $|\alpha|$.  Concerning the interpretation of these 
results the missing detector smearing for the shower simulation has to be kept in mind. 
However, in Ref.~\cite{Abe:1994nj} estimates for the size of the detector effects
are given, showing that the impact of the finite detector resolution is much smaller 
than the size of the physical effects.  The generic features of the two observables 
presented here are not dependent on detector effects, and they are well described 
by the new shower formulation. 

The conclusion of this is that the proposed parton shower algorithm with its notion
of emitter--spectator dipoles associated with the color flow of the event and using 
transverse momenta as evolution variable accounts for soft colour coherence and yields 
a very satisfying description, both on the qualitative and the quantitative level. 
It can be anticipated that such non-trivial quantum phenomena are of large importance
at the LHC, since the phase space for jet production is much larger and hard jets are 
produced copiously.  For a solid description of QCD therefore the systematic and 
correct inclusion of these effects is paramount.

%% file: cssummary.tex
\section{Conclusions and outlook}
\label{sec:cssummary}

In this publication a parton shower model based on Catani-Seymour dipole 
subtraction kernels has been presented, which was proposed for the first 
time in \cite{Nagy:2005aa,Nagy:2006kb}.  In the present implementation, 
the original proposal is extended to cover also initial-state splittings, 
finite parton masses, and QCD radiation off SUSY particles.  

Choices concerning the evolution parameter of the parton shower and the 
various scales entering running coupling constants, PDFs, etc.\ have been 
detailed, fixing the full algorithm.  The kinematics of massive splittings 
has been presented in some detail, and the corresponding massless limits 
have been discussed.  By direct comparison with some benchmark processes, 
at first order in $\as$, the differences of the parton shower approximation 
with respect to exact results have been worked out.  It has been shown that 
indeed the parton shower algorithm presented here reproduces the soft and 
collinear limits of the exact matrix elements and that differences between 
both results are non-singular terms only.  Some first results with this 
new parton shower formulation have been presented and show very encouraging 
agreement with other models and with experimental data.  

In the near future, this new algorithm will be fully incorporated into the 
\Sherpa framework and it will be made publicly available in the next releases 
of the code.  This will also involve a more careful tuning of the shower 
parameters and the inputs of the hadronisation models provided by or linked 
to \Sherpa, which surely will further improve the agreement with data.  
Planned is a detailed comparison against another new shower ansatz that is 
based on splitting colour dipoles \cite{adicic}, and that is also being 
developed in the \Sherpa framework at present.  In addition, a full merging 
with multi-leg matrix elements in the spirit of \cite{Catani:2001cc} will 
be implemented.  It can furthermore be anticipated that this new shower 
implementation will lend itself to incorporation of MC@NLO-techniques 
\cite{Frixione:2002ik,Nason:2006hf}.

%% file: csacknowledgments.tex
\section*{Acknowledgments}
\label{sec:acknowledgments}

\noindent
We would like to thank Davison Soper and Zoltan Nagy for inspiration and fruitful 
discussions.  We are indebted to the other members of the \Sherpa-team, and in 
particular Jan Winter, Tanju Gleisberg and Stefan H\"oche for helpful conversation.  
We are grateful to Keith Hamilton and Michael Kobel for carefully reading the 
manuscript.  

\noindent
Financial support by MCnet (contract number MRTN-CT-2006-035606) and BMBF is 
gratefully acknowledged.  

\noindent
Plots have been generated using the \Compare tool \cite{COMPARE} based on \Root 
\cite{Brun:1997pa}.  Feynman diagrams have been drawn using the packages feynMF 
\cite{Ohl:1995kr} and {A\scalebox{0.9}{XODRAW}\xspace} \cite{Vermaseren:1994je}.

\newpage